\documentclass[11pt]{article}

\usepackage{mathrsfs}
\usepackage{amsmath,latexsym,epsf}
\usepackage{epsfig,amsfonts,cite}
\usepackage{amssymb}

\begin{document}

\title{Critical dynamics in thin films}

\author{A.~Gambassi and S.~Dietrich \\
  {\small\it Max-Planck Institut f\"ur Metallforschung,}  \\
  {\small\it Heisenbergstr.~3, D-70569 Stuttgart, Germany,}  \\
  {\small\it and Institut f\"ur Theoretische und Angewandte Physik,
Universit\"at Stuttgart,} \\
  {\small\it Pfaffenwaldring 57, D-70569 Stuttgart, Germany}   \\
  {\small {\tt e-mail: gambassi@mf.mpg.de, dietrich@mf.mpg.de}}\\
}

\maketitle
\thispagestyle{empty}
\vspace{0.2cm}
\begin{abstract}
Critical dynamics in film geometry is analyzed
within the field-theor\-etical approach. In particular we consider the
case of purely relaxational dynamics (Model A) and Dirichlet
boundary conditions, corresponding to the so-called ordinary surface
universality class on both confining boundaries. The general
scaling properties for the 
linear response and correlation functions and for dynamic Casimir
forces are discussed. Within the Gaussian
approximation 
we determine the analytic expressions for the
associated universal scaling functions and study quantitatively 
in detail their qualitative
features as well as their 
various limiting behaviors close to the bulk critical point.
In addition we consider the effects of time-dependent fields on the
fluctuation-induced dynamic Casimir force and determine analytically
the corresponding 
universal scaling functions and their asymptotic behaviors for two
specific instances of instantaneous perturbations.
The universal aspects of
nonlinear relaxation from an initially ordered
state are also discussed emphasizing the different crossovers that
occur during this evolution. 
The model considered is relevant to the critical dynamics of actual
uniaxial ferromagnetic films with symmetry-preserving conditions at
the confining surfaces and for Monte Carlo simulations of spin system
with Glauber dynamics and free boundary conditions.
\\[1cm]
Dynamic critical phenomena -- confined geometry -- finite-size scaling
-- magnetic properties of films -- critical Casimir force
\end{abstract}

\clearpage

%
%

\newcommand{\dd}{{\rm d}}

\newcommand{\reff}[1]{(\ref{#1})}
\newcommand{\ve}[1]{{\bf #1}}
\newcommand{\fss}{{\sc FSS}}

\newcommand{\pe}[1]{#1_{\bot}}
\newcommand{\pa}[1]{#1_{\|}}
\newcommand{\spint}[1]{\ensuremath{\int {\rm d}^{#1}\ve{x} \, {\rm d}t}}
\newcommand{\imint}[2]{\ensuremath{\int \! \frac{ {\rm
        d}^{#1}\ve{#2}}{(2\pi)^{#1}} \, \frac{{\rm d}\omega}{2\pi}}} 
\newcommand{\G}[2]{\ensuremath \Gamma_{#1\,#2}} 
\newcommand{\rt}{\ensuremath T}
\newcommand{\m}{\ensuremath \mathfrak{m}}
\newcommand{\h}{\ensuremath \mathfrak{h}}
\newcommand{\oo}{\ensuremath \mathscr{O}}
\newcommand{\CC}{\ensuremath \mathscr{C}}
\newcommand{\PP}{\ensuremath \mathcal{P}}
\newcommand{\RR}{\ensuremath \mathscr{R}}
\newcommand{\TT}{\ensuremath \mathscr{T}}
\newcommand{\UU}{\ensuremath \mathscr{U}}
\newcommand{\FF}{\ensuremath \mathscr{F}}
\newcommand{\sn}{\mbox{sn}}
\renewcommand{\th}{\mbox{th}}

\renewcommand{\Im}{{\rm Im}}

\section{Introduction}
\label{preintro}
The microscopic understanding of collective dynamic phenomena in
condensed matter poses
one of the most 
difficult challenges for 
statistical physics. 
Accordingly the theory of these phenomena is in a significantly less
mature state than for static properties in thermal equilibrium; also
the corresponding experimental knowledge is very limited.
At present theoretical insight into collective dynamics can be gained
either by simulations or
by studying numerically or analytically 
rather simplified models of actual condensed matter systems. 
Whereas in the former case the limitations on system size and time
scales are very severe, in the latter case 
one has to
be careful in accounting within the model for all those aspects of the actual
systems 
which are relevant for the collective behavior under study.
Understanding the link between the microscopic physical parameters of
the system and those defining the effective dynamic models is also a
crucial issue. 
Due to these difficulties 
only in few cases one is able to provide theoretical predictions that
can be {\it quantitatively} compared with experiments -- provided that those
can be carried out in the first place. Nevertheless
there are instances in which a {\it universal} collective
behavior emerges which is largely independent of the microscopic
details of the system and, as a consequence, also 
of the specific model used to describe it. These highly valuable
circumstances arise  naturally upon approaching a critical point,
where the system undergoes a 
continuous, i.e., second-order phase transition. For universal
critical properties such as critical exponents, scaling functions, and
amplitude ratios, 
one is usually able to provide theoretical
predictions that can be tested {\it quantitatively} by comparison 
with experimental data.
In view of the universality of the critical properties, which is
justified by the framework of renormalization-group theory and
supported by experimental evidence,
it is possible to study the collective behavior in
terms of suitable field-theoretical models, 
based on minimalistic fixed-point equations of motions following from
Landau type fixed-point Hamiltonians.
This approach has been carried out successfully during the last
decades in order to study static and dynamic
critical properties of systems both in the 
bulk and in the presence of surfaces. In many cases the agreement between
such field-theoretical predictions and (mainly 
Monte Carlo) simulations or experimental data is striking.

The collective behavior of a system close to its critical point 
can be described in terms of
the order parameter whose actual nature depends specifically on the system. 
Indeed, as long as one is interested in its behavior at length
and time scales much larger than the microscopic ones, an effective
Hamiltonian 
can be used which reflects the internal symmetries of the underlying
microscopic system and which depends only on the order parameter and
potentially a few other slow modes.
 
Within this framework one can determine 
the actually observed
non-analytic behavior of thermodynamic quantities and structure
factors upon approaching the critical point. Moreover some of the
quantities characterizing such non-analyticities (e.g., critical
exponents or amplitude ratios) turn out to be universal in the sense that
they depend only on general features of the effective Hamiltonian such
as the spatial dimension and internal symmetries but that they are
independent of the details of the actual system. 
The numerical values of the universal properties
and of the universal scaling functions characterize the so-called
universality classes~\cite{ZJ-book}.
On time scales much larger than the microscopic ones it is possible to
describe the dynamics close to critical points in terms of
stochastic evolution equations for the order parameter such that its
resulting equilibrium distribution is given by the effective
Hamiltonian of the universality class which the system belongs
to~\cite{HH-77}.
This approach allows one to compute systematically 
the non-analytic behaviors
observed in dynamical quantities, e.g., in the
low-frequency limit of the dynamic structure factor. In turn the
associated universal quantities define the dynamic universality
class. 
One finds that each static universality class consists of several
dynamic sub-universality classes which differ, e.g., by different
conserved quantities, but nonetheless exhibit the same static
universal properties.
As an example,
the static universality class of the phase transitions in
uniaxial ferromagnets is the same as that of binary liquid mixtures although
their universal dynamic behavior is captured by two different dynamic
universality classes.
Various analytical methods, in particular the
renormalization-group theory, have been developed and applied to provide
predictions for universal quantities. (See, e.g.,
Ref.~\cite{PV-02} for a recent review of the results obtained for the most
relevant static universality classes of critical phenomena in the
bulk. A recent summary of bulk 
critical dynamics can be found in Ref.~\cite{FM-02}.) 

Within this framework it is possible to account for the effects of
surfaces on the critical behavior. Indeed real systems are always
bounded by surfaces or interfaces between different phases which break
translational invariance and thus are
expected to influence the physical properties including universal features. 
In particular it turns out that compared with the bulk the
critical behavior is locally altered within a distance from
the surface of the order of the bulk correlation length. The resulting
critical behavior depends only on general properties of the surface
and in turn it can be classified in terms of different 
surface universality classes branching from the same bulk
universality class and which are in general characterized by their own 
surface critical exponents different from the corresponding bulk ones 
(see Refs.~\cite{Binder-83,Diehl-86} 
for comprehensive reviews). 
On the other hand it turns out that there are no independent dynamic
critical surface exponents~\cite{DD-83}.

In addition to the local effects near surfaces, the properties of a system are
influenced by its finite size when the
correlation length $\xi$ becomes comparable with the typical sample size
$L$. Depending on the specific system and its geometry this can result
even in a
suppression of the phase transition or, generally, 
in a shift of the critical point and of coexistence curves 
which depends on $L$ and vanishes in
the bulk limit $L\rightarrow\infty$~\cite{Barber-83}. 
The scaling behavior that is
observed upon approaching the critical point is expected to involve
$L/\xi$ and its theoretical understanding is based on the finite-size
scaling theory~\cite{Barber-83,Cardy-88,BDT-00}. 
As a consequence of confinement and boundary
conditions fluctuation-induced effective forces on the confining
surfaces arise known as thermodynamic Casimir
forces (see, e.g., Refs.~\cite{BDT-00,KD-92,KD-92bis,Krech-94,SHD-03}
and references therein).

Thin films provide the simplest geometry for studying theoretically  
the effects of confinement on 
phase transitions; moreover they are particularly relevant
experimental realizations of finite-size systems.
Thin film are characterized by a finite width, which in the present
context is taken to be much
larger than the typical microscopic scale, and a macroscopicly large 
lateral extension, i.e., much larger than the
correlation length. 
Thin films of magnetic materials, confined fluids, and wetting
films represent specific systems with such a geometry 
which are indeed investigated experimentally. Their
static critical properties have been theoretically and
experimentally investigated in the past for different universality classes
and boundary conditions (see Refs.~\cite{KD-99,Dohm-93} 
and references therein).

Dynamics in confined geometry is, instead, a less explored subject,
both at the critical point and below.
Novel phenomena have been observed in the dynamics of 
phase separation~\cite{Onuki} occurring in the {\it two-phase region}
of the phase diagram of confined binary liquid mixtures, after a
quench from the homogeneous state. In particular the interplay between
surface-directed spinodal decomposition (see, e.g., Ref.~\cite{P-05})
and confinement has been studied  numerically in Ref.~\cite{DPHB-05}
for a symmetric binary mixture with purely diffusive dynamics (Model B
in the notion of Ref.~\cite{HH-77}), a simplified form of the
actual dynamics of fluid mixtures (Model H~\cite{HH-77}).
At the {\it critical point}, which is the focus of the present study, 
most of the theoretical results 
have been obtained for the case of a finite hypercubic geometry with
periodic boundary conditions and purely dissipative 
dynamics~\cite{Gold-86,NZJ-86,Niel-87,Diehl-87,KDS-96,Oerding-95,ZJ-book} 
(Model A in the notion of Ref.~\cite{HH-77}) or dynamics
coupled to a conserved density~\cite{KD-98} (Model C in the notion of
Ref.~\cite{HH-77}). The dynamic structure factor, the spin transport,
and the thermal conductivity of the
three-dimensional XY model on a cubic lattice with periodic boundary
conditions have been studied by means of Monte Carlo simulations~\cite{KL-99}. 
In all the cases mentioned above, 
due to the translational invariance, there are no surface, i.e.,
spatially varying
effects. 
Monte Carlo simulations have been performed to study the thermal
conductivity of the planar magnet lattice model in a bar-like geometry 
($H\times H\times L$ with $L\gg H$) with open boundary 
conditions~\cite{NM-01}, 
aiming for a comparison with experimental results for $^4$He at the
superfluid transition confined to an array of
pores~\cite{He-pores-ex}. The same problem has been
addressed within the field-theoretical approach, studying Model F
dynamics~\cite{HH-77,Dohm-91} in a $L\times L\times\infty$ geometry
with Dirichlet boundary conditions (DBC, i.e., vanishing surface fields) 
for the order parameter~\cite{TD-03}. In both cases the agreement 
with experimental data is quite good.
Critical dynamics in the film geometry has not yet been studied
systematically, in spite of available experimental data for
some specific systems. 
In particular, the thermal conductivity of
$^4$He close to the normal-superfluid transition and in confined
geometry has been investigated
experimentally in some detail.
Field-theoretical methods have been employed to analyze the so-called
thermal boundary resistance (Kapitza resistance) between the
superfluid $^4$He and the wall confining the system. This can be
carried out
by considering Model F dynamics in a semi-infinite space with 
DBC~\cite{Dohm-93,br-He}. In spite of these results,
for a specific surface quantity, the theoretical prediction for the
full finite-size behavior of the thermal resistance across a film is
still lacking. Moreover, recent experimental
findings~\cite{KMFGMLLA-02}
are in disagreement with the field-theoretical predictions
of Ref.~\cite{br-He}. It has been argued that this might be a
consequence of the choice of DBC
being inappropriate for describing 
helium confined to a film.
Some other transport properties have also been investigated
theoretically for the film geometry. 
In particular, in
Ref.~\cite{Bhat-96} the effects of confinement on the critical
diffusivities have been studied within the decoupled-mode
approximation~\cite{Ferrell-70} for the dynamic universality class of
liquid-vapor phase transitions in a one-component fluid 
(Model H in the notion of Ref.~\cite{HH-77}) and for
the superfluid transition (described by Model E~\cite{HH-77}) with DBC
at the confining plates. The diffusion constant
considered is the one associated with the density current and the
superfluid flow, respectively.
In Ref.~\cite{BB-98} the finite-size behavior of the ultrasonic
attenuation that is observed 
upon approaching the critical point of the superfluid
transition in $^4$He has been studied. The sound velocity can be
related to the frequency-dependent specific heat (see
Ref.~\cite{BB-98} and references therein), allowing for a quite direct
field-theoretical analysis within the Gaussian model. Instead of the
full Model F dynamics, which is the appropriate one to describe the
superfluid transition,
it is possible to deal approximately 
with this problem by considering the simpler
Model A dynamics of the superfluid order parameter. Then, by
applying Dirichlet boundary conditions, the scaling functions for
the ultrasonic attenuation have been computed, resulting in a good
agreement with available experimental data~\cite{BB-98}.

Recent efforts~\cite{BAF-03,NG-03} address theoretically some aspects of
non-equilibrium (critical) dynamics of a scalar fluctuating field
$\phi$ 
(which can be, e.g., the order parameter of an Ising ferromagnet or
the deformation of an elastic membrane) in
film geometry with DBC. The 
dynamics of $\phi$ is assumed to be purely dissipative (as in Model A),
the effective Hamiltonian is taken as a 
Gaussian, and the immobile confining walls to be
actually ``immersed'' in the fluctuating medium. 
However, different from the
usual Model A, the fluctuations of the field are taken either to be due
to external forces~\cite{BAF-03} or as quasi-equilibrium 
thermal noise generated by a space- and time-dependent
temperature profile~\cite{NG-03}. The main focus of these analyses is the
computation of the Casimir-like non-equilibrium 
fluctuation-induced force that acts
between the confining walls for different instances of driving forces
and temperature profiles, whereas little attention is paid to the
actual dynamics of the field $\phi$ in the space delimited by the walls.

In spite of these results a systematic investigation of the
critical dynamics in film geometry is still lacking. In view of the
rapidly developing experimental techniques able to resolve space- and
time-dependent quantities on the proper mesoscopic scale it is 
important to provide theoretical predictions
for experimentally accessible quantities such, e.g, time-dependent
response and correlation functions, dynamics of fluctuation-induced
forces, etc.

In the following we set up the field-theoretical description of the
purely relaxational dynamics (Model A in the notion of
Ref.~\cite{HH-77}) in film geometry with Dirichlet boundary
conditions. We consider the static universality class of
systems with a $N$-component order parameter and $O(N)$-symmetric
interactions, such as the Ising model ($N=1$) or the isotropic
XY ($N=2$) and Heisenberg ($N=3$) models. Although it is possible to
study 
the relaxational dynamics for general
$N$ by kinetic Monte Carlo simulations, for Model A 
$N=1$ is the only physically relevant case, experimentally
realized in anisotropic magnets. For systems with $N>1$ the actual
dynamics requires a description in terms of more 
complex models~\cite{HH-77,FS-94}.
Moreover also the proper description of 
mixing-demixing transitions in binary liquid mixtures
and liquid-vapor transitions in one-component fluids, whose static
universal properties are given by the case $N=1$, calls for
different dynamical models and boundary
conditions~\cite{HH-77,Onuki}.
Accordingly, the model we consider here is relevant for the dynamics
of actual uniaxial magnetic films (without energy conservation~\cite{HH-77})  
with symmetry-preserving boundary conditions and for
Monte Carlo simulations with Glauber dynamics of $O(N)$ order parameter
models with free boundary conditions.
Keeping in mind these caveats the present analysis provides a
theoretical framework which might nonetheless turn out to be useful
also for the more complicated dynamical models.
Explicit expressions for the universal scaling functions are obtained
within mean-field theory (Gaussian model),
whose actual behavior (independent of $N$) turns out to be
already quite rich. We discuss in detail the
surface behavior close to the confining walls and the temporal
crossover between different regimes. The analysis carried out here
lends itself for possible, future extensions within the 
field-theoretical approach 
to include the effect of fluctuations beyond mean-field theory.

After this introduction the paper is organized as follows. 
In Sec.~\ref{intro} we recall some general scaling properties which
will be useful in the following.
In Sec.~\ref{sec-model}  the model is described
and the universal scaling functions 
for the dynamic Gaussian linear response and correlation functions
are derived and studied in detail in Sec.~\ref{sec-linbeh}.
In addition we discuss the effects of time-dependent fields on
the fluctuation-induced Casimir force acting on the confining walls
of the system, providing analytic expressions for the associated
Gaussian scaling functions. 
In Sec.~\ref{sec-nlin} the nonlinear relaxation of the order parameter
is analyzed at the bulk critical point of the model and the crossover
between bulk-like, surface-like, and eventual linear relaxation is
emphasized. 
In Sec.~\ref{sec-sum} we provide a summary of the main
results. Most of the details of the computations are
reported in the Appendices~\ref{app-staticcorr}, \ref{app-formulas},
and~\ref{app-tI}--\ref{app-expansion}. 
Instead, 
in Appendix~\ref{app-dynampratios} the computation of the bulk
universal amplitude ratio associated with the divergence of the
relaxation time is reported, whereas in Appendix~\ref{app-OPprofile}
we determine and discuss the useful analytic expression for the static order
parameter profile across the film 
in the low-temperature phase and for Dirichlet boundary conditions, 
which, to our knowledge, has never been
reported in the literature.

\section{General scaling properties}
\label{intro}
We consider a confined system in $d$ dimensions with film geometry
$\infty^{d-1}\times L$ and Dirichlet-Dirichlet boundary conditions
(corresponding to the so-called ordinary-ordinary surface universality 
class~\cite{Diehl-86}) and
purely dissipative relaxational dynamics (Model A of Ref.~\cite{HH-77}).

For future reference we introduce here some of the notations 
used in the following. We define the reduced temperature
\begin{equation}
\tau = \frac{T-T_{c,b}}{T_{c,b}}
\end{equation}
where $T$ is the temperature and $T_{c,b}$ is the transition
temperature in the bulk. With $\xi$ and $\rt_R$ we denote the true
correlation length and the true relaxation time, respectively, as the
characteristic length and time scales defined via the exponential
decay of two-point bulk correlation functions in thermal equilibrium.
Upon approaching the critical point
($\tau\rightarrow 0$) both $\xi$ and $\rt_R$ diverge with the following
leading singularities:  
\begin{equation}
\xi(\tau \rightarrow 0^\pm) = \xi^\pm_0 |\tau|^{-\nu}
\label{defxi}
\end{equation}
and
\begin{equation}
\rt_R(\tau \rightarrow 0^\pm) 
= \rt_0^\pm |\tau|^{-\nu z} = \rt_0^\pm (\xi/\xi^\pm_0)^z
\label{defrt}
\end{equation}
where $\nu$ and $z$ are universal standard bulk critical exponents whereas 
$\xi_0^\pm$ and
$\rt_0^\pm$ are non-universal amplitudes depending on
the microscopic details of
the system. Within mean-field theory (MFT) corresponding to $d> 4$ one
has $\nu = 1/2$ and $z=2$,
whereas, for the Ising universality class $(N=1)$ in $d=3$, $\nu =
0.6301(4)$~\cite{PV-02} and $z \simeq 2.02$ (see Ref.~\cite{CMPV-03}
and references therein for a summary of the various estimates of $z$).  
The values of $\xi_0^\pm$ and $\rt_0^\pm$ are different for $\tau
\rightarrow 0^+$ and $\tau\rightarrow 0^-$, forming universal
amplitude ratios $\xi_0^+/\xi_0^-$
and $\rt_0^+/\rt_0^-$ with $\xi_0^+/\xi_0^- = \sqrt{2}$ within MFT, whereas
$\xi_0^+/\xi_0^- = 1.896(10)$~\footnote{%
This quantity is denoted by $U_{\xi_{\rm gap}}$ in Ref.~\cite{PV-02}. The numerical value quoted here, obtained by 
combining high-temperature
expansions with a parametric representation of the equation of
state, is taken from
Tab.~11 therein. %
}
for the Ising universality class in $d=3$. For the second-moment
correlation length similar results can be found in the 
literature~\cite{PHA-91,PV-02}.
In
Appendix~\ref{app-dynampratios} the ratio $\rt_0^+/\rt_0^-$ is
computed to first order in $\epsilon = 4 - d$ for the 
Ising universality class of Model A, leading to $\rt_0^+/\rt_0^-=2$
within MFT and $\rt_0^+/\rt_0^- = 3.3(4)$ in $d=3$~\footnote{%
From Eq.~\reff{ratioexp} one has $\rt_0^+/\rt_0^- = 2(1+ 2\epsilon\ln
2/3) + O(\epsilon^2)$. In order to gain 
a rough estimate of this ratio for
$\epsilon = 1$ the $[0,1]$ and $[1,0]$ Pad\'e approximants can be
used, yielding the value 3.3(4).
}.

The bulk order parameter $m$ vanishes for $\tau \rightarrow 0^-$ as
\begin{equation}
m = \m_0 (-\tau)^\beta\;,
\label{defm}
\end{equation} 
with the universal exponent $\beta$ and the non-universal amplitude
$\m_0$; within MFT $\beta =1/2$, whereas for the Ising universality
class in $d=3$, $\beta = 0.3265(3)$~\cite{PV-02}.

In the following we will be concerned with quantities $\oo$ defined in the
film geometry. They generally depend on a set $\{{\bf x},t\}$ 
of spatial coordinates and times, 
on the temperature (expressed in terms of $\tau$),
and on the film thickness $L$. Since upon approaching the critical point 
the dominant length and time scales are given by $\xi$ and $\rt_R$,
respectively, the following scaling behavior is expected in the critical
region $|\tau| \ll 1$:
\begin{equation}
\begin{split}
\oo(\{{\bf x},t\};\tau\gtrless 0,L) =& \mathfrak{o}_\oo^\pm
\left(\frac{\xi}{\xi_0^\pm}\right)^{-\Delta_\oo} F^{(1)}_{\oo,\pm} (\{{\bf
x}/\xi,t/\rt_R\};L/\xi) \\
=& \mathfrak{o}^\pm_\oo
\left(\frac{L}{\xi_0^\pm}\right)^{-\Delta_\oo} F^{(2)}_{\oo,\pm} (\{{\bf
x}/L,(t/\rt_0^\pm) (\xi^\pm_0/L)^z\};L/\xi)
\end{split}
\label{genscaling}
\end{equation}
where $\mathfrak{o}^\pm_\oo$ are non-universal constants which have the
same engineering dimension as the observable $\oo$ and can be
expressed in terms of $\xi_0^+$, $\m_0$,  $\rt_0^+$, and universal
amplitude ratios.
$\Delta_\oo$ is the scaling dimension of the quantity $\oo$, and 
$F^{(i)}_{\oo,\pm}$, $i=1,2$, are universal 
scaling functions. 
The second line in Eq.~\reff{genscaling} is the scaling form,
equivalent to the first one, in which we shall present our results. For
the two-point correlation function $C$ (see Sec.~\ref{sec-model}) 
one has $\Delta_C = d-2+\eta$ (which agrees with the static two-point
correlation function and defines the static bulk critical exponent
$\eta$ with $\eta=0$ within MFT, whereas $\eta=0.0364(5)$~\cite{PV-02} for the
three-dimensional Ising universality class)
whereas for the response function $R$,
$\Delta_R = \Delta_C + z = d-2+\eta+z$. 
For the
magnetization one has $\Delta_m = \beta/(\nu z)$ 
($\Delta_m = \Delta_C/(2z)$ if hyperscaling holds). 
Crossovers between surface and bulk singular behaviors
characterized by surface and bulk  critical exponents are
related to the singular behavior of the scaling functions
$F^{(i)}_{\oo,\pm}$ 
if
${\bf x}$ approaches the confining walls (see
Sec.~\ref{sec-nlin}). In this respect one has to keep in mind that the
scaling properties (Eq.~\reff{genscaling}) only hold in the scaling
limit, i.e., distances between two spatial points, distances from
confining walls, and time differences must be sufficiently large
compared to microscopic scales.
The field-theoretical approach has been proven to be
a powerful tool to compute both the exponents and the scaling functions
appearing in Eq.~\reff{genscaling} for various measurable
quantities. In the following we present the mean-field form
(i.e., tree-level approximation in the field-theoretical language
which is valid for $d=4$ up to logarithmic corrections) 
of the scaling functions $F^{(2)}_\oo$ for various observables. In many
cases a reasonably good agreement between experimental or
simulation data and field-theoretical computation is already obtained by
using mean-field scaling functions combined with higher-order estimates for
critical exponents entering into their scaling arguments.
%
%


\section{The Model}
\label{sec-model}

\subsection{Definition}

The time evolution of a $N$-component field $\varphi({\bf x},t) =
(\varphi_i({\bf x},t), i=1,\ldots, N)$ under 
purely dissipative relaxation dynamics~(Model A of Ref.~\cite{HH-77}) is
described by the stochastic Langevin equation
\begin{equation}
\label{lang}
\partial_t \varphi ({\bf x},t)=-\Omega
\frac{\delta \cal{H}[\varphi]}{\delta \varphi({\bf x},t)}+\zeta({\bf x},t) \; ,
\end{equation}
where $\Omega$ is a kinetic coefficient,
$\zeta({\bf x},t)$ a zero-mean stochastic Gaussian noise with correlations
\begin{equation}
\langle \zeta_i({\bf x},t) \zeta_j({\bf x}',t')\rangle= 2 \Omega \,
\delta({\bf x}-{\bf x}') \delta (t-t')\delta_{ij},
\label{noise}
\end{equation}
and $\cal{H}[\varphi]$ is the static Hamiltonian. The universal
properties near the critical point of a second-order phase transition
are captured by the Landau-Ginzburg form \cite{BJW-76,LRD-79,ZJ-book,Diehl-86}
\begin{equation}
{\cal H}[\varphi] = \int_V\!\dd V \left[
\frac{1}{2} (\nabla \varphi )^2 + \frac{1}{2} r_0 \varphi^2
+\frac{1}{4!} g_0 \varphi^4 \right] + \int_{\partial V}\!\dd^{d-1}
{\bf x}_\| \,\frac{c}{2}\varphi^2,\label{lgw}
\end{equation}
where $r_0\propto T$ is a parameter that takes the value $r_{0,{\rm
crit}}$  for $T=T_{c,b}$ ($r_{0,{\rm
crit}} = 0$ within MFT) and $g_0>0$ is the coupling constant
providing stability for $\tau<0$.
The surface term implies the boundary conditions $\varphi = c^{-1}
\partial_{x_\bot}\varphi$ at $x_\bot = 0,\, L$ such that the
fixed-point value $c=\infty$ leads to Dirichlet boundary conditions as
considered in the following; we do not consider surface field
contributions $h_s \int_{\partial V} \!\dd^{d-1}
{\bf x}_\|\;\varphi$.
We use the notations 
$V={\mathbb R}^{d-1}\times[0,L]$ and $\dd V = \dd^{d-1}{\bf
x}_\|\dd x_\bot$, where the position vector 
${\bf x} = ({\bf x}_\|,x_\bot)$ is decomposed into
the $d-1$-dimensional component ${\bf x}_\|$ parallel to the confining planar
walls and a one-dimensional one $x_\bot$ perpendicular to
them. 

Instead of solving the Langevin equation for $\varphi[\zeta]$ and then
averaging over the noise distribution $P[\zeta]$, 
the equilibrium  correlation and
response functions can be directly obtained by means of 
a suitable field-theoretical action $S[\varphi,\tilde\varphi]$
\cite{ZJ-book,BJW-76,LRD-79}
so that, for an observable ${\mathcal O}[\varphi]$,
\begin{equation}
\langle{\mathcal O}\rangle \equiv \int[{\rm d}\zeta]\; {\mathcal
O}[\varphi[\zeta]] P[\zeta] = \int [{\rm d}\varphi{\rm
d}\tilde\varphi]\; {\mathcal O} e^{-S[\varphi,\tilde{\varphi}]} \,.
\end{equation}
(Note that within the conventions we adopt, $\int [{\rm
d}\varphi{\rm d}\tilde\varphi]\; e^{-S[\varphi,\tilde{\varphi}]} =
1$~\cite{LRD-79}.)
For the Langevin equation (\ref{lang}) with the Gaussian noise
(Eq.~\reff{noise}) the
field-theoretical action is given by \cite{ZJ-book,BJW-76,LRD-79}
\begin{equation}
S[\varphi,\tilde{\varphi}]= \int\! \dd t \int_V\!\dd V
\left[\tilde{\varphi} \partial_t\varphi+
\Omega \tilde{\varphi} \frac{\delta \mathcal{H}[\varphi]}{\delta \varphi}-
\tilde{\varphi} \Omega \tilde{\varphi}\right],\label{msrh}
\end{equation}
where $\tilde{\varphi}({\bf x},t)$ is an auxiliary field, conjugate to
an external bulk field $h$ which linearly couples to the order
parameter $\varphi$ so that ${\cal H}[\varphi,h] = {\cal H}[\varphi] -
\int \dd V h \varphi$.
As a consequence, for an observable ${\cal O}$ the
following relation for the linear response to the field $h$
holds:
\begin{equation}
\left.{\delta \langle {\cal O} \rangle_h \over \delta h({\bf x},t)}\right|_{h=0} =
\Omega \langle\tilde\varphi({\bf x},t){\cal O}\rangle_{h=0} \ ,
\label{Rdef}
\end{equation}
where $\langle\cdot\rangle_h$ is the average taken with respect to the action
$S[\varphi,\tilde\varphi;h]$ associated with ${\mathcal
H}[\varphi,h]$.
In view of Eq.~\reff{Rdef}, $\tilde{\varphi}({\bf x},t)$ is called
response field. In the following we will be mainly concerned with the
response of the order parameter field to the external perturbation
$h$, given by
\begin{equation}
R({\bf x}_1, t_1;{\bf x}_2,t_2) = \left.{\delta \langle \varphi({\bf x}_2,t_2) \rangle_h \over \delta h({\bf x}_1,t_1)}\right|_{h=0} =
\Omega \langle\tilde\varphi({\bf x}_1,t_1)\varphi({\bf x}_2,t_2)\rangle_{h=0}.
\label{Rdefbis}
\end{equation}
Causality implies that $\langle  \varphi({\bf x}_2,t_2) \rangle_h$
does not depend on $h({\bf x}_1,t_1)$ whenever $t_1 > t_2$, i.e., the
order parameter at a given time does not depend on possible
perturbations at later times. Accordingly $R({\bf x}_1, t_1;{\bf
x}_2,t_2)$ vanishes for $t_2<t_1$.

The effect of confining walls in the case we are interested in amounts
to the Dirichlet boundary condition for the field $\varphi$ (i.e., infinite
surface enhancement $c$; see above and Ref.~\cite{Diehl-86}),
\begin{equation}
\varphi({\bf x}_{\mathcal B},t) = 0 \,,
\qquad \forall t ,
\label{BC1}
\end{equation}
where we denote by ${\bf x}_{\mathcal B}$ the 
position vector on the boundary $\partial V$.
The Gaussian (i.e., $g_0=0$) equation of motion for the field
$\tilde\varphi$ given by $-\partial_t\tilde\varphi + \Omega
(-\Delta\tilde\varphi + r_0 \tilde\varphi) = 0$ and Eq.~\reff{BC1} yield
 the boundary condition~\footnote{%
Indeed $\int_0^L\!\dd x_\bot\tilde\varphi\partial^2_{x_\bot}\varphi =
\int_0^L\!\dd x_\bot\varphi\partial^2_{x_\bot}\tilde\varphi +
(\tilde\varphi\partial_{x_\bot}\varphi -
\varphi\partial_{x_\bot}\tilde\varphi)|_{x_\bot=0}^L$, where
$f|_{x_\bot=0}^L \equiv f (x_\bot = L) - f(x_\bot = 0)$.
}.
\begin{equation}
\tilde\varphi\partial_{x_\bot}\varphi|_{x_\bot = 0}^L = 0 \, ,
\end{equation}
which is fulfilled by imposing
\begin{equation}
\tilde\varphi({\bf x}_{\mathcal B},t) = 0 \,,
\qquad \forall t .
\label{BC2}
\end{equation}
In order to diagonalize the Gaussian part of Eq.~\reff{msrh} it is
useful to decompose both fields $\varphi$ and $\tilde\varphi$
in terms of eigenfunctions of the Laplacian $\Delta$ fulfilling the 
boundary conditions \reff{BC1} and
\reff{BC2}~\cite{DD-83,KD-92} according to
\begin{equation}
\phi({\bf x},t) = \sum_{n=1}^\infty \int_{{\bf p},\omega}\! e^{i({\bf p}\cdot{\bf x_\|}
- \omega t)}\hat\phi_n({\bf p},\omega)\Phi_n(x_\bot;L), 
\label{repr}
\end{equation}
where $\phi =\varphi,\tilde\varphi$, and
\begin{equation}
\int_{{\bf p},\omega} \equiv \int_{{\mathbb R}^{d-1}}\!\frac{\dd^{d-1}{\bf
p}}{(2\pi)^{d-1}}\int_{\mathbb R}\!\frac{\dd \omega}{2\pi} 
\end{equation}
with ${\bf p}$ as the $d-1$-dimensional momentum parallel to the
confining walls. The transverse momentum takes, instead, discrete values 
$k_n = \pi n/L$ with $n=1,2,\ldots$. The eigenfunctions
$\Phi_n(x_\bot;L)$ are given by
\begin{equation}
\Phi_n(x_\bot;L) = \sqrt{2/L} \sin(k_n x_\bot) ,
\label{eigenf}
\end{equation}
so that $\Phi_n(x_\bot=0;L)=\Phi_n(x_\bot=L;L)=0$. Note that for the total
momentum ${\ve q}_n \equiv ({\ve p}, k_n)$ one has  $|{\ve q}_n| \ge
\pi/L$ and thus the homogeneous fluctuation mode is suppressed by the 
boundary conditions. In terms of the functions introduced above, the
Gaussian part $S_0$ of Eq.~\reff{msrh} can be written as
\begin{equation}
S_0[\hat\varphi,\hat{\tilde\varphi}] = \frac{1}{2}\sum_{n=1}^\infty\!\int_{{\bf
p},\omega}\! (\hat\varphi_n(-{\ve p},-\omega),\hat{\tilde\varphi}_n(-{\ve p},-\omega)){\mathbb
M}{\hat\varphi_n({\ve p},\omega) \choose\hat{\tilde\varphi}_n({\ve
p},\omega) } \;,
\end{equation}
where the inverse propagator ${\mathbb M}$ is given by the matrix 
\begin{equation}
{\mathbb M}\equiv \left( 
\begin{array}{cc}
0 & - i\omega + \Omega ({\ve q}_n^2 + r_0)\\
i\omega + \Omega ({\ve q}_n^2 + r_0)& -2\Omega
\end{array}
\right) \; .
\label{invprop}
\end{equation}
Within MFT, the two-point response and correlation
function $R^{(0)}$ and $C^{(0)}$, respectively, are determined by
${\mathbb M}^{-1}$:
\begin{equation}
\begin{split}
R^{(0)}_{i_1i_2n_1n_2}({\bf p}_1,\omega_1;{\bf p}_2,\omega_2) &\equiv
\Omega \langle\hat{\tilde\varphi}_{i_1n_1}({\ve p}_1,\omega_1)\hat\varphi_{i_2n_2}({\ve
p}_2,\omega_2)  \rangle_{g_0=0}\\
&= (2\pi)^d\delta^{(d-1)}({\bf
p}_1+{\bf
p}_2)\delta(\omega_1+\omega_2)\delta_{i_1i_2}\delta_{n_1n_2}R^{(0)}({\bf
q}_{n_2},\omega_2)
\end{split}
\end{equation}
where $i_k$ and $n_k$ indicate the field component and the
Fourier mode according to Eq.~\reff{repr}, respectively. For the
correlation function one finds 
\begin{equation}
\begin{split}
C^{(0)}_{i_1i_2n_1n_2}({\bf p}_1,\omega_1;{\bf p}_2,\omega_2) &\equiv \langle \hat\varphi_{i_1n_1}({\ve p}_1,\omega_1)\hat\varphi_{i_2n_2}({\ve p}_2,\omega_2)\rangle_{g_0=0}\label{defCorr1}\\
 &= (2\pi)^d\delta^{(d-1)}({\bf
p}_1+{\bf
p}_2)\delta(\omega_1+\omega_2)\delta_{i_1i_2}\delta_{n_1n_2}C^{(0)}({\bf
q}_{n_2},\omega_2)
\end{split} 
\end{equation}
while $\langle \hat{\tilde\varphi}_{i_1n_1}({\ve
p}_1,\omega_1)\hat{\tilde\varphi}_{i_2n_2}({\ve
p}_2,\omega_2)\rangle_{g_0=0}=0$ due to causality.
From Eq.~\reff{invprop} one obtains 
\begin{equation}
R^{(0)}({\bf q},\omega) = \frac{\Omega}{-i\omega +\Omega({\bf q}^2 +
r_0)} \label{resp} 
\end{equation}
and
\begin{equation}
C^{(0)}({\bf q},\omega) = \frac{2\Omega}{\omega^2 +[\Omega({\bf q}^2 +
r_0)]^2} \; .
\label{corr}
\end{equation}
Note that $R^{(0)}$ and $C^{(0)}$ are the mean-field 
response and correlation function, respectively, 
of the system in the bulk. 
Within MFT the presence of the
boundaries is accounted for by the spatial dependence of the
eigenfunctions $\Phi_n(x_\bot;L)$ (see Eq.~(\ref{eigenf})) and by the
quantization of the allowed momenta.

For later purposes 
it will be useful to 
provide for these
functions their representation in terms of time and transversal
coordinates $x_{i\bot}$.
Following Eq.~\reff{repr} we define
\begin{equation}
\phi({\bf p},x_\perp,t) = \sum_{n=1}^\infty \int\!\frac{\dd\omega}{2\pi} 
e^{- i\omega t}\hat\phi_n({\bf p},\omega)\Phi_n(x_\bot;L), 
\end{equation}
and
\begin{equation}
\phi({\bf p},x_\perp,\omega) = \sum_{n=1}^\infty \hat\phi_n({\bf p},\omega)\Phi_n(x_\bot;L),
\end{equation}
so that, according to Eq.~\reff{defCorr1},
\begin{equation}
\begin{split}
C^{(0)}_{i_1i_2}({\bf p}_1,x_{1\bot},t_1;{\bf p}_2,x_{2\bot},t_2) &\equiv
\langle\varphi_{i_1}({\ve p}_1,x_{1\bot},t_1)\varphi_{i_2}({\ve
p}_2,x_{2\bot},t_2)  \rangle_{g_0=0}\\
 & = (2\pi)^{d-1}\delta^{(d-1)}({\bf
p}_1+{\bf
p}_2)\delta_{i_1i_2}C^{(0)}({\bf p}_2,x_{1\bot},x_{2\bot},t_2-t_1) 
\end{split}
\end{equation}
and
\begin{equation}
\begin{split}
C^{(0)}_{i_1i_2}({\bf p}_1,x_{1\bot},\omega_1;{\bf p}_2,x_{2\bot},\omega_2) &\equiv \langle\varphi_{i_1}({\ve p}_1,x_{1\bot},\omega_1)\varphi_{i_2}({\ve
p}_2,x_{2\bot},\omega_2)  \rangle_{g_0=0} \\
& = (2\pi)^d\delta^{(d-1)}({\bf
p}_1+{\bf
p}_2)\delta(\omega_1+\omega_2)\delta_{i_1i_2}
C^{(0)}({\bf p}_2,x_{1\bot},x_{2\bot},\omega_2) 
\end{split}
\end{equation}
with an analogous expression for the response function.
Using the above formulae we obtain (recalling that ${\bf q}_n\equiv
({\bf p},\pi n/L)$)
\begin{equation}
C^{(0)}({\bf p},x_{1\bot},x_{2\bot},t) = \sum_{n=1}^\infty
\Phi_n(x_{1\bot};L)\Phi_n(x_{2\bot};L)\int\!\frac{\dd\omega}{2\pi}
e^{-i\omega t} C^{(0)}({\bf q}_n,\omega)
\label{reprC3}
\end{equation}
and
\begin{equation}
C^{(0)}({\bf p},x_{1\bot},x_{2\bot},\omega) = \sum_{n=1}^\infty
\Phi_n(x_{1\bot};L)\Phi_n(x_{2\bot};L) C^{(0)}({\bf q}_n,\omega),
\label{reprC4}
\end{equation}
with $C^{(0)}({\bf q},\omega)$ given in Eq.~\reff{corr}. In 
Appendix~\ref{app-staticcorr} 
we show how one recovers the known results
for the equal-time correlation function in the film geometry discussed in
Refs.~\cite{KD-99} 
and~\cite{KED-95} for various boundary conditions 
and in Ref.~\cite{Diehl-86}, Sec.~IVA, in the case of a 
film with equal boundary
conditions. Analogous
relations hold for the response function. 

\subsection{The fluctuation-dissipation theorem}
\label{subsec-FDT}

For small fluctuations within equilibrium dynamics 
the two-point response and correlation functions are not independent
quantities.
The relation between them is provided by the
fluctuation-dissipation theorem (FDT):
\begin{equation}
\frac{\dd C(t)}{\dd t} = - R(t) \;,\qquad \mbox{for}\qquad t>0\;,
\label{fdt}
\end{equation}
where with $C(t)$  and $R(t)$ we indicate summarily 
the time-dependent two-point 
correlation function and the response function, respectively. 
As equilibrium quantities their dependence on two time variables 
reduces to a dependence on the time difference only. Indeed the
theorem is a consequence of the time-translation invariance and
time-reversal symmetry of equilibrium dynamics. 
Keeping in mind that correlations vanish in the
long-time limit, one thus has
\begin{equation}
C(t) = \int_t^\infty\!\dd s \; R(s) \;
,\qquad\mbox{for}\qquad t>0 \;.
\label{fdtint}
\end{equation}
Time-reversal symmetry in the equilibrium
state implies $C(t)=C(-t)$, which, combined with
Eq.~\reff{fdtint}, allows one to determine completely the correlation
function from the response function. Let us recall that the causality
of the response function (linear or not) implies 
$R(t)\propto \theta(t)$ where $\theta(t) = 1$ for $t>0$ and $0$
otherwise. For later purposes it is useful to express the
FDT also in other forms. 
Defining the Fourier transform of $C(t)$ as
\begin{equation}
C(\omega) = \int_{-\infty}^{+\infty}\!\!\! \dd t\; e^{i\omega t} C(t)
\label{deftFourier}
\end{equation}
and with an analogous definition for $R(\omega)$ the
FDT (Eq.~\reff{fdt}) can be written also as
\begin{equation}
C(\omega) = \frac{2}{\omega} \Im R(\omega)
\label{fdtomega}
\end{equation}
where $\Im$ indicates the imaginary part of the
expression. Equation~(\ref{fdtint}) and causality yield another
useful form of the theorem:
\begin{equation}
C(t=0) = R(\omega = 0) .
\label{fdtCR}
\end{equation} 

It is straightforward to verify that $R^{(0)}$ and $C^{(0)}$ given in
Eqs.~(\ref{resp}) and (\ref{corr}) as well as Eqs.~(\ref{reprC3}) and
(\ref{reprC4}) satisfies the
FDT (\ref{fdtomega}). Moreover it can be
shown that, as expected on general grounds,
this is true also if the effect of fluctuations are taken
into account, i.e., beyond MFT. 

\subsection{Some mean-field results}

According to the general scaling properties discussed in Sec.~\ref{intro},
within the Gaussian approximation 
one can identify the correlation length $\xi$ and the 
relaxation time $\rt_R$ from Eqs.~\reff{resp} and~\reff{corr}:
\begin{equation}
\xi(\tau > 0) = r_0^{-1/2} 
\label{MFxi}
\end{equation}
and
\begin{equation}
\rt_R(\tau > 0) = (\Omega r_0)^{-1}\;,
\label{MFrt}
\end{equation}
respectively. Thus within MFT 
$\nu=1/2$ and $z=2$ as expected, and the
kinetic coefficient 
$\Omega$ in Eq.~\reff{lang} can be expressed in terms of the
experimentally accessible (non-universal) amplitudes $\xi_0$ and
$\rt_0$ (see Eqs.~\reff{defxi} and~\reff{defrt}) as
\begin{equation}
\Omega = \frac{{\xi^+_0}^2}{\rt^+_0}\;
\label{omega}
\end{equation}
and
\begin{equation}
r_0 = \frac{\tau}{(\xi_0^+)^2}\;.
\label{r0}
\end{equation}
Moreover, for $T<T_{c,b}$ the mean-field equation of state for the
bulk order parameter $m$ leads to
\begin{equation}
m = (-6r_0/g_0)^{1/2}\;,
\end{equation}
i.e., $\beta = 1/2$,
and
\begin{equation}
g_0 = 6 (\m_0\xi^+_0)^{-2} \;.
\label{defg0}
\end{equation}
The relations in Eqs.~(\ref{MFxi})-(\ref{defg0}) hold for the present
continuum model (Eqs.~(\ref{lang})-(\ref{lgw})).

\section{Linear behavior}
\label{sec-linbeh}

In this section we study in some detail the behavior of the response and 
correlation function for the model introduced in Sec.~\ref{sec-model}.
By virtue of the FDT linear response and
correlation functions are not independent but the correlation function can
be obtained from the response function and vice versa. We shall first
focus on the response function in Subsec.~\ref{sec-response} and then in 
Subsec.~\ref{sec-correlation} we shall determine correlation functions by
applying the FDT.

\subsection{Scaling forms for the response and correlation functions}

\label{subsec-genscalform}

For future reference we provide here the general scaling forms for
some of the quantities we shall discuss in the following. As already
stated in Sec.~\ref{intro} (see Eq.~\reff{genscaling}), scaling
occurs upon approaching the critical point. In the
specific case of the two-point response function in the 
$({\bf p},x_\bot,\omega)$-representation one has
\begin{equation}
R({\bf p},x_{1\bot},x_{2\bot},\omega) = \hat{\mathfrak{o}}^\pm_R
\left(\frac{L}{\xi_0^\pm} \right)^{1-\eta} \!\!\!\!\RR_\pm({\bf
p}L,x_{1\bot}/L,x_{2\bot}/L,\omega \rt_0^\pm(L/\xi_0^\pm)^z,L/\xi),
\label{scalRomega}
\end{equation}
where $\hat{\mathfrak{o}}^\pm_R$ are non-universal amplitudes which we
fix to be equal to the corresponding bulk ones. 
The functions $\RR_\pm$ are universal scaling functions. 
In the $({\bf p},x_\bot,t)$-representation, this reads
\begin{equation}
R({\bf p},x_{1\bot},x_{2\bot},t) = 
\frac{\hat{\mathfrak{o}}^\pm_R}{\rt^\pm_0}
\left(\frac{L}{\xi_0^\pm} \right)^{1-\eta-z} \!\!\!\!{\bar\RR}_\pm({\bf
p}L,x_{1\bot}/L,x_{2\bot}/L,(t/\rt_0^\pm)(\xi_0^\pm/L)^z,L/\xi)\;,
\label{scalRt}
\end{equation}
where the universal functions $\bar\RR_\pm$ are the Fourier transforms 
of $\RR_\pm$ with respect to their fourth argument. 

Analogously, for the correlation function in the $({\bf
p},x_\bot,\omega)$-representation one has
\begin{equation}
C({\bf p},x_{1\bot},x_{2\bot},\omega) = 
 \hat{\mathfrak{o}}^\pm_C
\left(\frac{L}{\xi_0^\pm} \right)^{1-\eta+z} {\CC_\pm}({\bf
p}L,x_{1\bot}/L,x_{2\bot}/L,\omega \rt_0^\pm(L/\xi_0^\pm)^z,L/\xi)\;,
\label{scalComega}
\end{equation}
where $\hat{\mathfrak{o}}^\pm_C$ are non-universal amplitudes, which again
we fix to be equal to the corresponding bulk ones, and 
the functions ${\CC_\pm}$ are universal scaling functions. 
In the $({\bf p},x_\bot,t)$-representation, this reads
\begin{equation}
C({\bf p},x_{1\bot},x_{2\bot},t) = 
\hat{\mathfrak{o}}_C^\pm\frac{1}{\rt_0^\pm}
\left(\frac{L}{\xi_0^\pm} \right)^{1-\eta} {\bar\CC_\pm}({\bf
p}L,x_{1\bot}/L,x_{2\bot}/L,(t/\rt_0^\pm)(\xi_0^\pm/L)^z,L/\xi)\;,
\label{scalCt}
\end{equation}
where the universal functions $\bar\CC_\pm$ are the Fourier transforms 
of $\CC_\pm$ with respect to their fourth argument. 
In view of the
previous scaling forms it is convenient to introduce the following
suitable set of dimensionless
scaling variables, defined as ${\bf \bar p} = {\bf p} L$, $\bar
x_{i\bot} = x_{i\bot}/L$, $\bar t^\pm = (t/\rt_0^\pm)(\xi_0^\pm/L)^z$,
$\bar \omega^\pm = \omega \rt_0^\pm(L/\xi_0^\pm)^z$, and $\bar L = L/\xi$.

As stated above, concerning the non-universal amplitudes 
$\hat{\mathfrak{o}}_R^\pm$ and 
$\hat{\mathfrak{o}}_C^\pm$ we consider the corresponding correlation
functions (Eqs.~\reff{scalRomega} and~\reff{scalCt}) in the bulk. 
The critical structure factor in the bulk is given by
\begin{equation}
C_{\rm crit}^{\rm bulk}({\bf q}, t=0) = \frac{D_\infty }{q^{2-\eta}}\;,
\label{defDinf}
\end{equation}
which defines the non-universal amplitude $D_\infty$ (see, e.g., 
Ref.~\cite{PHA-91}). 
Here and in the
following  using the subscript 
``crit'' means that the 
function corresponds to $\tau=0$, i.e., to bulk criticality.
Beyond MFT (in this case two-scale universality
holds~\cite{PHA-91}) $D_\infty$ can be expressed
in terms of the universal amplitude ratios $Q_3$, $R_c$, and $R_\xi^+$
(see Ref.~\cite{PHA-91} for
their definitions and numerical values) and the non-universal bulk
amplitudes $\m_0$ and $\xi_0^+$ (see also Ref.~\cite{KD-99}), as
\begin{equation}
D_\infty = \frac{Q_3 R_c}{(R_\xi^+)^d} \m_0^2(\xi_0^+)^{d-2+\eta}\;.
\label{Dinftyuniratio}
\end{equation}
From Eq.~\reff{defDinf}, via a Fourier transform in one of the $d$
dimensions, one finds
\begin{equation}
C_{\rm crit}^{\rm bulk}
({\bf p}, x_{1\bot}, x_{2\bot}=x_{1\bot}, t=0) = {\mathcal
G}_V p^{-1+\eta}
\label{amplitudeCbulk}
\end{equation}
where ${\mathcal G}_V$ has been introduced in Appendix A of 
Ref.~\cite{KD-99} and is given by
\begin{equation}
{\mathcal G}_V = \frac{D_\infty}{2\sqrt{\pi}}
\frac{\Gamma(1/2-\eta/2)}{\Gamma(1-\eta/2)} \;.
\end{equation}
In view of Eqs.~\reff{amplitudeCbulk} and~\reff{scalCt} this leads to
\begin{equation}
\hat{\mathfrak{o}}^\pm_C = {\mathcal G}_V
\rt_0^\pm(\xi_0^\pm)^{1-\eta} = \frac{1}{2\sqrt{\pi}}
\frac{\Gamma(1/2-\eta/2)}{\Gamma(1-\eta/2)} \frac{Q_3
R_c}{(R_\xi^+)^d} \m_0^2\rt_0^\pm(\xi_0^\pm)^{d-1}\;.
\label{defoC}
\end{equation}
The FDT (see Eq.~\reff{fdtint}) establishes
the following relation between the correlation and response function:
\begin{equation}
C({\bf p}, x_{1\bot}, x_{2\bot}, t=0) = 
R({\bf p}, x_{1\bot}, x_{2\bot}, \omega=0)
\end{equation}
so that in the bulk one has
\begin{equation}
R_{\rm crit}^{\rm bulk}({\bf p}, x_{1\bot}, x_{2\bot}=x_{1\bot}, \omega=0) 
= {\mathcal G}_V p^{-1+\eta} \;.
\end{equation}
Comparing this equation with Eq.~\reff{scalRomega} leads to
\begin{equation}
\hat{\mathfrak{o}}^\pm_R = {\mathcal G}_V (\xi_0^\pm)^{1-\eta} = 
\frac{1}{2\sqrt{\pi}}
\frac{\Gamma(1/2-\eta/2)}{\Gamma(1-\eta/2)} \frac{Q_3
R_c}{(R_\xi^+)^d} \m_0^2(\xi_0^\pm)^{d-1}\;.
\label{defoR}
\end{equation}
Thus the non-universal amplitudes of $R$ and $C$ are determined
by the experimentally accessible non-universal bulk amplitudes $\m_0$,
$\xi_0^+$, and $\rt_0^+$. This fixes the normalization of the scaling
functions ${\bar\RR_\pm}$ and ${\bar\CC_\pm}$. 
For the present model and within 
mean-field theory these non-universal amplitudes are
\begin{equation}
D_\infty^{(0)} = 1
\end{equation}
and
\begin{equation}
{\mathcal G}_V^{(0)} = \frac{1}{2}
\end{equation}
so that
\begin{equation}
\hat{\mathfrak{o}}^{\pm(0)}_C = \frac{\rt_0^\pm \xi_0^\pm}{2}
\label{amplitudeC}
\end{equation}
and
\begin{equation}
\hat{\mathfrak{o}}^{\pm(0)}_R = \frac{\xi_0^\pm}{2} \;. 
\label{amplitudeR}
\end{equation}
Here and in the following with the superscript $(0)$ we indicate the mean-field value
of the quantities which the superscript refers to.
The
FDT (see Eq.~\reff{fdtint}) provides, together with Eqs.~(\ref{defoC})
and (\ref{defoR}),
the following relation between the scaling functions $\bar \RR_\pm$ and
$\bar \CC_\pm$ (and thus between  $\RR$ and
$\CC$):
\begin{equation}
\bar \CC_\pm({\bf \bar p},\bar x_{1\bot},\bar x_{2\bot},\bar t,\bar
L) = \int_{|\bar t|}^\infty \!\! \dd\bar s\, \bar\RR_\pm({\bf \bar p},\bar
x_{1\bot},\bar x_{2\bot},\bar s,\bar L).
\end{equation}

In the following we shall be mainly concerned with the case $\tau >
0$. In order to avoid a clumsy notation we thus shall omit in the
following the specification
$\pm$ from scaling forms and amplitudes. 

In Subsec.~\ref{sec-response} we shall discuss the behavior of the
response function in the semi-infinite geometry close to a confining
wall (see, c.f., Eq.~\reff{RSDsemiinf}). It can be obtained from the scaling
function Eq.~\reff{scalRt} in the limit $L\rightarrow\infty$ with
$x_{i\bot}$, $\xi$, and $t$ fixed. Thus one expects a
well-defined limit for the response function, i.e.,
\begin{equation}
\begin{split}
&{\bar\RR}({\bf
p}L,x_{1\bot}/L,x_{2\bot}/L,(t/\rt_0)(\xi_0/L)^z,L/\xi) \\
&\quad\quad 
\xrightarrow[L\rightarrow\infty]{}
\left(\frac{L}{\xi}\right)^{-(1-\eta-z)} {\bar\RR}^{\infty/2}({\bf
p}\xi,x_{1\bot}/\xi,x_{2\bot}/\xi,(t/\rt_0)(\xi_0/\xi)^z)
\end{split}
\label{scalRsemiinf}
\end{equation}
where $ {\bar\RR}^{\infty/2}$ is the scaling function for the
semi-infinite geometry. By using the short-distance
expansion~\cite{Diehl-86} one easily concludes that for
$x_{i\bot}\rightarrow 0$ (i.e., $x_{i\bot}\ll \xi,|{\bf p}|^{-1},
\xi_0(t/\rt_0)^{1/z}$)
\begin{equation}
\begin{split}
&
{\bar\RR}^{\infty/2}({\bf
p}\xi,x_{1\bot}/\xi,x_{2\bot}/\xi,(t/\rt_0)(\xi_0/\xi)^z)\\
&\hspace{1cm} 
\xrightarrow[x_{i\bot}\rightarrow 0]{}
\left(\frac{x_{1\bot}}{\xi}\frac{x_{2\bot}}{\xi}
\right)^{(\beta_1-\beta)/\nu} {\bar\RR}^{\infty/2}_W({\bf
p}\xi, (t/\rt_0)(\xi_0/\xi)^z)
\end{split}
\label{scalingSDRt}
\end{equation}
where $\beta_1$ is the critical exponent for the surface
magnetization~\cite{Diehl-86} ($\beta_1=1$ at the ordinary transition
within mean-field approximation, whereas 
$\beta_1\simeq 0.77(2)$~\cite{Diehl-86} for the
three-dimensional Ising universality class). 
Considering the case ${\bf p} = {\bf 0}$ and
$T\rightarrow T_{c,b}$ (i.e., $\xi \rightarrow\infty$) one expects
\begin{equation}
\lim_{y \rightarrow 0}  {\bar\RR}^{\infty/2}_W({\bf
0}, y) = {\mathcal D} y^{-2 (\beta_1-\beta)/(\nu z) + (1-\eta-z)/z}\;,
\label{scalingSDRtgen}
\end{equation}
where ${\mathcal D}$ is a universal constant. 
Thus, for $T=T_{c,b}$ and $x_{i\bot}\ll \xi_0(t/\rt_0)^{1/z}$
\begin{equation}
R_{\infty/2}({\bf p}={\bf 0},x_{1\bot},x_{2\bot},t\rightarrow\infty)
=
{\mathcal D} \frac{\hat{\mathfrak{o}}_R}{\rt_0} \left(
\frac{\rt_0}{t}\right)^{2(\beta_1-\beta)/(\nu z) - (1-\eta-z)/z} 
\left(\frac{x_{1\bot}}{\xi_0}\frac{x_{2\bot}}{\xi_0}
\right)^{(\beta_1-\beta)/\nu}
\label{scalingSDRtfinal}
\end{equation}
where $R_{\infty/2}$ is the response function in the semi-infinite
geometry.

In Sec.~\ref{sec-correlation} we shall discuss in detail the
mean-field behavior of the correlation function $C({\bf
p},x_\bot,x_\bot,\omega)$ in planes parallel to the
confining walls. For this function we present in the following
some scaling properties valid beyond the mean-field approximation. 

From a short-distance expansion one concludes that the scaling
function $\CC$ (see Eq.~\reff{scalComega}) 
of the correlation function $C$ behaves as
\begin{equation}
\begin{split}
&
\CC({\bf p}L, x_{1\bot}/L, 
x_{2\bot}/L,\omega\rt_0(L/\xi_0)^z,L/\xi) \\
&\hspace{1cm} 
\xrightarrow[x_{i\bot}\rightarrow 0]{}
\left(\frac{x_{1\bot}}{L}\frac{x_{2\bot}}{L}\right)^{(\beta_1-\beta)/\nu}\CC_W({\bf p}L,\omega\rt_0(L/\xi_0)^z,L/\xi) 
\end{split}
\label{scalComegaWall}
\end{equation}
for $x_{i\bot} \ll L, \xi_0(\omega \rt_0)^{-1/z}, \xi, |{\bf
p}|^{-1}$, 
where $\CC_W$ is
the universal scaling function associated with the behavior close to
the walls (in this case the one located at $x_\bot = 0$). The behavior of
$\CC_W$ allows one to define the
universal constant
\begin{equation}
{\mathcal A}^W \equiv \CC_W({\bf 0},0,0) \;,
\label{defA}
\end{equation}
that appears in the scaling behavior 
\begin{equation}
C_{\rm crit}({\bf p}={\bf 0},x_\bot,x_\bot,\omega =0) 
= \hat{\mathfrak{o}}_C\,{\mathcal A}^W
\left(\frac{L}{\xi_0}\right)^{1-\eta+z}
\left(\frac{x_\bot}{L}\right)^{2(\beta_1-\beta)/\nu} 
\label{CCW000}
\end{equation}
of the critical (i.e., $\tau=0$) correlation function for $x_\bot \ll L$. 
Analogously we can define the following universal constants:
\begin{equation}
\lim_{w\rightarrow \infty} \CC_W({\bf 0},0,w) = {\mathcal
A}^W_{\infty} w^{2(\beta_1-\beta)/\nu - (1-\eta+z)}\;,
\label{CW00x}
\end{equation}
\begin{equation}
\lim_{v\rightarrow \infty} \CC_W({\bf 0},v,0) = {\mathcal
B}^W_{\infty} v^{2(\beta_1-\beta)/(\nu z) - (1-\eta+z)/z} \;,
\label{CW0x0}
\end{equation}
and
\begin{equation}
\lim_{{\bf u}\rightarrow \infty} \CC_W({\bf u},0,0) = {\mathcal
C}^W_{\infty} |{\bf u}|^{2(\beta_1-\beta)/\nu  - (1-\eta+z)}
\label{CWx00}
\end{equation}
entering into the scaling functions
\begin{equation}
C({\bf p}={\bf 0}, x_\bot, x_\bot,\omega = 0) = \hat{\mathfrak{o}}_C  {\mathcal
A}^W_{\infty} \left(\frac{\xi}{\xi_0}\right)^{1-\eta+z}\left(\frac{x_\bot}{\xi}\right)^{2(\beta_1-\beta)/\nu}
\label{CCW00x}
\end{equation}
for $x_\bot\ll \xi \ll L$,
\begin{equation}
C_{\rm crit}({\bf p}={\bf 0}, x_\bot, x_\bot,\omega) = \hat{\mathfrak{o}}_C  {\mathcal
B}^W_{\infty} (\omega\rt_0)^{-(1-\eta+z)/z+2(\beta_1-\beta)/(\nu
z)}\left(\frac{x_\bot}{\xi_0}\right)^{2(\beta_1-\beta)/\nu}
\label{CCW0x0}
\end{equation}
for $(\xi_0/L)^z \ll \omega\rt_0 \ll (\xi_0/x_\bot)^z$, and
\begin{equation}
C_{\rm crit}({\bf p}, x_\bot, x_\bot,\omega=0) = 
\hat{\mathfrak{o}}_C  {\mathcal
C}^W_{\infty} (|{\bf p}|\xi_0)^{-(1-\eta+z)+2(\beta_1-\beta)/\nu}\left(\frac{x_\bot}{\xi_0}\right)^{2(\beta_1-\beta)/\nu}
\label{CCWx00}
\end{equation}
for $1/L \ll |{\bf p}| \ll 1/x_\bot$, respectively.

From the previous equations we 
recover the values of well-known surface critical
exponents for the semi-infinite geometry,
i.e., $\sigma_{\tau}^{(s)}$, $\sigma_\omega^{(s)}$, and
$\sigma_{p}^{(s)}$~\cite{DD-83,D-90}. They describe the
divergence of the two-point correlation function parallel to the
surface, so that for ${\bf p}={\bf 0}$ and $\omega = 0$ it diverges $\sim
\tau^{-\sigma^{(s)}_\tau}$, for $\tau=0$ and $\omega = 0$ 
it diverges $\sim |{\bf
p}|^{-\sigma^{(s)}_p}$, whereas for ${\bf p}={\bf 0}$ and $\tau = 0$
it 
diverges for $\omega\rightarrow 0$ 
as $\omega^{-\sigma^{(s)}_\omega}$~\cite{D-90}. 

As far as the behavior of the correlation function in the middle of the film,
i.e., $C({\bf p},L/2,L/2,\omega)$ is concerned we define (see Eq.~\reff{scalComega})
\begin{equation}
\CC({\bf p}L,1/2,1/2,\omega\rt_0(L/\xi_0)^z,L/\xi) = \CC_I({\bf p}L,\omega\rt_0(L/\xi_0)^z,L/\xi) 
\label{CCI}
\end{equation}
where $\CC_I$ is a universal scaling function. As already discussed
for $\CC_W$ we can define also from $\CC_I$ the following universal
constants:
\begin{align}
 \CC_I({\bf 0},0,0) &= {\mathcal A}^I \;,\label{CI000}\\
\lim_{w\rightarrow \infty} \CC_I({\bf 0},0,w) &= {\mathcal A}_\infty^I
w^{-(1-\eta+z)}\;,\label{CI00x}\\
\lim_{v\rightarrow \infty} \CC_I({\bf 0},v,0) &= {\mathcal B}_\infty^I
v^{-(1-\eta+z)/z}\;,\label{CI0x0}
\end{align}
and
\begin{equation}
\lim_{{\bf u}\rightarrow \infty} \CC_I({\bf u},0,0) = {\mathcal C}_\infty^I
|{\bf u}|^{-(1-\eta+z)}\;.\label{CIx00}
\end{equation}
These constants enter into the following scaling functions:
\begin{align}
C_{\rm crit}({\bf p}={\bf 0},L/2,L/2,\omega = 0) &= {\mathcal A}^I
\hat{\mathfrak{o}}_C 
\left(\frac{L}{\xi_0}\right)^{1-\eta+z}\;, \label{CCI}\\
C({\bf p}={\bf 0},L/2,L/2,\omega = 0) &=
\hat{\mathfrak{o}}_C {\mathcal A}^I_\infty
\left(\frac{\xi}{\xi_0}\right)^{1-\eta+z} \!\quad\! \mbox{for} \!\quad\!
\xi\ll L\;,\label{CCI00x}\\
C_{\rm crit}({\bf p}={\bf 0},L/2,L/2,\omega) &=
\hat{\mathfrak{o}}_C {\mathcal B}^I_\infty
(\omega \rt_0)^{-(1-\eta+z)/z} \!\!\quad\! \mbox{for} \!\quad\!
\omega\rt_0\ll \left(\frac{\xi_0}{L}\right)^z\!\!,\label{CCI0x0}
\end{align}
and
\begin{equation}
C_{\rm crit}({\bf p},L/2,L/2,\omega = 0) =
\hat{\mathfrak{o}}_C {\mathcal C}^I_\infty
({|{\bf p}|\xi_0})^{-(1-\eta+z)} \quad \mbox{for} \quad
|{\bf p}| \gg \frac{1}{L}\;,\label{CCIx00}
\end{equation}
respectively.
In Sec.~\ref{sec-correlation} we shall confirm these scaling forms
within mean-field approximation and determine also the mean-field
values of the universal constants involved.

\subsection{Response Function}
\label{sec-response}

\begin{figure*}
\begin{center}
\epsfig{file=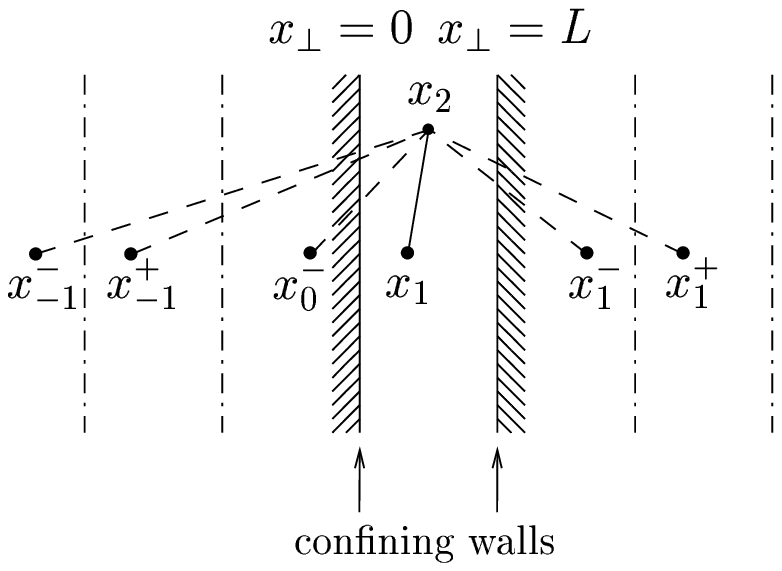,width=0.50\textwidth} 
\end{center}
\caption{Set of images corresponding to a perturbation actually
applied at $x_1$. $x_2$ is the point at which the effects of the
perturbation are observed. Image sources are obtained by successive
reflections of the real and image sources 
with respect to the confining walls. At each reflection the sign of
the contribution to the total response changes, starting from a positive
sign for the actual source.}
\label{images}
\end{figure*}

Our aim here is to discuss, in different representations, the response
function introduced in Sec.~\ref{sec-model}.
Combining the analogue of Eq.~\reff{reprC3} for the response function
and taking into account the explicit expression in Eq.~\reff{resp} we
have
\begin{equation}
\int_{-\infty}^{+\infty}\!\!\frac{\dd\omega}{2\pi}
e^{-i\omega t} R^{(0)}({\bf q}_n,\omega) = \theta(t)\, \Omega\, e^{-\Omega({\bf
q}_n^2 + r_0)t}
\end{equation}
and thus
\begin{equation} 
R^{(0)}({\bf p},x_{1\bot},x_{2\bot},t) = 
\theta(t)\,\Omega\, e^{-\Omega({\bf
p}^2+r_0)t} \sum_{n=1}^\infty \Phi_n(x_{1\bot};L)\Phi_n(x_{2\bot};L)
e^{-\Omega(\pi n/L)^2 t} .
\label{Rgeneral}
\end{equation}
Using the results of Appendix~\ref{app-formulas} we can write
\begin{equation}
R^{(0)}({\bf p},x_{1\bot},x_{2\bot},t) = 
\theta(t)\frac{\xi_0^z}{\rt_0}\frac{1}{L}e^{-[({\bf
p} L)^2 +(L/\xi)^2](t/\rt_0)(\xi_0/L)^z}
\Psi(x_{1\bot}/L,x_{2\bot}/L,(t/\rt_0)(\xi_0/L)^z)
\label{Rgenpsi}
\end{equation}
with the mean-field expression of the 
scaling function $\Psi$ given in Eq.~\reff{defpsi}, which, within MFT, 
does not depend on ${\bf p}$ or $\xi$. Within the same approximation
the scaling variable $(t/\rt_0)(\xi_0/L)^z$ is given by 
\begin{equation}
\bar t\equiv (t/\rt_0)(\xi_0/L)^z = \Omega t/L^2 \; .
\label{omegascal}
\end{equation}
In favor of a compact notation in the following we use 
this abbreviation
keeping in mind that it can be replaced by
the r.h.s. of Eq.~\reff{omegascal}.
The scaling
properties of the response function clearly emerge from this
expression. Comparing with the general scaling form Eq.~\reff{scalRt}
it is easy to see that (${\bf \bar p} = {\bf p}L$, $\bar x_{i\bot} =
x_{i\bot}/L$, $\bar L = L/\xi$)
\begin{equation}
\bar \RR^{(0)}({\bf \bar p}, \bar x_{1\bot}, \bar x_{2\bot},\bar
t,\bar L) = 2 e^{-({\bf
\bar p}^2 +\bar L^2)\bar t}
\Psi(\bar x_{1\bot},\bar x_{2\bot},\bar t\,)
\label{RPsi}
\end{equation}
where we used Eq.~\reff{amplitudeR}. 
Moreover, one can easily recover the result for the semi-infinite
geometry. Indeed, using Eq.~\reff{onewall} we find that for
$\bar x_{i\bot}\ll 1$ (i.e., close to the near
wall at $x_\bot = 0$) and $\bar t \ll 1$
(so that the influence from the wall at $x_\bot = L$ can be neglected 
near $x_\bot=0$) 
\begin{equation}
\begin{split}
&
R^{(0)}({\bf p},x_{1\bot}\ll L,x_{2\bot}\ll L,t) =\\
&\hspace{3cm}
\theta(t)\frac{\xi_0^2}{\rt_0}\frac{1}{L} \frac{e^{- ({\bar{\bf
p}}^2+\bar L^2)\bar t}}{\sqrt{4\pi\bar t}}
\left[ e^{-(\bar x_{1\bot}-\bar x_{2\bot})^2/(4\bar t\,)} -
e^{-(\bar x_{1\bot}+\bar x_{2\bot})^2/(4\bar t\,)} \right]
\end{split}
\label{Ronewall}
\end{equation}
in agreement with Eqs.~(II.18) and (II.19) in
Ref.~\cite{DD-83}. Note that the previous expression is indeed
independent of $L$, as expected for the limit we are considering.
Equation~\reff{onewall} provides also a representation of the response
function in terms of the bulk response to a set of image
excitations. Let us recall that the response function in the bulk is
given by
\begin{equation}
R^{(0)}_{\rm bulk}({\bf p},x_{1\bot},x_{2\bot},t) =  
\theta(t)\frac{\xi_0}{\rt_0}e^{-({\bf
p}^2\xi_0^2+\xi_0^2/\xi^2)t/\rt_0} \frac{1}{\sqrt{4\pi t/\rt_0}}
e^{-(x_{1\bot}-x_{2\bot})^2/(4\xi_0^2 t/\rt_0)}\; .
\end{equation}
According to Eqs.~\reff{Rgenpsi} and~\reff{onewall} one has
\begin{equation}
R^{(0)}({\bf p},x_{1\bot},x_{2\bot},t) = \sum_{n=-\infty}^{+\infty}
[R^{(0)}_{\rm bulk}({\bf p},x^+_{n\bot},x_{2\bot},t) - R^{(0)}_{\rm
bulk}({\bf p},x^-_{n\bot},x_{2\bot},t)]
\label{imageR}
\end{equation}
where $x^+_{n\bot} \equiv x_{1\bot} + 2 n L$ and $x^-_{n\bot}\equiv -
x_{1\bot} + 2 n L$. The set $\{x^+_{n\bot}\}_{n\neq 0}$ represents the
positions at which the ``positive'' images are located, 
whereas $\{x^-_{n\bot}\}_{n}$ gives the positions of the ``negative''
ones. This construction is illustrated in Fig.~\ref{images}. By virtue
of the FDT Eq.~(\ref{imageR}) is also valid for the correlation
function, with $R$ replaced by $C$ in the expression. Moreover it 
is an extention to the dynamics of the analogous
formula known for the static correlation function (see, e.g., Subsec. IVA
of Ref.~\cite{Diehl-86} and Ref.~\cite{BAF-03}). 

%
%

Let us consider the long-time limit of the response
function in Eq.~\reff{Rgeneral}. For 
$\bar t \gg 1$, i.e., when the effect of confinement is no longer 
negligible, the sum in Eq.~\reff{Rgeneral} is dominated by the lowest
mode of the system, i.e., by $n=1$. Thus one has
\begin{equation}
R^{(0)}({\bf p},x_{1\bot},x_{2\bot},t\rightarrow\infty) = 
\theta(t)\frac{\xi_0^2}{\rt_0}e^{-({\bar{\bf p}}^2+\bar L^2)\bar t} \left[\Phi_1(x_{1\bot};L)\Phi_1(x_{2\bot};L)
e^{-\pi^2\bar t}  + O(e^{-4\pi^2\bar t})\right],
\label{Rproject}
\end{equation}
i.e., the linear response to an
external perturbation decays in time {\it exponentially} with a factor
$\exp[-(\bar{\bf q}_1^2 + \bar L^2)\bar t]$
where, according to our notation,
${\bf q}_n = ({\bf p}, \pi n/L)$. This exponential decay also holds
for the two-point correlation functions. Even at $T_{c,b}$
and for excitations which do not break the translational invariance in
lateral direction ${\bf x}_\|$, i.e., for ${\bf p}=0$, there is an exponential
decay due to $|{\bf q}_1|\ge\pi/L$. This is the result of the
combined effect of confinement {\it and} Dirichlet boundary conditions
which suppress homogeneous modes (with
vanishing total momentum ${\bf q}_0 = {\bf 0}$) in the system.
In the case of periodic and Neumann-Neumann boundary conditions for
this confined system we expect an {\it algebraic} 
decay as function of
time for the critical response to an external perturbation with ${\bf
p}={\bf 0}$, given that the mode with ${\bf q}_0 = {\bf 0}$
does occur in the corresponding spectra. On the other hand an
algebraic decay can be recovered also in the case we are
considering. Indeed for $T < T_{c,b}$ the dimensionless variable $\bar
L^2 = (L/\xi)^2$ appearing in the previous equations has to be
replaced by $-1/2 (L/\xi)^2$ within MFT leading to the decay 
$\sim\exp[-(\bar{\bf q}_1^2 - \bar L^2/2)\bar t]$. Accordingly, upon
decreasing the temperature $T$ a critical value $T_c(L)<T_{c,b}$ exist for
which this exponent vanishes. For $T=T_c(L)$
the bulk correlation length $\xi$ attains the value $\xi_c =
L/\sqrt{2}\pi$. This corresponds to the critical-point shift in film
geometry, that can also be determined from the onset of
a non-trivial order-parameter profile (see Appendix~\ref{app-OPprofile}).  
From this point of view the fact that at $T=T_{c,b}$ the
response and correlation functions decay exponentially reflects that 
$T_{c,b}$ is above the critical temperature of
the system in the film, i.e., located within the disordered phase
of the film.
In the case of a semi-infinite system the asymptotic decay of
the response function
is indeed algebraic even for Dirichlet boundary conditions, 
as one can see directly from Eq.~\reff{Ronewall}
for ${\bf p}={\bf 0}$, $T=T_{c,b}$, and $t/\rt_0 \gg x^2_{i\bot}/\xi_0^2$:
\begin{equation}
R_{\infty/2}^{(0)}({\bf p}={\bf
0},x_{1\bot},x_{2\bot},t\rightarrow\infty) = 
\frac{1}{\sqrt{4\pi}} \frac{\xi_0}{\rt_0} \left( \frac{t}{\rt_0}\right)^{-3/2}
\frac{x_{1\bot}}{\xi_0}\frac{x_{2\bot}}{\xi_0} 
[1 + O((x_{i\bot}/\xi_0)^2(\rt_0/t))].
\label{RSDsemiinf}
\end{equation}
This expression is in agreement with the general scaling form
in Eq.~\reff{scalingSDRtfinal} with the mean-field values of the
exponents and amplitudes (see Eqs.~\reff{omega}
and~\reff{amplitudeR}), and the universal constant ${\mathcal D}$ (see
Eq.~\reff{scalingSDRtgen}) takes the value
\begin{equation}
{\mathcal D}^{(0)} = \frac{1}{\sqrt{\pi}} \;.
\label{RD0}
\end{equation}
This can also be interpreted as reflecting 
the fact that in the semi-infinite
geometry there is no critical point shift. The same conclusion can be
reached in the case of the film geometry with periodic or
Neumann-Neumann boundary conditions which, as mentioned previously,
within mean-field theory do not lead to a critical point shift in films.

From Eq.~\reff{Rproject} we can see that for asymptotically large
times, the spatial dependence of the response function in the film
geometry is given by
$\sin(\pi x_{1\bot}/L)\sin(\pi x_{2\bot}/L)$.

According to Eq.~\reff{Rdef} the linear response function $R$ represent the
order-parameter profile due to a $\delta$-like perturbation
applied at an early time. Of course, being derived in linear
approximation, this function is useful only as long as the subsequent
values assumed by the order parameter are small enough compared to
nonlinear terms. The case of nonlinear relaxation will be discussed in 
Sec.~\ref{sec-nlin}. 

\begin{figure*}
\epsfig{file=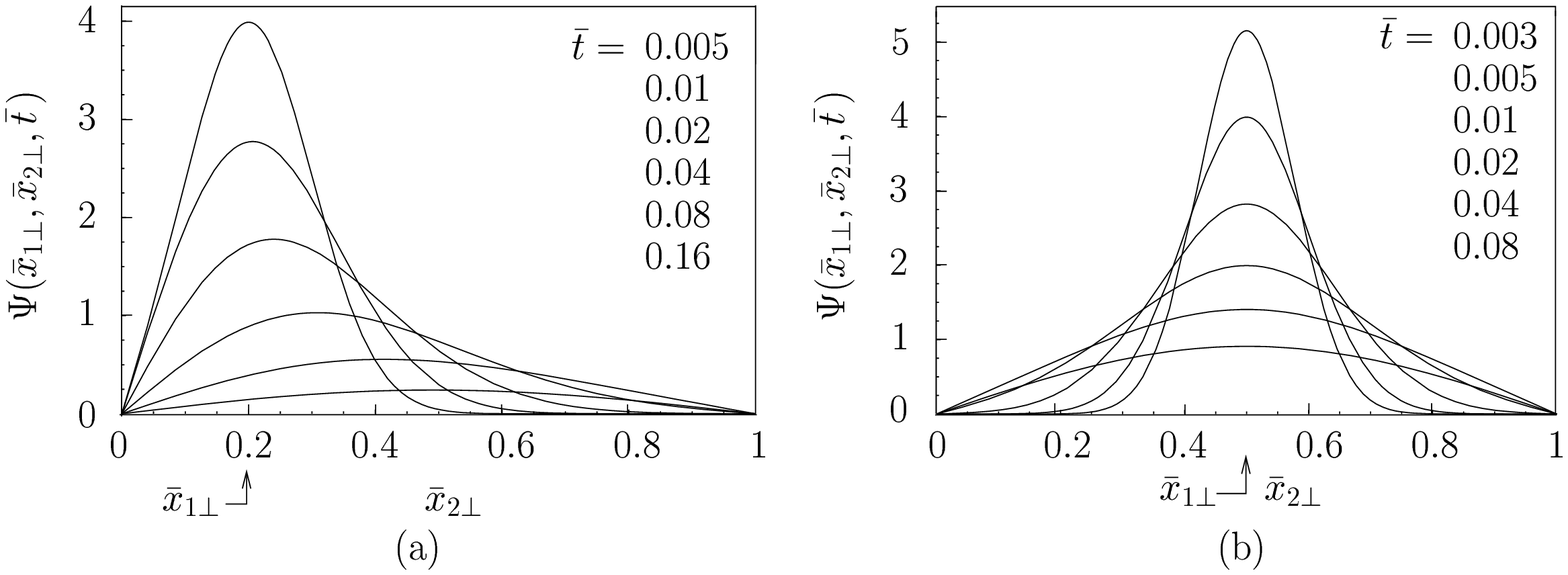,width=1\textwidth} 
\caption{Time evolution of the scaling function $\Psi(\bar x_{1\bot},\bar
x_{2\bot},\bar t\,)$ which enters into the expression of the response
function in Eq.~\reff{Rgenpsi} (see also Eqs.~\reff{scalRt} and~\reff{RPsi}) with $\bar x_{i\bot} = x_{i\bot}/L$ and
$\bar t = (t/\rt_0)(\xi_0/L)^z$. In (a) the 
relaxation follows an excitation at
the point $\bar x_{1\bot} = 0.2$ and in (b) at the point $\bar x_{1\bot}
= 0.5$. Reduced times $\bar t$ listed in (a) and (b) 
refer to the various curves shown from top to bottom.}
\label{Rreal}
\end{figure*}

In Fig.~\ref{Rreal} the function $\Psi(\bar x_{1\bot}, \bar
x_{2\bot},\bar t\,)$, i.e., $R^{(0)}({\bf
p},x_{1\bot},x_{2\bot},t)$ up to a spatially constant prefactor (see Eqs.~\reff{scalRt}, \reff{Rgenpsi}, and~\reff{RPsi}),
is shown for two
values of $\bar x_{1\bot}$, i.e., the point at which the perturbation has
been applied at time $t=0$. In accordance with our previous observation, the
response function for $\bar t \gg 1$ turns into a sine function with
period $\bar x_{2\bot} = 2$.
We observe clearly that there is a qualitative change
in the shape of the responding order-parameter profile as time increases. In
particular the inflection points, which are present just after the
perturbation has been applied~\footnote{%
According to Eq.~\reff{onewall}, for $\bar t \ll 1$, $\Psi(\bar
x_{1\bot}, \bar x_{2\bot}, \bar t\,)$ has a Gaussian form as function
of $\bar x_{2\bot}$ for fixed $\bar x_{1\bot}$ and vice versa.
}, disappear in the
long-time limit, after having reached the closest surfaces.

\begin{figure*}
\begin{center}
\epsfig{file=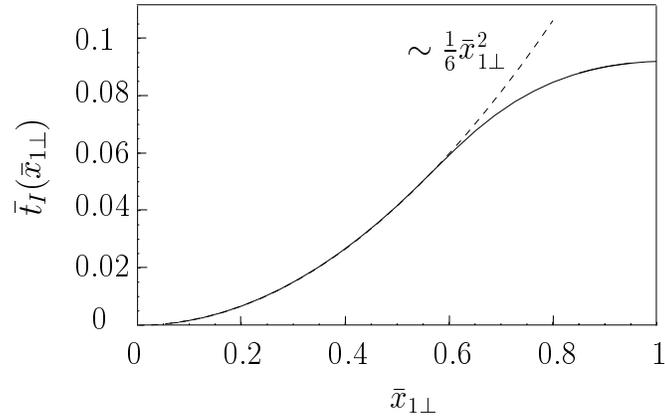,width=0.55\textwidth} 
\end{center}
\caption{Time $t_I(\bar x_{1\bot})$ at which the inflection
in the response function disappears as a function of the position
$\bar x_{1\bot}$ at which the perturbation is applied. The dashed curve
$\bar t(\bar x_{1\bot}) = \bar x_{1\bot}^2/6$
is the quadratic behavior expected for small $\bar x_{1\bot}$ which
actually describes the curve even for a wide range of values of 
$\bar x_{1\bot}$.}
\label{tflex}
\end{figure*}

In Fig.~\ref{tflex} we report 
the time $\bar t_I(\bar
x_{1\bot})$ at which the inflection point (of $\Psi$ as a function of
$\bar x_{2\bot}$) close to the wall at $\bar x_\bot
=0$ disappears. Given the symmetry of the problem the analogous time for
the inflection close to the wall at $\bar x_{\bot} = 1$ is simply
given by $\bar t_I(1-\bar x_{1\bot})$. The behavior of 
$\bar t_I(\bar x_{1\bot}\rightarrow 0)$ 
can be easily predicted by taking into account 
that in the semi-infinite geometry, for a given
$x_{1\bot}$, we expect a finite non-zero value $t_I$. According to this
argument, in
the limit $L\rightarrow\infty$ with fixed $x_{1\bot}$, the relation
\begin{equation}
\bar t_I = F_{t_I}(\bar x_{1\bot})
\label{defFI}
\end{equation}
should become independent of $L$, i.e., (see Eq.~\reff{omegascal})
\begin{equation}
F_{t_I}(y\rightarrow 0) \sim y^2 
\end{equation}
and thus (see Fig.~\ref{tflex})
\begin{equation}
\bar t_I(\bar x_{1\bot}\rightarrow 0) \sim \bar x^2_{1\bot}
\; .
\label{tIorig}
\end{equation}
As discussed in Appendix~\ref{app-tI} it is possible to determine
analytically, within MFT, 
the proportionality factor in Eq.~\reff{tIorig}, which turns out 
to be $1/6$. Moreover $F_{t_I}(y=1)$ can also be determined.

Interestingly, within mean-field theory it is possible to prove
(c.f., Subsec.~\ref{sec-casimir}) that 
$t_I(\bar x_{1\bot})$ has also the meaning of being 
the time at which the fluctuation-induced force acting on the
confining walls is maximal.

Figures~\ref{Rreal}(a) and~\ref{Rreal}(b) clearly show that for $\bar
t$ small enough the order parameter profile is well localized around
the point at which the perturbation has been applied, in accordance
with the expectation that the effects of the perturbation 
reach the different points of the system only with a certain delay.
On the other hand, irrespective of how small $\bar t$
is, the order parameter is non-zero in the {\it whole} range
$0<\bar x_{2\bot} < 1$ (even {\it everywhere} in the case of unbounded
geometries), as one realizes from Eqs.~\reff{RPsi}
and~\reff{defpsi}. This absence of a {\it finite} front propagation
speed is a consequence of the coarse-grained
description underlying the field-theoretical approach, in which the
{\it microscopic} time and length scales are assumed to be negligible
compared to the {\it mesoscopic} ones, to the effect that the
microscopic dynamics, which of course exhibits a speed limit for the
front propagation, appears to be actually arbitrarily fast. 
(This is analogous to the case of random-walk models of free
diffusion and their corresponding continuum descriptions.)


For studies of the dynamical properties in the film geometry by means of
elastic scattering experiments 
one is interested in the two-point correlation
function in the $({\bf p},x_\bot,\omega)$-representation. (The
corresponding static
properties of thin films near continuous phase transitions 
have been studied theoretically in Ref.~\cite{KD-99}.) By applying
the FDT we can compute the corresponding
function once the
expression for the response function is known in the same
representation. In doing so we can take advantage of the analytical
results known for the static correlation function (see
Ref.~\cite{KED-95} and Appendix~\ref{app-staticcorr}) 
given that, apart from a factor $\Omega^{-1}$, the expression for $R^{(0)}$ in
Eq.~\reff{resp} is related to that for the static correlation
function (i.e., $\int\!\dd\omega/(2\pi) C^{(0)}({\bf q},\omega) =
1/({\bf q}^2 + \xi^{-2})$)
by means of a formal shift $\xi^{-2}\mapsto \xi^{-2} - i\omega/\Omega$.
Using Eq.~\reff{Cstatstand} we find~\footnote{%
Note that fixing the branch 
of the square root defining $a = \sqrt{{\bf p}^2 + \xi^{-2} -
i\;\omega/\Omega}$ is irrelevant, because Eq.~\reff{Cstat} is
symmetric in $a$. 
\label{footnote3}
}
\begin{equation}
\label{Romega}
\begin{split}
R^{(0)}({\bf p},x_{1\bot},x_{2\bot},\omega) &= 
L\frac{\sinh(a x_\bot^</L)\sinh[a
(L- x_\bot^>)]}{a L
\sinh(a L)}\;,\\ 
a^2 &\equiv {\bf p}^2 + \xi^{-2} - i\omega/\Omega\;,
\end{split}
\end{equation}
where $x_\bot^> = {\rm max}\{x_{1\bot},x_{2\bot}\} =
(x_{1\bot}+x_{2\bot}+|x_{1\bot}-x_{2\bot}|)/2$, and
$x_\bot^< = {\rm min}\{x_{1\bot},x_{2\bot}\} =
(x_{1\bot}+x_{2\bot}-|x_{1\bot}-x_{2\bot}|)/2$.
Equation~\reff{Romega} agrees 
with what was found in Ref.~\cite{Bhat-96}, Eq.~(7), and 
Ref.~\cite{NG-03}, Eq.~(13).
Using Eq.~\reff{amplitudeR} 
one can write this expression in the scaling form 
given in Eq.~\reff{scalRomega}: 
\begin{equation}
\RR^{(0)}({\bf \bar p},\bar x_{1\bot},\bar
x_{2\bot},\bar\omega,\bar L) = 2 \frac{\sinh(\bar a
\bar x_\bot^<)\sinh[\bar a
(1- \bar x_\bot^>)]}{\bar a 
\sinh \bar a } \;, \label{scalR}
\end{equation}
where $\bar x_\bot^{<,>} = x_\bot^{<,>}/L$,
\begin{equation}
\bar a^2 \equiv {\bf \bar p}^2 + \bar L^2 - i\bar
\omega \;,\quad \bar{\bf p} = {\bf p} L, \quad \bar L = L/\xi,
\quad\mbox{and}\quad \bar\omega =
(\omega \rt_0)(L/\xi_0)^z.
\label{abar}
\end{equation}

\subsection{Casimir force}
\label{sec-casimir}

In a confined system the spectrum of the allowed critical fluctuations of the
order parameter is modified compared to the bulk case, depending on
the specific boundary conditions (i.e., surface universality
classes) and on the film thickness $L$. This leads to a finite-size
contribution to the free energy. Accordingly, by varying $L$
one observes that the confining walls are subject to an
$L$-dependent effective force $F$ per cross-section area of the film 
and per $k_{\rm
B}T_{c,b}$ (where $k_{\rm B}$ is the Boltzmann constant)
which is the statistical analogue~\cite{Krech-94,BDT-00} of
the Casimir force of quantum electrodynamics. 
In the case of the film geometry with the confining plates belonging
to $(a,b)$-surface universality classes
[the case we are currently interested in is the ordinary-ordinary
$(O,O)$ one] and in the {\it st}atic case
$F$ has been shown to scale as function of the thermodynamic
parameters (up to contributions from additive renormalizations) 
as~\cite{KD-92bis}
\begin{equation}
F(\tau,h,L) = \frac{1}{L^d}\,
\FF^{\rm (st)}_{a,b}((h/\h_0)(L/\xi_0)^{\beta\delta/\nu},L/\xi) 
\label{scal_cas}
\end{equation} 
where the universal exponent $\delta$ and the nonuniversal amplitude 
$\h_0$ are defined through the critical equation of state
\begin{equation}
h = \h_0 \frac{m}{\m_0}\left|\frac{m}{\m_0}\right|^{\delta-1}\;,
\label{defh0}
\end{equation}
which express the dependence of the bulk order parameter $m$ on the 
external field $h$ along the bulk critical isotherm $T=T_{c,b}$.
Within MFT one has $\delta = 3$.
Because of the two scale-factor universality (see, e.g,
Ref.~\cite{PHA-91}) the nonuniversal amplitude $\h_0$ is related to
the nonuniversal amplitudes $\m_0$ and $\xi_0$ introduced in
Sec.~\ref{intro} via 
\begin{equation}
\h_0 = \frac{R_\chi(R_\xi^+)^d}{R_c} (\xi_0^+)^{-d} \m_0^{-1}\,,
\label{defh0amp}
\end{equation}
where $R_c$, $R_\xi^+$ [see also Eq.~(\ref{Dinftyuniratio})], and
$R_\chi$ are universal amplitude ratios whose actual definitions and
values can be found in Ref.~\cite{PHA-91}. In Eq.~(\ref{scal_cas})
the function $\FF^{\rm (st)}_{a,b}$ 
is a universal scaling function with 
$\FF^{\rm (st)}_{a,b}(0,0)=(d-1)\Delta_{a,b}$. 
The amplitude $\Delta_{a,b}$ is the so-called universal
Casimir amplitude corresponding to $(a,b)$ surface universality
classes. 
The effective force $F$ between the
confining walls is attractive when $F<0$ and repulsive
otherwise. In this sense $F$ can be viewed as a special case of
the more general case of fluctuation-induced effective interaction between
different objects immersed in a critical medium.
In the following we consider only the case $(O,O)$ of ordinary-ordinary
boundary conditions and therefore we shall omit the specification $(a,b)$
from the scaling functions and amplitudes.
Within the Gaussian approximation $\Delta$ is
given by~\cite{KD-92}
\begin{equation}
\Delta = -\frac{1}{(4 \pi)^{d/2}} N \, \Gamma(d/2)\zeta(d)
\label{deltaOO}
\end{equation}
where $\Gamma(z)$ and $\zeta(z)$ denote the Gamma function and
Riemann's zeta function, respectively.
The universal amplitude $\Delta$ and the universal scaling function $\FF^{\rm
(st)}$ can be determined by computing the singular part of
the free energy of the system in confined geometry,
based on the 
Hamiltonian~\reff{lgw}. An alternative approach, more suited for
extensions to dynamics, is based on the connection between the Casimir
force and the expectation value of the
stress-tensor $T_{\mu\nu}$. 
(We refer the reader to the literature~\cite{Krech-94}
for details.) In particular one finds that
the force density per $k_{\rm B}T_{c,b}$ 
on one of the confining walls $\partial V$ is given by
\begin{equation}
F = \left.\langle T_{\bot\bot} \rangle\right|_{\partial V}
\end{equation}
where $T_{\bot\bot}$ is suitably defined in terms of the order
parameter field $\varphi$. For the case we are interested in one has,
in the absence of surface fields 
(see, e.g., Ref.~\cite{Krech-94}),
\begin{equation}
F_{l(r)}({\bf x}_\|) = \left.\langle T_{\bot\bot}\rangle\right|_{\partial V} 
= \frac{1}{2}\left.\langle\left(\frac{\partial
\varphi({\bf x})}{\partial x_\bot}\right)^2\rangle\right|_{{\bf x}\in \partial V} \,.
\label{casim-st}
\end{equation}
Note that in the {\it static} case, for which the stress-tensor is a
conserved quantity (i.e., $\partial \langle T_{\mu\nu}\rangle/\partial
x_\mu = 0$, where $\mu,\nu = \bot,\|$),  
$\langle T_{\bot\bot}\rangle$ is actually independent
of the point of evaluation, including the surfaces.  
In the {\it dynamic} case, in general the force density on the left
wall ($l$) is different from the one on the right wall ($r$) and it may vary
spatially along the walls. The mean force per area on the wall
$l$($r$) is given
by
\begin{equation}
F_{l(r)} = \frac{1}{A}\int_A {\rm d}^{d-1} {\bf x}_\|\; F_{l(r)}({\bf x}_\|)\;.
\end{equation}
The relation between the thermodynamic Casimir force (defined from the
finite-size behavior of the free energy, see, e.g.,
Ref.~\cite{Krech-94}) and the
expectation value of the stress tensor is based on the fact that the
{\it equilibrium} distribution function of the order parameter field
$\varphi$ is proportional to 
$\exp\{-{\mathcal H}[\varphi]\}$. In the case we are
interested in, ${\mathcal H}[\varphi]$ is the Landau-Ginzburg
Hamiltonian in Eq.~\reff{lgw}. (In turn, the specific expression of
the stress-tensor in terms of $\varphi$ (Eq.~\reff{casim-st}) is
determined by ${\mathcal H}$.)
When studying critical dynamics and the effects of time-dependent
external fields, such a connection is no longer evident because
equal-time correlation functions are
generated through the distribution
$\propto \exp\{-{\mathcal H}[\varphi]\}$ only
asymptotically for large times, i.e., long after any perturbation has
been switched off.
In principle it is not even obvious how to define a {\it
thermodynamic} Casimir force when studying dynamics, because strictly
speaking in this case the {\it equilibrium} free energy loses its
significance. 
Therefore we {\it assume} as the definition of the dynamic Casimir force
the time-dependent expectation value of the
equilibrium stress-tensor, which renders the static Casimir force in
thermal equilibrium (see also Refs.~\cite{NG-03,BAF-03}).
Heuristically, this amounts to assume that at each time, analogous to
thermal equilibrium,
there is an ``energy cost'' $A\delta\! L \langle T_{\bot\bot}\rangle$ 
associated with the displacement $\delta\! L$ of one of the confining
walls, i.e., $\langle
T_{\bot\bot}\rangle$ provides the {\it local
pressure}~\cite{NG-03,BAF-03}.
It is desirable to establish a clearer connection between this
definition of the dynamic Casimir force and the force that can be
measured directly in actual experiments and molecular dynamics
simulations.
The previous definition is particularly suited for field-theoretical
analysis. On the other hand it does not lend itself straightforwardly
for the study of the dynamic force via Monte Carlo simulations. 
First, the explicit expression of the stress tensor in terms of the
order parameter field can be determined in general only 
perturbatively in terms of the coupling constant $g_0$. 
Second, one has to construct the lattice version of the stress tensor in terms
of the microscopic degrees of freedom (e.g., spin variables); this
construction is in general not free from ambiguities. An alternative
approach to this problem has been taken in Ref.~\cite{DK-04} to study
the statistical fluctuations of the Casimir force~\cite{BAFG-02} via
Monte Carlo simulations. However, so far this approach could be
implemented only for periodic boundary conditions.

We now consider the case in which the film, thermodynamically close to
$T_{c,b}$, is perturbed by a
time-dependent external field $h({\bf x},t)$. For the ensuing {\it
dy}namic force density per $k_{\rm B} T_{c,b}$ exerted on one of
the confining walls one expects a scaling behavior, as in
Eq.~\reff{scal_cas},
\begin{equation}
\begin{split}
&F_{l(r)}({\bf x}_\|, \tau,L,t, \{h({\bf x},t)\}) 
= \\
&\qquad\frac{1}{L^d}\,
\FF^{\rm (dy)}_{a,b}({\bf x}_\|/L,L/\xi,  (t/\rt_0)(\xi_0/L)^z,
\{(h({\bf
x}/L,(t/\rt_0)(\xi_0/L)^z)/\h_0)(L/\xi_0)^{\beta\delta/\nu}\}) 
\end{split}
\label{scal_cas_dyn}
\end{equation} 
in terms of the scaling field
$\bar
h$ defined via 
\begin{equation}
h({\bf x},t) = \h_0 (L/\xi_0)^{-\beta\delta/\nu} \bar
h({\bf x}/L,(t/\rt_0)(\xi_0/L)^z). 
\end{equation}
In Eq.~\reff{scal_cas_dyn} we have
assumed that $L$ does not vary as function of time. 
For a time-independent,
spatially homogeneous
external field $h$, $\FF^{\rm (dy)}_{a,b}$ reduces to $\FF^{\rm
(st)}_{a,b}$ introduced in Eq.~\reff{scal_cas}.
As explained in Appendix~\ref{app-casimir}, within the Gaussian approximation 
it is possible to compute the force $F$
exactly for a general
external applied field $h({\bf x},t)$. In order to elucidate some
qualitative features of the dynamics of the Casimir force after the
perturbation, we consider
the particular case in which the field is instantaneously 
applied at a given time $t_1$ and then immediately
switched off again, i.e., $h({\bf x},t) = h({\bf x})\delta(t-t_1)$ with
$h({\bf x})$ localized in the interior of the film. After the perturbation
the response starts to propagate in the film. At the very early stage
the response has practically not yet reached the confining walls, so that the
force acting on them is basically the equilibrium 
one corresponding to a vanishing magnetic field. In course of time 
the perturbation
induced by the field hits the confining walls and correspondingly the
force exerted on them changes. Finally, because of the relaxational
character of the dynamics, 
in the limit of long times the perturbation induced by $h$
disappears and accordingly the effective force reaches again its equilibrium
value.   
\subsubsection{Planar perturbation}
In order
to illustrate such a behavior we consider the case in which the
perturbation does not break the translational symmetry along the
confining walls and is localized in the plane $x_\bot = x_{1\bot}$,
i.e., $h({\bf x},t) = h_W \delta(x_\bot - x_{1\bot})\delta(t-t_1)$. 
From Eqs.~\reff{casimir-h-bis} and~\reff{scalh-hat} one finds for the
left wall (the upper index $^{(0)}$ indicates the Gaussian approximation)
\begin{equation}
\FF^{(\rm dy)(0)}_l(\bar L,\bar t,\hat h_W) = 
\FF^{(\rm st)(0)}(0,\bar L)
+ \frac{1}{8}\hat h_W^2
\left[\partial_{\bar x_{2\bot}}\bar \RR^{(0)}(\bar
{\bf p}={\bf 0},\bar x_{1\bot},\bar x_{2\bot}, \bar t -\bar t_1,\bar L)|_{\bar x_{2\bot}=0}
\right]^2,
\label{casimir-h-spec-wall}
\end{equation}
where the scaling function $\bar \RR^{(0)}$ 
of the response function is given by Eq.~\reff{RPsi} and
\begin{equation} 
\hat h_W  =
\xi_0^{(d+2)/2}\left(\frac{L}{\xi_0}\right)^{\beta\delta/\nu-z-1}
\frac{1}{\xi_0\rt_0} h_W
\label{defhWhat}
\end{equation}
is the scaling variable associated with $h_W$. 
In the present
case the functional dependence of the force density on $h({\bf x},t)$
reduces to the dependences on $h_W$, $x_{1\bot}$, and $t_1$ and there
is no spatial dependence of the force on the lateral coordinates.
According to Eq.~\reff{casimir-h-spec-wall}, 
the response of the Casimir force to an external field is
related to the square of the spatial derivative of the response function evaluated 
at one of
the confining walls. As expected, the Casimir force depends
only on the time $\delta\bar t = \bar t - \bar t_1$ 
elapsed since the application of the external
perturbation. 

In the following we 
discuss in more detail the relaxation of the Casimir force at
the bulk critical point $T=T_{c,b}$, i.e., for $\bar L =0$. In this case
$\bar \RR^{(0)}(\bar
{\bf p}={\bf 0},\bar x_{1\bot},\bar x_{2\bot}, \delta\bar t,\bar L = 0)
= 2 \Psi(\bar x_{1\bot},\bar x_{2\bot}, \delta\bar t\,)$ [see
Eq.~\reff{RPsi}] and plots of the function $\Psi$ are provided in
Fig.~\ref{Rreal} for various values of its scaling arguments. 
In the present context the square of the derivative of these graphs at
$\bar x_{2\bot}=0$ matters; $\bar x_{1\bot}$ is the position of the
applied perturbation.
Moreover,
the static contribution in Eq.~\reff{casimir-h-spec-wall} is simply given
by the Casimir amplitude $\Delta$ [see Eq.~\reff{deltaOO}], i.e., 
$\FF^{(\rm st)(0)}(0,\bar L = 0) = (d-1)\Delta$. Note that $\Delta <0$
for the case we are interested in~\cite{KD-92}, 
whereas the contribution stemming
from the field perturbation is always 
positive. Therefore the overall sign of the effective force on the
left wall depends on the strength $\hat h_W$ of the applied field and on
the actual position where it has been applied; in
particular it may become positive, i.e., repulsive within a time
window during the relaxation process.
Given the expression of $\Psi$ in Appendix~\ref{app-formulas}
[see Eq.~\reff{defpsi}] it is possible to compute analytically the
critical scaling function in Eq.~\reff{casimir-h-spec-wall}. According to
the qualitative behavior described above and according to Fig.~\ref{Rreal}, 
$|\partial_{\bar x_{2\bot}}\bar \RR^{(0)}(\bar
{\bf p}={\bf 0},\bar x_{1\bot},\bar x_{2\bot}, \bar t,\bar
L = 0)|_{\bar x_{2\bot}=0}| = 2 |\partial_{\bar x_{2\bot}} \Psi(\bar
x_{1\bot},\bar x_{2\bot},\bar t\,)|_{\bar x_{2\bot}=0}|$ displays, as
function of time,
a maximum for $\bar t = \bar t_M(\bar x_{1\bot})$ which 
is implicitly defined by the condition
\begin{equation}
\partial_{\bar t} \partial_{\bar x_{2\bot}} \Psi(\bar
x_{1\bot},\bar x_{2\bot},\bar t\,)|_{\bar x_{2\bot}=0, \bar t = \bar
t_M(\bar x_{1\bot})} = 0 \,.
\label{cond_max}
\end{equation} 
On the other hand, from the definition of $\Psi$ in
Eq.~\reff{somma}, together with Eq.~\reff{eigenf}
it is easy to realize that $\partial_{\bar t}  \Psi(\bar
x_{1\bot},\bar x_{2\bot},\bar t\,) = \partial^2_{\bar
x_{1\bot}}  \Psi(\bar
x_{1\bot},\bar x_{2\bot},\bar t\,) =  \partial^2_{\bar
x_{2\bot}} \Psi(\bar
x_{1\bot},\bar x_{2\bot},\bar t\,)$. 
Accordingly $\partial_{\bar t} \partial_{\bar x_{2\bot}} \Psi(\bar
x_{1\bot},$ $\bar x_{2\bot},\bar t\,) = \partial^3_{\bar x_{2\bot}}  \Psi(\bar
x_{1\bot},\bar x_{2\bot},\bar t\,)$ and the necessary condition
\reff{cond_max} can be written as
\begin{equation}
\partial^3_{\bar x_{2\bot}} \Psi(\bar
x_{1\bot},\bar x_{2\bot},\bar t\,)|_{\bar x_{2\bot}=0, \bar t = \bar
t_M(\bar x_{1\bot})} = \Psi^{(0,3)}(\bar x_{1\bot},0,t_M(\bar
x_{1\bot})) = 0 \,.
\label{cond_max_bis}
\end{equation} 
where we use the notation introduced after Eq.~\reff{D2Psi} in
Appendix~\ref{app-tI}. Comparing Eq.~\reff{cond_max_bis} with
Eq.~\reff{impleqtI} one sees that $\bar t_M(\bar x_{1\bot})$ and $\bar
t_I(\bar x_{1\bot})$ (Eq.~\reff{defFI}) 
satisfy the same equation. On the basis of the
qualitative behavior of $\Psi$ one expects the solution to be unique
and thus $\bar t_M(\bar x_{1\bot}) = t_I(\bar x_{1\bot})$,
i.e., the time at which $|\partial_{\bar x_{2\bot}} \Psi(\bar
x_{1\bot},\bar x_{2\bot},\bar t\,)|_{\bar x_{2\bot}=0}|$ displays a
maximum for a fixed value of $\bar x_{1\bot}$ equals the time
$\bar t_I(\bar x_{1\bot})$
at which the inflection point of $\Psi(\bar
x_{1\bot},\bar x_{2\bot},\bar t\,)$ as a function of $\bar x_{2\bot}$
reaches the surface at $\bar x_{2\bot}=0$ (see Fig.~\ref{tflex}).

\begin{figure*}
\begin{center}
\epsfig{file=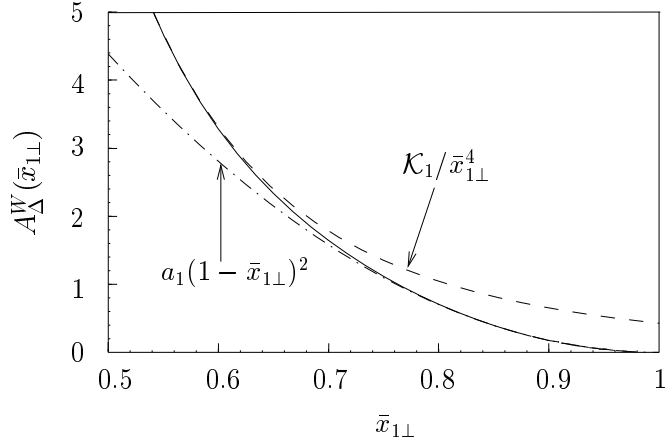,width=0.55\textwidth} 
\end{center}
\caption{Dependence of the amplitude $A^W_\Delta(\bar
x_{1\bot})$, which determines the critical Casimir force maximum $(d-1)\Delta +
\hat h_W^2 A^W_\Delta(\bar x_{1\bot})$ on the left wall
[Eq.~\reff{AAh}], on the normal distance $\bar x_{1\bot}$ where the
planar perturbation is applied.
The asymptotic behaviors for $\bar x_{1\bot} \rightarrow 0$ and $\bar
x_{1\bot} \rightarrow 1$ are indicated as dashed and dashed-dotted
lines, with 
${\mathcal K}_1 = 27 e^{-3}/\pi\simeq  0.427889$ and $a_1\simeq 17.541$,
respectively. Note that already for $\bar
x_{1\bot}\lesssim 0.6$ the asymptotic behavior ${\mathcal K}_1/\bar
x^4_{1\bot}$ for $\bar x_{1\bot}\rightarrow 0$ provides a rather good
estimate of the actual value $A^W_\Delta(\bar x_{1\bot})$.
}
\label{plotAmph}
\end{figure*}

\begin{figure*}
\begin{center}
\epsfig{file=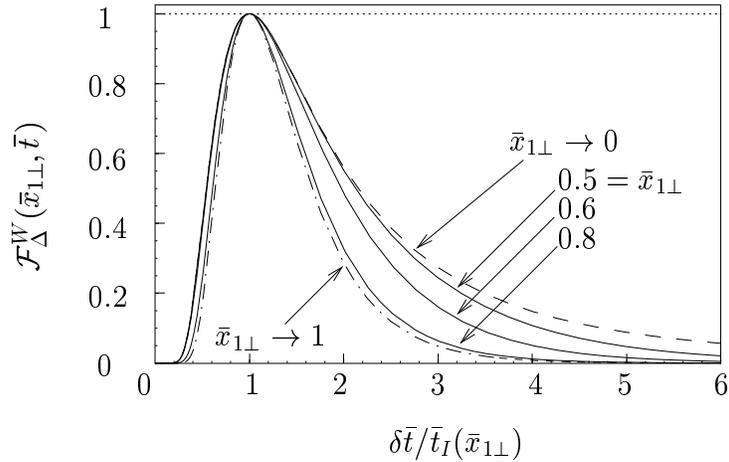,width=0.6\textwidth} 
\end{center}
\caption{Time dependence of the normalized dynamic part of
the critical Casimir force $\FF^W_\Delta(\bar x_{1\bot},\bar t\,) = [\FF^{(\rm dy)(0)}_l(0,\bar t,\hat h_W) - \FF^{(\rm
st)(0)}(0,0)]/[\hat h_W^2 A^W_\Delta(\bar x_{1\bot})]$ for various values
of the position $\bar
x_{1\bot}$ where the perturbation is applied. 
For $\bar t = \bar t_I(\bar x_{1\bot})$ the Casimir force reaches its
maximum (compare Fig.~\ref{tflex}).
The dashed and dash-dotted lines indicate the limiting shapes of the
curves for $\bar x_{1\bot}\rightarrow 0$ and $\bar
x_{1\bot}\rightarrow 1$, respectively. For $\bar t \gg 1/(3\pi^2)$ the
actual curves decay as $\sim e^{-2\pi^2\bar t}$.
}
\label{plotscalh}
\end{figure*}

Motivated by Eq.~\reff{casimir-h-spec-wall} we define
\begin{equation}
\begin{split}
A^W_\Delta(\bar x_{1\bot}) &= \frac{\FF^{(\rm dy)(0)}_l(0,\bar t_I(\bar
x_{1\bot}),\hat h_W) - \FF^{(\rm st)(0)}(0,0)}{\hat h_W^2} \\
&= \frac{1}{2}
\left[\partial_{\bar x_{2\bot}} \Psi(\bar x_{1\bot},\bar x_{2\bot},
\bar t_I(\bar x_{1\bot}))|_{\bar x_{2\bot}=0}
\right]^2 \,
\end{split}
\label{AAh}
\end{equation}
so that $(d-1)\Delta + \hat h_W^2 A^W_\Delta(\bar x_{1\bot})$ is the
maximum value of the force to which the left wall at $\bar x_\bot = 0$ is
subject if the field 
is applied at the normal distance $\bar x_{1\bot}$. 
Figure~\ref{plotAmph} shows the dependence of the maximum
of the field-induced 
force on the position $x_{1\bot}$ of the perturbation whereas
in Fig.~\ref{plotscalh} we show the time dependence of the normalized
dynamic part of the force 
\begin{equation}
\FF^W_\Delta(\bar x_{1\bot},\bar t\,) = 
\frac{\FF^{(\rm dy)(0)}_l(0,\bar t,\hat h_W) - \FF^{(\rm
st)(0)}(0,0)}{\hat h_W^2 A^W_\Delta(\bar x_{1\bot})}
\label{def-Fdelta}
\end{equation}
for various values of $\bar x_{1\bot}$.
The asymptotic behaviors of $A_\Delta^W(\bar x_{1\bot})$ for $\bar
x_{1\bot}\rightarrow 0$ and $\bar x_{1\bot}\rightarrow 1$
are given by
$
A_\Delta^W(\bar x_{1\bot}\rightarrow 0) = {\mathcal K}_1/\bar
x_{1\bot}^4 
$
with ${\mathcal K}_1 = 27 e^{-3}/\pi\simeq 0.427889$
and
$
A_\Delta^W(\bar x_{1\bot}\rightarrow 1) = a_1 (1-\bar x_{1\bot})^2
$
with $a_1 \simeq 17.541$,
respectively (see Appendix~\ref{sub-asyA}). 
As Fig.~\ref{plotAmph} clearly shows, they provide very
good approximations to the actual function $A_\Delta^W$ already for
$\bar x_{1\bot}\lesssim 0.6$ and $\bar x_{1\bot}\gtrsim 0.75$,
respectively. 
As expected, when the plane in which the external field
is applied approaches the distant wall (i.e., $\bar
x_{1\bot}\rightarrow 1$), the actual response of the system is
reduced due to the Dirichlet boundary condition there and, due to 
the relaxational character of the dynamics, the ensuing
perturbation affects only slightly the wall at $\bar x_\bot
=0$. This qualitative behavior clearly emerges from
Fig.~\ref{plotAmph}.
On the other hand, if the external field is applied close to that wall
where the response is monitored, 
the spatial variation of the response function as a
function of $\bar x_{2\bot}$ is sufficiently pronounced 
to induce a strong force on
the close wall. However, this force quickly decays with time, and indeed 
$\bar t_I(\bar x_{1\bot}\rightarrow 0)\rightarrow 0$ (see Fig.~\ref{tflex}).
The time dependence of the normalized dynamic part of the
force $\FF^W_\Delta(\bar x_{1\bot},\bar t\,)$, reported in
Fig.~\ref{plotscalh}, depends only weakly on 
$\bar x_{1\bot}$ when plotted as a function of the scaled time $\bar
t/\bar t_I(\bar x_{1\bot})$. In 
Eqs.~\reff{FWx1} and~\reff{FWx0} we provide the asymptotic expressions
for $\FF^W_\Delta(\bar x_{1\bot}\rightarrow 1,\bar t\,)$ and
$\FF^W_\Delta(\bar x_{1\bot}\rightarrow 0,\bar t\,)$, respectively, 
shown in Fig.~\ref{plotscalh}. For $\bar x_{1\bot}$ fixed and within
mean-field theory $\FF_\Delta^W$ decays asymptotically as
$\FF_\Delta^W(\bar x_{1\bot},\bar t \rightarrow \infty)\sim e^{-
2\pi^2 \bar t}$ (see Eq.~\reff{FFWexpdecay}).
\subsubsection{Localized perturbation}
\begin{figure*}
\begin{center}
\epsfig{file=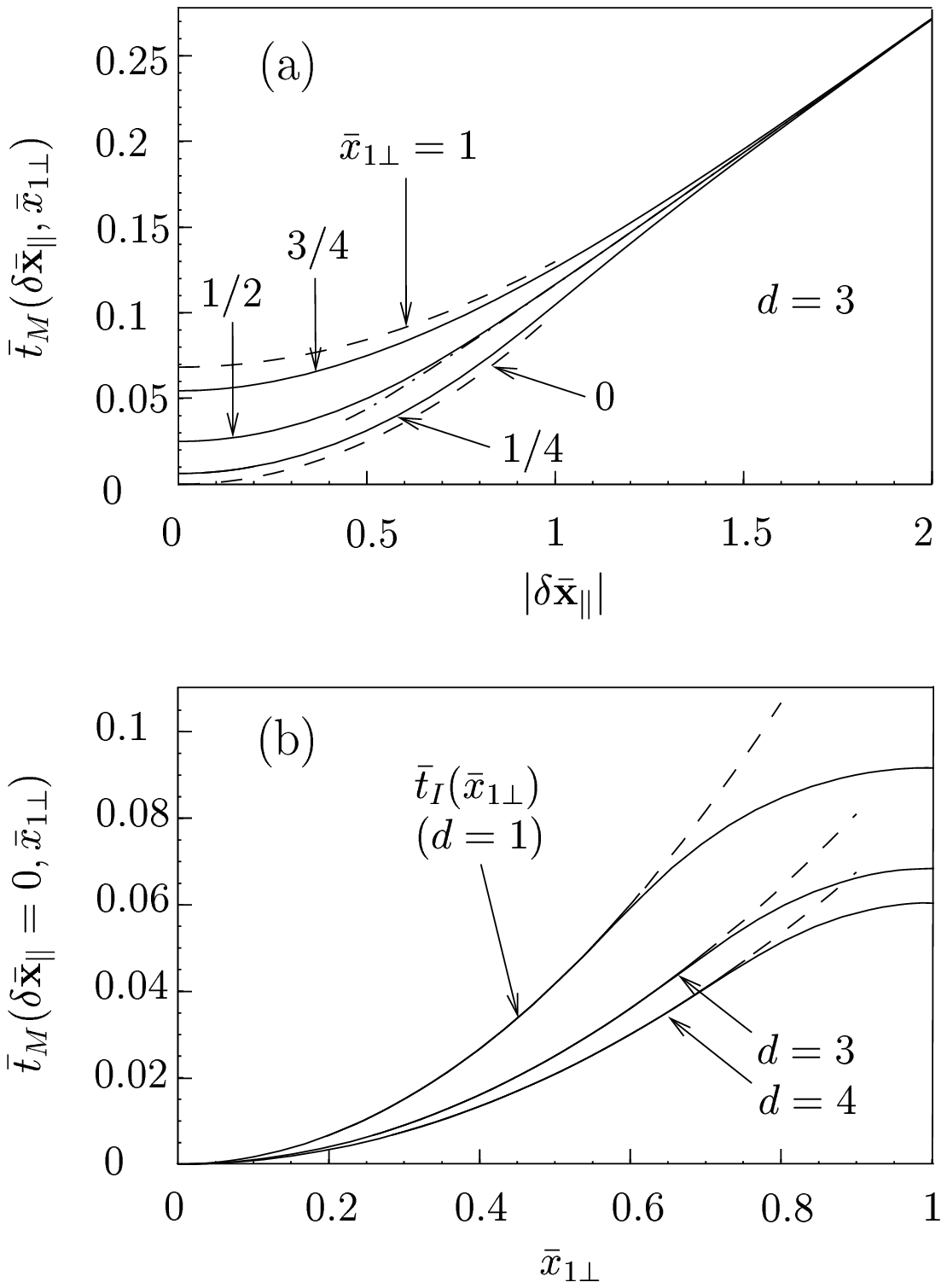,width=0.55\textwidth} 
\end{center}
\caption{Time $\bar t_M(\delta\bar{\bf x}_\|,\bar x_{1\bot})$, at which
the critical Casimir force at the point on the left wall at
$\bar{\bf x}_{1\|}+\delta\bar{\bf
x}_\|$ reaches its maximum, as a function of $|\delta \bar {\bf
x}_\||$ in $d=3$ (a). The external field has been applied at $({\bf
x}_{1\|},x_{1\bot})$.
The solid lines, from top to
bottom, refer to the cases $\bar x_{1\bot} = 3/4$, $1/2$, and $1/4$,
whereas the dashed lines show the asymptotic behaviors for $\bar
x_{1\bot} \rightarrow 0$ (Eq.~\reff{tMx0}) 
and $\rightarrow 1$ (Eq.~\reff{tMx1}) for 
$\pi^2\bar t_M \ll 1$. The dash-dotted line indicates the
approximation of $\bar t_M$ for large $|\delta {\bf x}_\||$ given by
Eq.~\reff{tM-large-xp}, which is already rather good for $|\delta {\bf
x}_\|| \gtrsim 1.0$. For large $|\delta \bar {\bf x}_\||$ the time
$\bar t_M$ becomes independent of $\bar x_{1\bot}$; it increases
linearly with the slope giving the inverse of the speed for the
propagation of the force maximum on the left wall (see Eq.~\reff{speed}).
In (b) we compare the time  $\bar t_M(\delta\bar{\bf x}_\|=0,\bar
x_{1\bot})$ for $d=3,4$ with $\bar t_I(\bar x_{1\bot})$ (see
Fig.~\ref{tflex}), formally corresponding to the case $d=1$. The
dashed curves indicate the quadratic behaviors expected for small
$\bar x_{1\bot}$ and arbitrary $d$ (see Eq.~\reff{tMx0}) which
actually describe the curves rather well 
even for a wide range of values of $\bar
x_{1\bot}$.}
\label{plottMpoint}
\end{figure*}

Here we discuss the case in which the external field 
is localized at a single
point $({\bf x}_{1\|},x_{1\bot})$ within the film, i.e., 
$h({\bf x},t) = h_P  \delta({\bf x}_\| - {\bf x}_{1\|}) \delta(x_\bot
- x_{1\bot})\delta(t-t_1)$. 
From Eqs.~\reff{casimir-h-bis} and~\reff{scalh-hat} one finds for the
left wall at $\bar x_{\bot} = 0$
\begin{equation}
\begin{split}
&
\FF^{(\rm dy)(0)}_l(\bar L,\bar t,\hat h_P) = \\
&\hspace{2cm}
\FF^{(\rm st)(0)}(0,\bar L) +
\frac{1}{2}\hat h_P^2\frac{e^{-(\delta\bar{\bf x}_\|)^2/(2
\delta \bar t\,) - 2 \bar L^2 \delta\bar t}}{(4  \pi \delta \bar t\,)^{d-1}}
\left[\partial_{\bar x_{2\bot}}\Psi (\bar x_{1\bot},\bar x_{2\bot}, \delta\bar t\,)|_{\bar x_{2\bot}=0}
\right]^2,
\end{split}
\label{casimir-h-spec-point}
\end{equation}
where $\delta\bar t = \bar t - \bar t_1$ and $\delta\bar{\bf x}_\| = \bar
{\bf x}_\| -  \bar {\bf x}_{1\|}$.
The function $\Psi$ is given by Eq.~\reff{defpsi} and
\begin{equation}
\hat h_P  =
\xi_0^{(d+2)/2}\left(\frac{L}{\xi_0}\right)^{\beta\delta/\nu-z-d}
\frac{1}{\xi_0^d\rt_0} h_P\;.
\label{defhPhat}
\end{equation}
In the following we discuss in more detail the relaxation of the
Casimir force at bulk criticality $T = T_{c,b}$, i.e., $\bar L =0$.
Different from the case of a planar perturbation, here at a given time 
the force varies laterally. It depends on the normal distance of the
epicenter from the wall under consideration and on the radial distance
$|\delta \bar{\bf x}_\||$ from the epicenter projected on this wall. 
The force decreases monotonically for increasing radial distances
$|\delta \bar{\bf x}_\||$.
The qualitative time dependence of the force, generated by a
point-like perturbation, 
is expected to be independent of the actual lateral position where it acts:
As in the case of planar perturbations, 
the force equals the equilibrium one for very short
and very long times with a maximum in between at 
$\bar t = \bar t_M(\delta\bar{\bf x}_\|,\bar x_{1\bot})$ measured from 
$\bar t_1$. Upon increasing
the lateral distance $|\delta \bar {\bf x}_\||$ from the source of the
perturbation $\bar t_M$ is expected to
increase, because the
perturbation has to cover a larger distance until it hits the wall at the
specified point. The asymptotic behavior of $\bar t_M$ for large
$|\delta\bar{\bf x}_\||$ can be inferred from Eq.~\reff{casimir-h-spec-point}
by taking into account that
\begin{equation} 
\Psi^{(0,1)}(\bar x_{1\bot},0,\bar t\,) \equiv \partial_{\bar x_{2\bot}}\Psi (\bar x_{1\bot},\bar x_{2\bot},
\bar t\,)|_{\bar x_{2\bot}=0} = 2\pi \sum_{n=1}^\infty n
e^{-\pi^2n^2\bar t} \sin(\pi n \bar x_{1\bot}) \;.
\label{Psi01asy-largex}
\end{equation} 
For $\pi^2 \bar t \gg
1$ the sum is dominated by its first term.  This allows one to
determine its extremum as function of $\bar t$. At
leading order the result is independent of $\bar x_{1\bot}$:
\begin{equation}
\bar t_M(|\delta\bar{\bf x}_\|| \gg 1,\bar x_{1\bot}) =
\frac{\sqrt{4\pi^2(\delta\bar{\bf x}_\|)^2 + (d-1)^2}-(d-1)}{4\pi^2} \;.
\label{tM-large-xp}
\end{equation}
It is remarkable that in spite of the diffusive propagation of the
perturbation (Eq.~\reff{casimir-h-spec-point}), the maximum of the
laterally varying force moves asymptotically with constant velocity: 
$\bar t_M(|\delta\bar{\bf x}_\|| \rightarrow \infty) = |\delta\bar{\bf
x}_\||/(2\pi)$ so that with Eq.~\reff{omegascal} this asymptotic speed
$v$ is given by
\begin{equation}
v = 2 \pi (\xi_0/\rt_0)(\xi_0/L)^{z-1}\;.
\label{speed}
\end{equation} 
Thus this speed decreases with increasing film thickness. 
It is
possible to provide an estimate of the typical value of $v$ by
considering the values of $\xi_0$ and $\rt_0$ that have been
experimentally determined. 
In Ref.~\cite{DV-05} the critical dynamics 
of a ultrathin film (bilayer) of iron grown on a tungsten substrate has been
investigated via the magnetic ac susceptibility. This system
should provide a realization of the two-dimensional Ising universality class
with Model A dynamics ($z\simeq 2.1$~\cite{CMPV-03,DV-05}). 
In particular $\rt^+_{0,{\rm exp}}$ for the exponential relaxation
time (see Appendix~\ref{app-dynampratios}) has been obtained from the
fit of experimental data, yielding
$\rt^+_{0,{\rm exp}} = 2.6\pm 0.6\times 10^{-10}$ s. For $\xi_0$ we
take $\xi_0 \simeq  3$\AA\ corresponding to the lattice constant, as
it is usually the case in magnetic materials~\cite{FS-94}.
Accordingly, for a thin film with $L = 50\, \xi_0$, one finds $v\simeq
0.1$ m/s.
By using the behavior of $\Psi^{(0,1)}$ for $\pi^2 \bar t
\ll 1$ reported in Eqs.~\reff{asy1} and~\reff{asy2} for $\bar
x_{1\bot}\rightarrow 1$ and  $\bar
x_{1\bot}\rightarrow 0$, respectively, one can determine the
corresponding behaviors of $\bar t_M$ under the assumption $\pi^2\bar
t_M \ll 1$ for which they are given by
\begin{equation}
\bar t_M = \frac{d+5+(\delta\bar{\bf x}_\|)^2 -
\sqrt{[d+5+(\delta\bar{\bf x}_\|)^2]^2 -  4(d+2)[(\delta\bar{\bf
x}_\|)^2 + 1]}}{4(d+2)} 
\label{tMx1}
\end{equation}
for $\bar x_{1\bot} \rightarrow 1$
and
\begin{equation}
\bar t_M = \frac{(\delta\bar{\bf x}_\|)^2 + \bar x_{1\bot}^2}{2(d+2)}\quad \mbox{for} \quad \bar x_{1\bot}
\rightarrow 0\;.
\label{tMx0}
\end{equation}
Comparing with the numerical determination of $\bar t_M(\delta \bar
{\bf x}_\|,\bar x_{1\bot})$ it turns out that Eq.~\reff{tMx0} provides
a good approximation for $\bar t_M$ up to few
percent in the region $|\delta \bar
{\bf x}_\||,\bar x_{1\bot} \lesssim 0.5$, which corresponds to $\bar
t_M \lesssim 0.1$. 
(Note that for $d=1$ and $\delta\bar{\bf x}_\| = 0$ we
formally recover the result for the planar perturbation, as can be seen
by comparing Eqs.~\reff{casimir-h-spec-wall} and~\reff{casimir-h-spec-point}.) 
In particular, from Eq.~\reff{tMx1} one finds
$
\bar t_M(\delta\bar{\bf x}_\| = 0,\bar x_{1\bot} \rightarrow 1) = (4 -
\sqrt{11})/10 \simeq  0.0683375
$
for $d=3$ and
$
\bar t_M(\delta\bar{\bf x}_\| = 0,\bar x_{1\bot} \rightarrow 1) = (9 -
\sqrt{57})/20 \simeq  0.0604236
$
for $d=4$ for which the Gaussian approximation
becomes exact apart from logarithmic corrections. 
In the case $\bar
x_{1\bot} \rightarrow 0$ one finds from Eq.~\reff{tMx0}, 
$\bar t_M(\delta\bar{\bf
x}_\| = 0,\bar x_{1\bot} \rightarrow 0) = \bar x_{1\bot}^2/[2(d+2)]$. 
Figure~\ref{plottMpoint}(a) shows $\bar
t_M(\delta\bar{\bf x}_\|,\bar x_{1\bot})$ in $d=3$ as a function of
$|\delta\bar{\bf x}_\||$
for certain values of
$\bar x_{1\bot}$. The asymptotic behaviors for $\bar x_{1\bot}
\rightarrow 1$ (Eq.~\reff{tMx1}) and for $\bar x_{1\bot}\rightarrow 0$ (Eq.~\reff{tMx0}) as functions of $|\delta \bar {\bf x}_\||$
are also shown as dashed lines up to a corresponding value of $\bar
t_M \simeq 0.1$ (fulfilling the condition $\pi^2 \bar t_M \lesssim 1$
under which Eqs.~\reff{tMx1} and~\reff{tMx0} have been derived). 
The dash-dotted line
in Fig.~\ref{plottMpoint}(a) indicates the leading linear behavior of 
$\bar t_M$ for large
$|\delta\bar{\bf x}_\||$ [see Eq.~\reff{tM-large-xp}], 
which provides a rather good approximation of the actual dependence
already for $|\delta\bar{\bf x}_\|| \gtrsim 1.0$. 
Figure~\ref{plottMpoint}(b) shows the comparison between $\bar
t_M(\delta\bar{\bf x}_\| = 0,\bar x_{1\bot})$ in $d=3,4$ and $\bar
t_I(\bar x_{1\bot})$ (see also Fig.~\ref{tflex}) [formally
corresponding to the case $d=1$]. The dashed curves indicate the
approximate expression Eq.~\reff{tMx0} 
valid for small $\bar x_{1\bot}$, actually providing a
rather good approximation already for $\bar x_{1\bot}\lesssim
0.6-0.8$. 
Figure~\ref{plottMpoint}(b) clearly indicates that $t_M$ decreases
upon increasing $d$. 
Indeed, for fixed $\bar x_{1\bot}$, the factor
$\bar t^{d-1}$ in the denominator of Eq.~\reff{casimir-h-spec-point}
increases the force at short
times compared to the case of the planar perturbation 
(Eq.~\reff{casimir-h-spec-wall}) where it is absent. 
This causes the maximum of the resulting force to occur at earlier
times $\bar t$, as indicated by Fig.~\ref{plottMpoint}(b).

\begin{figure*}
\begin{center}
\epsfig{file=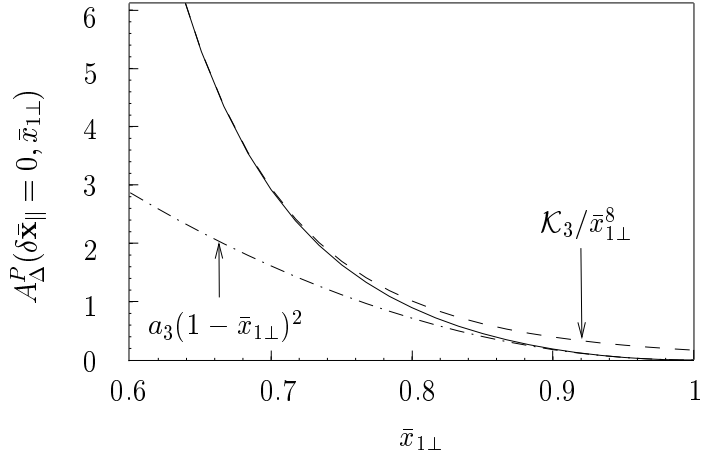,width=0.58\textwidth} 
\end{center}
\caption{%
Dependence of the amplitude $A^P_\Delta(\delta\bar{\bf
x}_\| = 0,\bar x_{1\bot})$, which determines the critical 
Casimir force maximum
$(d-1)\Delta + \hat h_P^2 A^P_\Delta(\delta\bar{\bf
x}_\| = 0,\bar x_{1\bot})$ at the point of the left wall
closest to the point within the film where the external field was
applied, on the normal distance of the perturbation from the left wall. 
The curves refer to the
three-dimensional case, although the qualitative behavior is the same
in $d=4$. The asymptotic behaviors for $\bar x_{1\bot}\rightarrow 0$
and $\bar x_{1\bot}\rightarrow 1$ (see Appendix~\ref{sub-asyA}) 
are indicated as dashed and
dashed-dotted lines, respectively, 
with ${\mathcal K}_3 = 5^5/(4\pi^2 e^5)\simeq
0.169773$ and $a_3 \simeq 17.927$. Note that, as in the case in
Fig.~\ref{plotAmph}, the asymptotic behavior ${\mathcal K}_3/\bar
x_{1\bot}^8$ for $\bar x_{1\bot}\rightarrow 0$ provides a rather good
estimate of the actual values 
already for $\bar x_{1\bot} \lesssim 0.65$.
}
\label{plotAx0point}
\end{figure*}

In analogy to Eq.~\reff{AAh}, we define as a measure of the maximum force
\begin{equation}
\begin{split}
A^P_\Delta(\delta\bar{\bf
x}_\|,\bar x_{1\bot}) &= \frac{\FF^{(\rm dy)(0)}_l(0,\bar t_M(\delta\bar{\bf
x}_\|, \bar
x_{1\bot}),\hat h_P) - \FF^{(\rm st)(0)}(0,0)}{\hat h_P^2}\\
& = \frac{1}{2}
\frac{e^{-(\delta\bar{\bf x}_\|)^2/[2
\bar t_M(\delta\bar{\bf
x}_\|, \bar
x_{1\bot})]}}{[4 \pi \, \bar t_M(\delta\bar{\bf
x}_\|, \bar
x_{1\bot})]^{d-1}}
\left[\partial_{\bar x_{2\bot}}\Psi (\bar x_{1\bot},\bar x_{2\bot}, \bar t_M(\delta\bar{\bf
x}_\|, \bar
x_{1\bot}))|_{\bar x_{2\bot}=0}
\right]^2\;.
\end{split}
\label{AAhpoint}
\end{equation}
We first discuss $A^P_\Delta(\delta\bar{\bf
x}_\| = 0,\bar x_{1\bot})$, i.e., the maximum of the field-induced
force at the point of the wall closest to the
source of the perturbation.
Figure~\ref{plotAx0point} shows the function $A_\Delta(\delta\bar{\bf
x}_\| = 0,\bar x_{1\bot})$ for $d=3$, together with its asymptotic behaviors
determined in  Appendix~\ref{sub-asyA}. According to
Eqs.~\reff{Apsmallxp0} and~\reff{Apxt1xp0} they are given by 
$A^P_\Delta(\delta\bar{\bf x}_\| = 0,\bar x_{1\bot}\rightarrow 0) =
{\mathcal K}_d/\bar x_{1\bot}^{2(d+1)}$, with ${\mathcal K}_d$ defined
in Eq.~\reff{defKd}, and  $A^P_\Delta(\delta\bar{\bf x}_\| = 0,\bar
x_{1\bot}\rightarrow 1) = a_d (1-\bar x_{1\bot})^2$, with $a_d$
defined by Eq.~\reff{defad}. The qualitative dependence of
$A^P_\Delta(\delta\bar{\bf x}_\| = 0,\bar x_{1\bot})$ on $\bar
x_{1\bot}$ is the same as that of $A^W_\Delta(\bar x_{1\bot})$. 

\begin{figure*}
\begin{center}
\epsfig{file=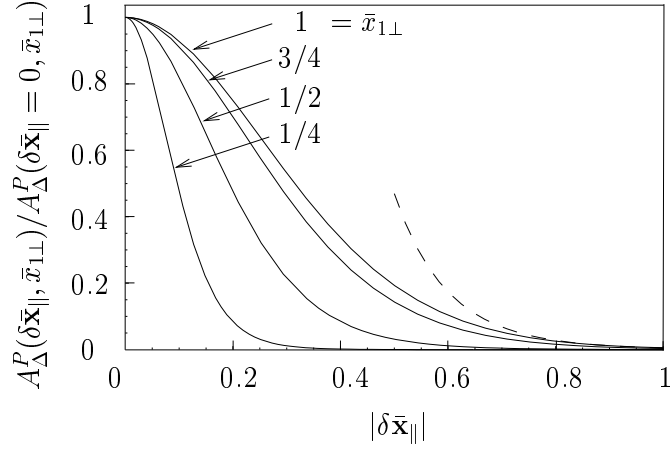,width=0.55\textwidth} 
\end{center}
\caption{%
Dependence of the normalized amplitude $A^P_\Delta(\delta\bar{\bf
x}_\|,\bar x_{1\bot})$ of the critical Casimir force maximum with 
$d=3$ on $|\delta\bar{\bf
x}_\||$ for various fixed $\bar x_{1\bot}$.
The dashed line is the asymptotic behavior for $|\delta \bar
{\bf x}_\|| \gg 1$ and $\bar x_{1\bot}\rightarrow 1$, as
obtained from Eq.~\reff{RatioAlargexp}. For $|\delta \bar {\bf x}_\||
\rightarrow \infty$ the curves vanish as $\sim \exp (- 2 \pi |\delta \bar
{\bf x}_\||)$ (Eq.~\reff{RatioAlargexp}).
}
\label{plotRatioA3d}
\end{figure*}

According to Eq.~\reff{casimir-h-spec-point}, at a given time $\delta
\bar t$ elapsed after the perturbation, the field-induced part of the
Casimir force is a decreasing function of $|\delta \bar {\bf
x}_\||$. Therefore the maximum amplitude $A_\Delta^P(\delta\bar{\bf
x}_\|,\bar x_{1\bot})$ decreases with increasing  
lateral distance  $|\delta \bar {\bf
x}_\||$ from the point where the field has been
applied. Figure~\ref{plotRatioA3d} shows the dependence of the maximum
force 
amplitude $A^P_\Delta(\delta\bar{\bf
x}_\|,\bar x_{1\bot})$ in $d=3$ on $\delta\bar{\bf
x}_\|$ for fixed values of $\bar x_{1\bot}$, normalized to its value
at $\delta\bar{\bf
x}_\| = 0$ (compare Fig.~\ref{plotAx0point}). 
The solid lines
correspond, from top to bottom, to $\bar x_{1\bot}=1$, $3/4$, $1/2$,
and $1/4$. For $\bar x_{1\bot},|\delta \bar {\bf x}_\|| \lesssim 0.5$
the approximate expression of $A_\Delta^P(\delta\bar{\bf
x}_\|,\bar x_{1\bot})/A_\Delta^P(\delta\bar{\bf
x}_\| = 0,\bar x_{1\bot}) \simeq [1 + (\delta \bar {\bf x}_\|)^2/\bar
x_{1\bot}^2]^{-(d+2)}$ (Eq.~\reff{Ratiosmallxpxt})
provides a very good approximation of the actual curves. 
For $|\delta \bar {\bf x}_\|| \rightarrow \infty$
this ratio decays $\sim \exp(-2\pi|\delta \bar {\bf x}_\||)$ 
(Eq.~\reff{RatioAlargexp}). In Fig.~\ref{plotRatioA3d} the dashed line
refers to this asymptotic behavior for 
$\bar x_{1\bot} \rightarrow 1$. It is interesting to note that for all
$\bar x_{1\bot}$ the maximum of the field-induced 
force decays rapidly with $|\delta\bar{\bf x}_\||$ and
is practically negligible compared with its value at $\delta\bar{\bf x}_\|
=0$ already for $|\delta\bar{\bf x}_\|| \simeq
1$.
\begin{figure*}
\begin{center}
\epsfig{file=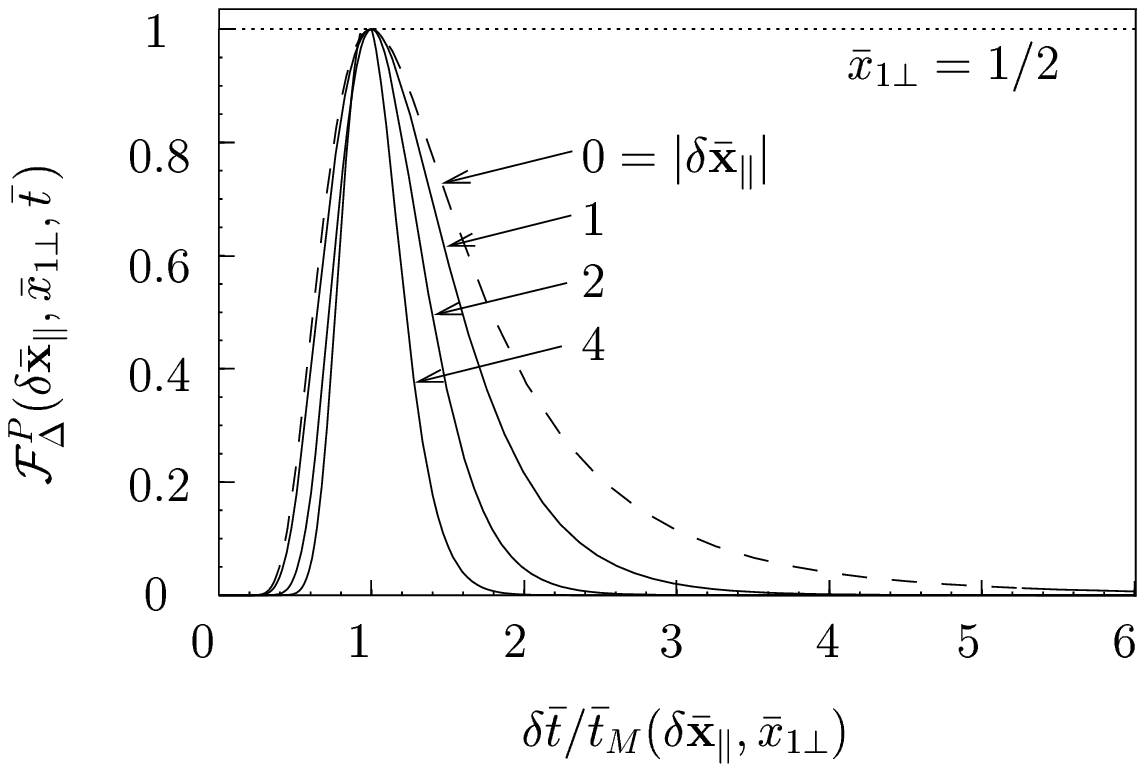,width=0.6\textwidth} 
\end{center}
\caption{%
Time dependence of the normalized dynamic part of
the critical Casimir force $\FF^P_\Delta(\bar x_{1\bot},\bar t\,) = [\FF^{(\rm
dy)(0)}_l(0,\bar t,\hat h_P) - \FF^{(\rm
st)(0)}(0,0)]/[\hat h_P^2 A^P_\Delta(\delta \bar{\bf x}_\|,\bar
x_{1\bot})]$ for various lateral distances $|\delta\bar {\bf x}_\||$
from the location of the perturbation at $\bar
x_{1\bot} =1/2$, with $d=3$.
For $\bar t = \bar t_M(\delta \bar {\bf x}_\|, \bar x_{1\bot})$ 
the Casimir force reaches its
maximum (compare Fig.~\ref{plottMpoint}). The different curves refer to 
$|\delta \bar {\bf x}_\|| = 0,1,2$ and $4$, from top to bottom. The
qualitative behavior of the curves is the same for other values of
$\bar x_{1\bot}$. For $\delta\bar t \rightarrow\infty$ the curve
vanish as $(\delta\bar t)^{-(d-1)} e^{-2\pi^2\delta\bar t}$. 
}
\label{scalTT3d}
\end{figure*}

Figure~\ref{scalTT3d} shows the time dependence of the normalized
dynamic part of the Casimir force 
\begin{equation}
\FF^P_\Delta(\delta \bar {\bf x}_\|,\bar x_{1\bot},\bar t\,) =
\frac{\FF^{(\rm dy)(0)}_l(0,\bar t,\hat h_P) - \FF^{(\rm
st)(0)}(0,0)}{\hat h_P^2 A^P_\Delta(\delta \bar{\bf x}_\|,\bar x_{1\bot})}
\end{equation}
for $d=3$, fixed $\bar x_{1\bot}=1/2$, and various
values of $|\delta\bar {\bf x}_\|| = 0$, $1$, $2$, and $4$, from top
to bottom. When the $\FF^P_\Delta$ is plotted as a function of $\bar
t/\bar t_M$ its shape does neither depend sensitively on the specific
value of $\bar x_{1\bot}$ nor on the value of
$|\delta \bar {\bf x}_\||$. For fixed $\delta \bar {\bf x}_\|$ and
$\bar x_{1\bot}$, $\FF_\Delta^P$ decays asymptotically as
$\FF_\Delta^P(\delta \bar {\bf x}_\|,\bar x_{1\bot},\bar t \rightarrow
\infty) \sim (\delta\bar t)^{-(d-1)} e^{-2\pi^2\,\delta\bar t}$ (see
Eqs.~\reff{casimir-h-spec-point} and~\reff{Psi01asy-largex}).

\subsection{Correlation function}
\label{sec-correlation}

\subsubsection{General expressions}

\begin{figure*}
\epsfig{file=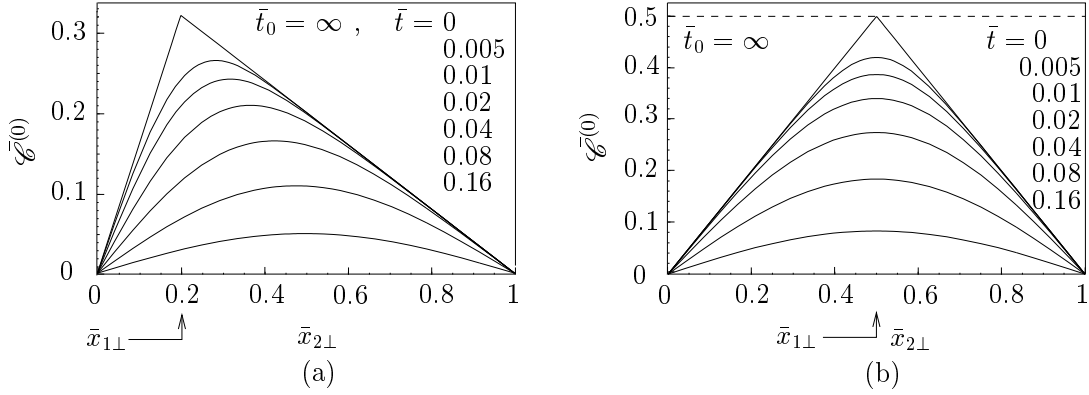,width=0.9\textwidth}
\caption{Time evolution of the mean-field scaling function $\bar \CC^{(0)}({\bf
\bar p},\bar x_{1\bot},\bar x_{2\bot},\bar t,\bar L)$ which enters
into the expression of the correlation function in Eq.~\reff{scalComega} 
with $\bar x_{i\bot} = x_{i\bot}/L$,  
$\bar t = (t/\rt_0)(\xi_0/L)^z$, $\bar L = L/\xi$, 
and $\bar t_0 = 1/({\bf \bar p}^2 +
\bar L^2)$ for ${\ve p}={\ve 0}$ and $T=T_{c,b}$, 
i.e., $\bar t_0 = \infty$. 
We show correlations between points ${\bar x_{2\perp}}$ and
$\bar x_{1\bot} = 0.2$ (a) and $\bar x_{1\bot} = 0.5$ (b). 
Reduced times $\bar t$ listed in (a) and (b) 
refer to the various curves shown from top to bottom. At $\bar t =0$,
the static correlation function
$\bar \CC^{(0)}$ exhibits a cusp at $\bar x_{1\bot} = \bar x_{2\bot}$.}
\label{Crealcrit}
\end{figure*}

\begin{figure*}
\epsfig{file=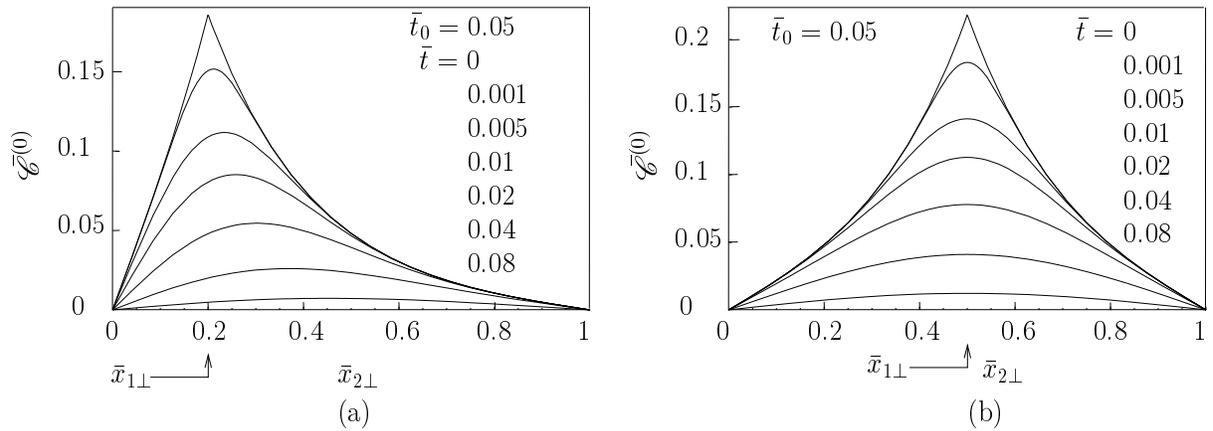,width=1\textwidth}
\caption{Same as in Fig.~\ref{Crealcrit} but for $\bar t_0 = 0.05$
which leads to reduced correlations compared to $\bar t_0 =
\infty$. The cusps at $\bar x_{1\bot} = \bar x_{2\bot}$ remain also
for $\bar t_0 < \infty$.}
\label{Creal}
\end{figure*}

As discussed in Subsec.~\ref{subsec-FDT}, 
the FDT relates the correlation function and
the response function via Eq.~\reff{fdt}.
Thus, taking into account Eq.~\reff{Rgenpsi}, one finds
that the correlation function can be cast into the scaling form given in
Eq.~\reff{scalCt} with
\begin{equation}
\bar\CC^{(0)}({\bf \bar p}, \bar x_{1\bot}, \bar x_{2\bot}, \bar t,
\bar L) = 2 \int_{|\bar t|}^\infty \dd \bar s\, e^{-\bar s/\bar
t_0({\bf \bar p},\bar L)} \Psi(\bar x_{1\bot},\bar x_{2\bot},\bar s)
\label{defC0Psi}
\end{equation}
where the value of the non-universal amplitude $\hat{\mathfrak{o}}_C$
is given by Eq.~\reff{amplitudeC} and where we have introduced 
\begin{equation}
\bar t_0 = \frac{1}{{\bf \bar
p}^2 + \bar L^2} \; . 
\label{deft0}
\end{equation}
The dependence of the scaling function $\bar\CC^{(0)}$ on $\bar t$ for
selected values of $\bar x_{1\bot}$ is shown in Fig.~\ref{Crealcrit}
at criticality and for ${\bf p}={\bf 0}$, i.e., $\bar t_0=\infty$, and
for $\bar t_0 <\infty$ in Fig.~\ref{Creal}. As expected,
$\bar\CC^{(0)}$ vanishes in the limit $\bar t \rightarrow\infty$. 
According to
Eqs.~(\ref{Cstatstand}) and (\ref{scalCt}) the scaling function $\bar\CC^{(0)}_{\rm st}({\bf \bar p},\bar x_{1\bot}, \bar x_{2\bot},
\bar L) = \bar\CC^{(0)}({\bf \bar p}, \bar x_{1\bot}, \bar x_{2\bot}, \bar t=0,
\bar L)$ of the
static correlation function
is given by
\begin{equation}
\bar\CC^{(0)}_{\rm st}({\bf \bar p},\bar x_{1\bot}, \bar x_{2\bot},
\bar L) = 
2 \frac{\sinh(\bar t_0^{\,-1/2}\bar x_\bot^<)\sinh[\bar t_0^{\,-1/2} (1-
\bar x_\bot^>)]}{ \bar t_0^{\,-1/2} \sinh (\bar t_0^{\,-1/2})} \,,
\end{equation}
which, independently of the value of $\bar t_0$, 
is characterized by a cusp-like singularity for $\bar
x_{1\bot} = \bar x_{2\bot}$, clearly displayed in Figs.~\ref{Crealcrit}
and~\ref{Creal}. In particular at bulk criticality, i.e., $\bar L =0$ and
for ${\bf \bar p}={\bf 0}$ (i.e., $\bar t_0 = \infty$), one finds
(compare Eq.~\reff{Romega})
\begin{equation}
\bar\CC^{(0)}_{\rm st}({\bf \bar p}={\bf 0},\bar x_{1\bot}, \bar x_{2\bot},
\bar L =0) =
2 \, \bar x_\bot^<(1-\bar x_\bot^>) \;,
\end{equation}
which corresponds to a  
triangularly shaped correlation function vanishing
at the boundaries [see Fig.~\ref{Crealcrit} and Eq.~\reff{Cstatstand}].
The two-point correlation function in thin films can be probed by
scattering X-rays or neutrons under grazing
incidence~\cite{DH-95,KD-99}. The dependence of the corresponding
scattering cross section on the lateral momentum transfer is dominated
by the singular behavior of the two-point correlation function in
planes parallel to the surfaces of the film. With future technologies
it might be possible to extend such kind of scattering experiment into
the time domain. Therefore we discuss the particular case $x_{1\bot} =
x_{2\bot}$, i.e., $C({\bf p},x_{\bot},x_{\bot},\omega)$.
The analysis of this particular case may also serve to enhance the
general physical insight into correlations in films.

In order to understand the behavior of this quantity we shall
consider  two relevant limits: (a) the behavior close to
the confining walls, i.e., $x_\bot\ll L$ (surface behavior; 
see Appendix~\ref{app-expansion}), and (b) the behavior in the
interior, i.e., $x_\bot = L/2$. Crossover effects are expected to
occur in between.

\subsubsection{Behavior near the wall}
\label{subsec_bntw}

\begin{figure*}
\begin{center}
\epsfig{file=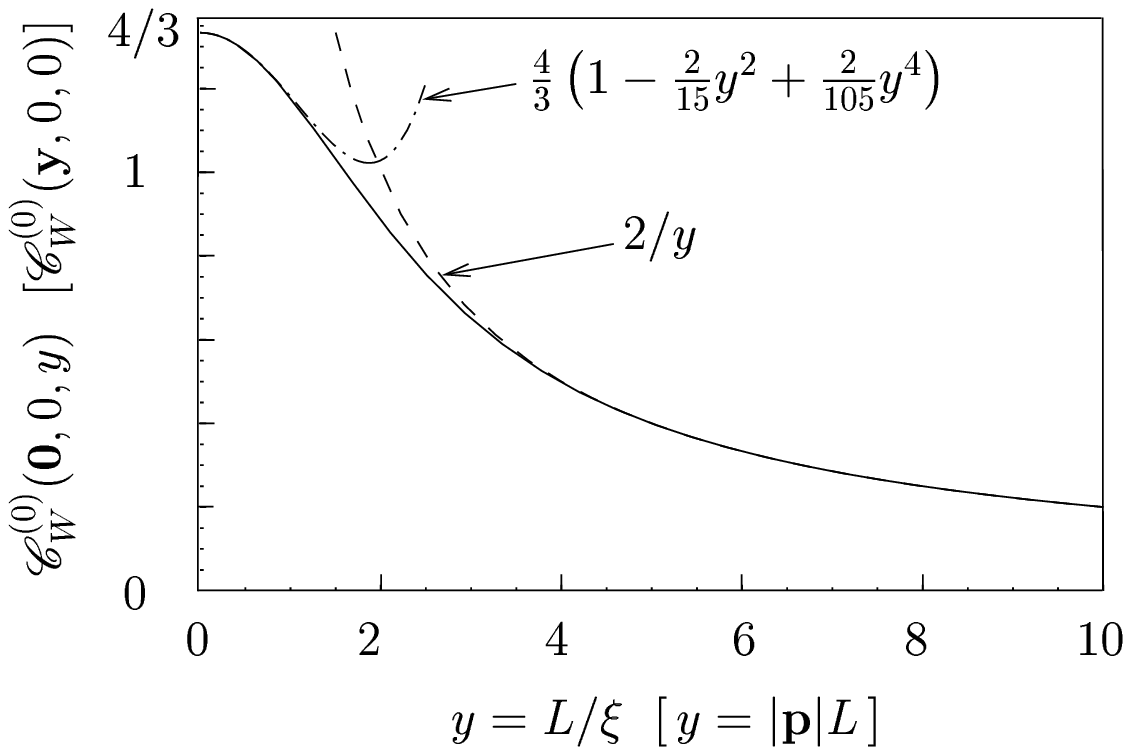,width=0.55\textwidth} 
\end{center}
\caption{Mean-field scaling function $\CC_W^{(0)}({\bf
0},0,y)$ $[\CC^{(0)}_W({\bf y},0,0)]$ which enters into the expression of
the correlation function  
$C({\bf p}={\bf 0},x_\bot,x_\bot,\omega=0) =
\hat{\mathfrak{o}}_C(L/\xi_0)^{1-\eta-z}(x_\bot/L)^{2(\beta_1-\beta)/\nu}\CC_W({\bf
0},0,y = L/\xi)$ 
$[C_{\rm crit}({\bf p},x_\bot,x_\bot,\omega=0) = \hat{\mathfrak{o}}_C(L/\xi_0)^{1-\eta-z}(x_\bot/L)^{2(\beta_1-\beta)/\nu}\CC_W({\bf
y}={\bf p}L,0,0)]$ for $L/x_\bot \gg 1, L/\xi$ $[L/x_\bot \gg 1, |{\bf
p}|L]$, so that $\CC_W^{(0)}({\bf
0},0,y\rightarrow 0)$ $[\CC^{(0)}_W({\bf y}\rightarrow {\bf 0},0,0)] = (4/3)
[1-(2/15) y^2 + (2/105) y^4 + O(y^6)]$ and  $\CC_W^{(0)}({\bf
0},0,y\rightarrow \infty)$ $[\CC^{(0)}_W({\bf y}\rightarrow \infty,0,0)]
= 2/y + O(1/y^2)$. Beyond mean-field theory the asymptotic behavior of 
$\CC_W({\bf 0},0,y = L/\xi)$ and $\CC_W({\bf
y}={\bf p}L,0,0)$ for
large $y$ is given by Eqs.~\reff{CW00x} and~\reff{CWx00}, respectively.  
The index $W$ of the scaling function here and in
Fig.~\ref{Csp0r0} indicates the behavior near the wall.}
\label{Csp0o0}
\end{figure*}

\begin{figure*}
\begin{center}
\epsfig{file=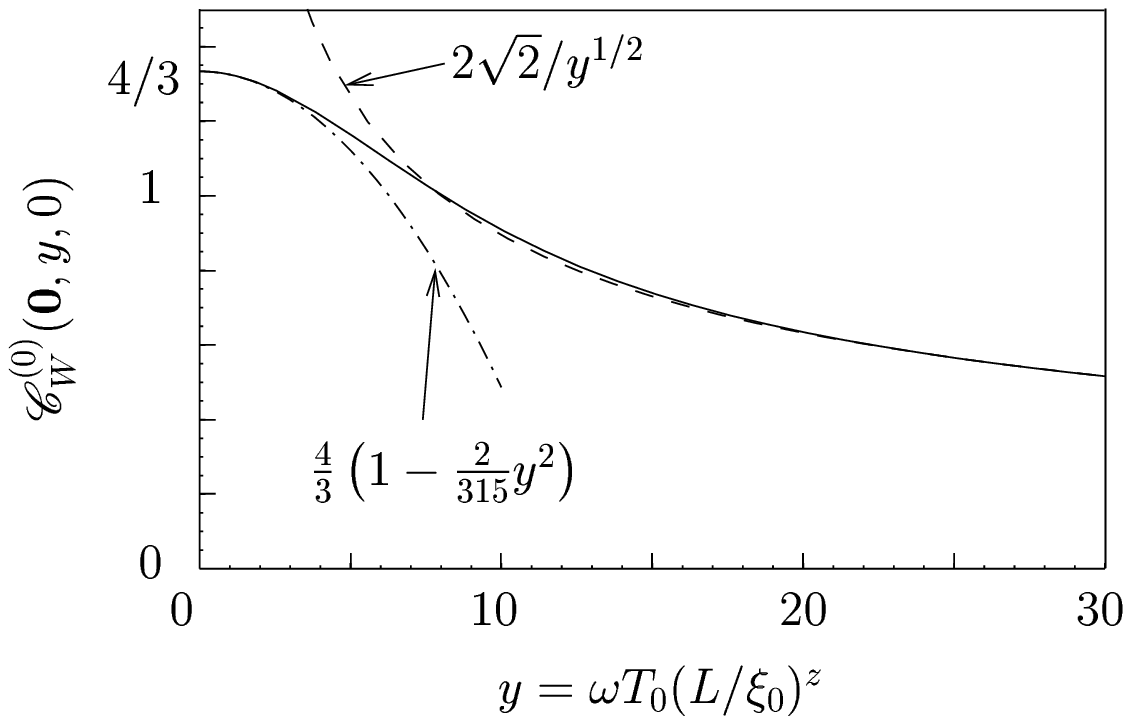,width=0.55\textwidth} 
\end{center}
\caption{Mean-field scaling function $\CC_W^{(0)}({\bf
0},y,0)$ which enters into the expression of the correlation function  
$C_{\rm crit}({\bf p}={\bf 0},x_\bot,x_\bot,\omega) =
\hat{\mathfrak{o}}_C(L/\xi_0)^{1-\eta-z}(x_\bot/L)^{2(\beta_1-\beta)/\nu}\CC_W({\bf
0},y=\omega \rt_0(L/\xi_0)^z,0)$ for  $T=T_{c,b}$ and 
$L/x_\bot \gg 1, \omega \rt_0(L/\xi_0)^z$, so that $\CC_W^{(0)}({\bf
0},y\rightarrow 0,0) = (4/3) [1-(2/315) y^2 + O(y^4)]$ and
$\CC_W^{(0)}({\bf 0},y\rightarrow \infty,0) = 2\sqrt{2}/y^{1/2} + O(1/y)$. 
Beyond mean-field theory the asymptotic behavior of 
$\CC_W({\bf 0},y = \omega\rt_0(L/\xi_0)^z,0)$ for 
large $y$ is given by Eq.~\reff{CW0x0}.} 
\label{Csp0r0}
\end{figure*}

To this end we rewrite Eq.~\reff{Romega}, use the
FDT (see Eq.~\reff{fdtomega}), cast it
into the scaling form given in Eq.~\reff{scalComega}, and use
Eq.~\reff{amplitudeC} for the nonuniversal amplitude:
\begin{equation}
\CC^{(0)}({\bf \bar p}, \bar x_{1\bot}, \bar x_{2\bot}, \bar\omega,
\bar L) = 
\frac{2}{\bar\omega} \Im \frac{ \cosh[\bar a(1-|\bar
x_{1\bot}-\bar x_{2\bot}|)] - \cosh[\bar a(1-\bar x_{1\bot}-\bar
x_{2\bot})]}{\bar a \sinh \bar a}
\label{scalC}
\end{equation}
where $\bar a$ has been defined in Eq.~\reff{abar}.
By using the previous equation we obtain
\begin{equation}
\CC^{(0)}({\bf \bar p}, \bar x_\bot, \bar x_\bot, \bar\omega,
\bar L) = \frac{2}{\bar\omega} \Im \frac{ \cosh \bar a - \cosh[\bar
a(1- 2\bar x_\bot)]}{\bar a \sinh \bar a }
\label{Cinplane}
\end{equation}
so that for $|\bar a|\bar x_\bot = |a|x_\bot \ll 1$ 
(i.e., sufficiently close to one of the walls)
\begin{equation}
\CC^{(0)}({\bf \bar p},\bar x_\bot,\bar x_\bot,\bar \omega) =
-4 \frac{{\bar x_\bot}^2}{\bar\omega}
\Im \left\{\bar a\coth \bar a (1+ O(|\bar a|\bar x_\bot))\right\}\;,
\end{equation}
in agreement with the general behavior of $\CC$ close to one
wall, given in Eq.~(\ref{scalComegaWall}) with
\begin{equation}
\CC^{(0)}_W({\bf \bar p},\bar \omega,\bar L) = -\frac{4}{\bar\omega}
\Im \left\{\bar a\coth \bar a  (1+ O(|\bar a|\bar x_\bot))\right\}\;.
\end{equation}
We can distinguish two regimes, i.e., $|\bar a| = |a|L\gg 1$ 
(relevant for discussing
the connection with the results for the semi-infinite geometry) and
$|\bar a| = |a|L\ll 1$.\\[0.2cm]
$\bullet\ \  |\bar a| = |a|L \gg 1$: 
In this case $\bar a\coth\bar a=\bar a(1+O(e^{-2\bar a}))$ 
(we choose the branch 
with positive real part in
the square root defining $\bar a$, see footnote~\ref{footnote3}) and thus
\begin{equation}
\CC^{(0)}_W({\bf\bar  p},\bar x_\bot,\bar x_\bot,\bar \omega) 
= - \frac{4}{\bar\omega}
\Im \left\{\bar a  (1+ O(|\bar a|\bar x_\bot,e^{-2\bar a}))\right\} \;.
\end{equation}
Using the definition of $\bar a$, one obtains
\begin{equation}
\begin{split}
&\CC^{(0)}_W({\bf \bar p},\bar x_\bot,\bar x_\bot,\bar \omega) =
2\sqrt{2} \frac{1}{\sqrt{{\bf
\bar p}^2 + \bar L^2 +\sqrt{({\bf \bar p}^2 + \bar L^2)^2 + \bar
\omega^2}}} \; ,\\
&\sqrt[4]{({\bf\bar p}^2 +\bar L^2)^2 + \bar \omega^2} \gg 1 \; \quad\mbox{and}\quad
\bar x_\bot \; \sqrt[4]{({\bf\bar p}^2 +\bar L^2)^2 + \bar \omega^2}
\ll 1 \;.
\end{split}
\label{CCCCW}
\end{equation}
Equation~\reff{CCCCW} renders the limiting behaviors listed in
Eqs.~(\ref{CCW00x}), (\ref{CCW0x0}), and (\ref{CCWx00}) with the
mean-field universal amplitudes
\begin{align}
{\mathcal A}^{W(0)}_{\infty} &= 2 \;, \label{ampAWinf0}\\
{\mathcal B}^{W(0)}_{\infty} &= 2\sqrt{2} \;, \\
{\mathcal C}^{W(0)}_{\infty} &= 2 \; \label{ampCWinf0},
\end{align}
and the mean-field values of the critical exponents.
Corrections to the formulae~(\ref{CCW00x}), (\ref{CCW0x0}), and
(\ref{CCWx00}) are exponentially small. \\[0.2cm]
$\bullet\ \  |\bar a| = |a|L \ll 1$:  In this case 
one has (see Appendix~\ref{app-expansion}, Eq.~\reff{Cinplaneexp}),
\begin{equation}
\label{Cgeneral}
\begin{split}
\CC^{(0)}({\bf\bar p},\bar x_\bot,\bar x_\bot,\bar\omega) =&
\frac{4}{3} \bar x_\bot^2 (1-\bar x_\bot)^2
\Bigg\{1-\frac{2}{15} (1 + 2\bar x_\bot - 2 \bar x_\bot^2)({\bf\bar p}^2 +
{\bar L}^2) \\
&+ \frac{2}{105}(1+2\bar x_\bot -\frac{\bar
x_\bot^2}{2}-3\bar x_\bot^3 +\frac{3}{2}\bar x_\bot^4)\left[ ({\bf\bar p}^2 + 
{\bar L}^2)^2 - \frac{1}{3}{\bar \omega}^2 \right]  +
\ldots \Bigg\}\;,\\
&\sqrt[4]{({\bf\bar p}^2 +{\bar L}^2)^2 + {\bar \omega}^2} \ll  1
\;,\quad \bar x_\bot \le 1 \;.
\end{split}
\end{equation}
Near  one wall, i.e., for $\bar
x_\bot \ll 1$ 
this expression leads to
\begin{equation}
\CC^{(0)}_W({\bf\bar p},\bar x_\bot,\bar x_\bot,\bar \omega) = 
\frac{4}{3}\left[1 -
\frac{2}{15}({\bf\bar p}^2 + {\bar L}^2) + \frac{2}{105}({\bf\bar p}^2 + 
{\bar L}^2)^2
- \frac{2}{315}{\bar \omega}^2 +\ldots\right]\; .
\label{CsurfaceW}
\end{equation}
From this expression we can identify the mean-field
value of the universal constant introduced in Eq.~\reff{defA}:
\begin{equation}
{\mathcal A}^{W(0)} = 4/3 \,.
\label{ampAW0}
\end{equation}
Figures~\ref{Csp0o0} and~\ref{Csp0r0}  display 
the functions $\CC^{(0)}_W({\bf\bar p}={\bf 0}, \bar
\omega=0, \bar L)$, $\CC^{(0)}_W ({\bf\bar p}, \bar
\omega=0, \bar L =0)$, and $\CC^{(0)}_W ({\bf\bar p} = {\bf 0}, \bar
\omega, \bar L =0)$, respectively, together with
their asymptotic behaviors given in Eqs.~\reff{CCW00x}--\reff{CCWx00}
and (\ref{CsurfaceW}).

Thus, in contrast to
the case of a semi-infinite geometry (in which the
validity of the behaviors in Eqs.~\reff{CCW00x}, \reff{CCW0x0}, and
\reff{CCWx00} 
extends down to $1/\xi=0$, $\omega=0$, or
$\ve{p}=\ve{0}$, respectively), 
the critical correlation
function in the film does not diverge upon approaching the origin of the $(\xi^{-1},{\bf
p},\omega)$-space. This is a consequence of the critical point shift
in the film geometry. Of course this divergence is correctly recovered in
the limit $L\rightarrow\infty$ (i.e., in the semi-infinite limit),
where this shift is absent.

\subsubsection{Film behavior}

Now we consider the behavior of $C^{(0)}({\bf
p},x_\bot,x_\bot,\omega)$ in the middle of the film, i.e., 
$C^{(0)}({\bf p},L/2,L/2,\omega)$. According to
Eqs.~(\ref{scalComega}) and (\ref{CCI}) this amounts to studying
$\CC^{(0)}_I({\bf\bar p},\bar\omega,\bar L)$. In that case we find
(see Eq.~(\ref{Cinplane}))
\begin{equation}
\CC^{(0)}_I({\bf\bar p},\bar\omega,\bar L) = 
\CC^{(0)}({\bf\bar p},1/2,1/2,\bar \omega, \bar L) = \frac{2}{\bar\omega}
\Im\frac{\tanh(\bar a/2)}{\bar a} \; .
\end{equation}
Again, we consider the two possible regimes $|\bar a| = |a|L\gg 1$ and
$|\bar a|=|a|L\ll 1$. \\[0.2cm]
$\bullet\ \  |\bar a| = |a|L \gg 1$:
In this regime one has
\begin{equation}
\CC^{(0)}_I({\bf\bar p},\bar\omega,\bar L) = \frac{2}{\bar\omega}
\Im \left\{ \frac{1}{\bar a}(1 + O(e^{-2|\bar a|}))\right\}
\end{equation}
so that 
\begin{equation}
\begin{split}
&\CC^{(0)}_I({\bf\bar p},\bar\omega,\bar L) = \sqrt{2} 
\frac{1}{\sqrt{\left({\bf\bar p}^2 + {\bar L}^2\right)^2+ {\bar
\omega}^2}\sqrt{{\bf\bar p}^2 + {\bar L}^2 +
\sqrt{\left({\bf\bar p}^2 + {\bar L}^2\right)^2+ {\bar \omega}^2}}}\; ,\\
&\sqrt[4]{({\bf\bar p}^2 +{\bar L}^2)^2 + {\bar\omega}^2} \gg 1\; .
\end{split}
\end{equation}
From this equation one easily recovers the limiting behaviors given in
Eqs.~(\ref{CI00x}), (\ref{CI0x0}), and~(\ref{CIx00}), with the
mean-field universal amplitudes
\begin{align}
{\mathcal A}^{I(0)}_\infty &= 1 \; \label{ampAIinf0}\\
{\mathcal B}^{I(0)}_\infty &= \sqrt{2} \;,\\
{\mathcal C}^{I(0)}_\infty &= 1 \;, \label{ampCIinf0}
\end{align}
and with the corresponding mean-field values of the critical
exponents. \\[0.2cm]
$\bullet\ \  |\bar a| = |a|L \ll 1$:
On the other hand, in the limit $|\bar a| = |a|L\ll 1$ one can make use of 
Eq.~\reff{Cgeneral} (valid in the case $|a|L,|a|x_\bot \ll 1$ and
arbitrary $x_\bot < L$), with $\bar x_\bot = 1/2$, leading to
\begin{equation}
\begin{split}
&\CC^{(0)}_I({\bf\bar p},\bar\omega,\bar L) = \frac{1}{12} 
\left[1 -
\frac{1}{5}({\bf\bar p}^2 + {\bar L}^2) + \frac{17}{560}({\bf\bar p}^2 + 
{\bar L}^2)^2
- \frac{17}{1680}{\bar \omega}^2 +\ldots\right]\;,\\
&\sqrt[4]{({\bf\bar p}^2 +{\bar L}^2)^2 + {\bar\omega}^2} \ll 1\; .
\end{split}
\label{Cbulk}
\end{equation}
From Eq.~(\ref{Cbulk}) one infers the universal amplitude 
\begin{equation}
{\mathcal A}^{I(0)} = \frac{1}{12} \label{ampAI0}
\end{equation}
introduced in Eq.~(\ref{CI000}).
As it is the case near a wall (see the remark at the end of
Subsec.~\ref{subsec_bntw}), also in the middle of the film the correlation
function does not diverge at the origin of the $({\ve
p},\xi^{-1},\omega)$-space; divergences appear only in the 
limit $L\rightarrow\infty$, for which the critical-point shift disappears
and $T_{c,b}$ coincides with the critical temperature of the system.
In particular, from Eq.~\reff{Cgeneral} one finds
\begin{equation}
\CC_{\rm crit}^{(0)}({\bf\bar p}={\bf 0}, \bar x_\bot,\bar x_\bot,\bar
\omega = 0) = \frac{4}{3}{\bar x_\bot}^2
(1-\bar x_\bot)^2 \;.
\end{equation}

\begin{figure*}[h!]
\begin{center}
\epsfig{file=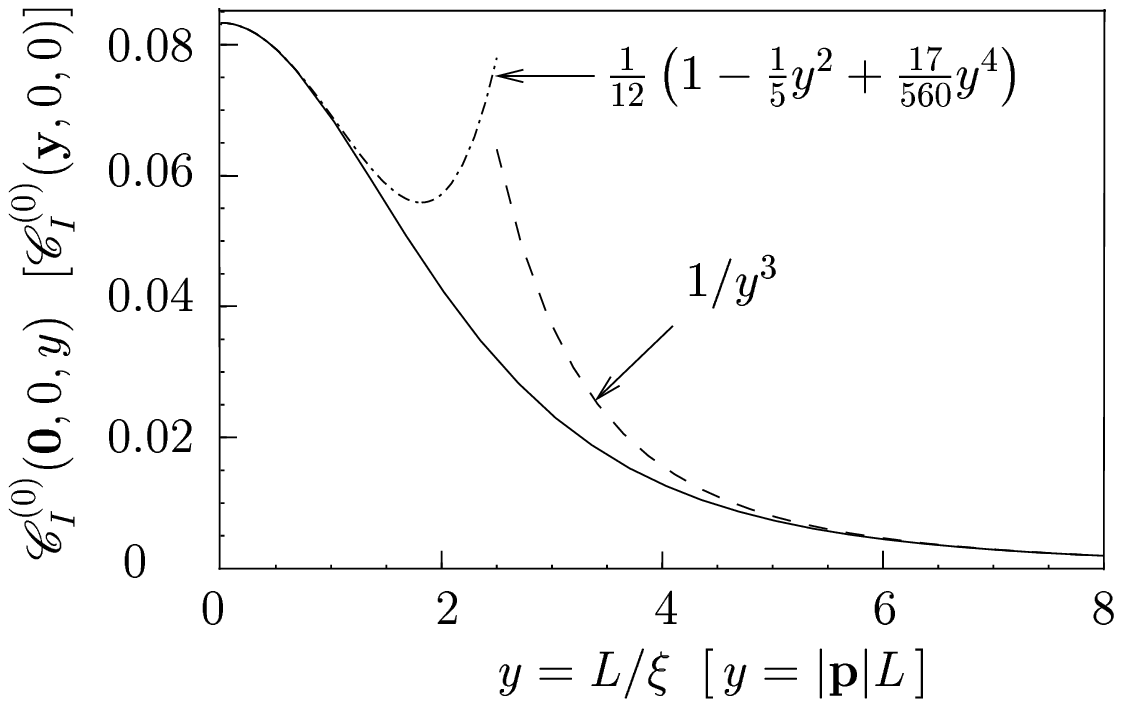,width=0.55\textwidth} 
\end{center}
\caption{Mean-field scaling function $\CC_I^{(0)}({\bf
0},0,y)$ $[\CC^{(0)}_I({\bf y},0,0)]$ which enters into the expression of
the correlation function  
$C({\bf p}={\bf 0},x_\bot=L/2,x_\bot=L/2,\omega=0) =
\hat{\mathfrak{o}}_C(L/\xi_0)^{1-\eta-z}\CC_I({\bf 0},0,y = L/\xi)$ 
$[C_{\rm crit}({\bf p},x_\bot=L/2,x_\bot=L/2,\omega=0) =
\hat{\mathfrak{o}}_C(L/\xi_0)^{1-\eta-z}\CC_I({\bf y}={\bf p}L,0,0)]$, 
so that $\CC_I^{(0)}({\bf
0},0,y\rightarrow 0)$ $[\CC^{(0)}_I({\bf y}\rightarrow {\bf 0},0,0)] = 
(1/12)[1-(1/5) y^2 + (17/560) y^4 + O(y^6)]$ and  $\CC_I^{(0)}({\bf
0},0,y\rightarrow \infty)$ $[\CC^{(0)}_I({\bf y}\rightarrow \infty,0,0)]
= 1/y^3 + O(1/y^4)$. Beyond mean-field theory the asymptotic behavior of 
$\CC_I({\bf 0},0,y = L/\xi)$ and $\CC_I({\bf
y}={\bf p}L,0,0)$ for
large $y$ is given by Eqs.~\reff{CI00x} and~\reff{CIx00}, respectively.  
The index $I$ of the scaling functions here and in
Fig.~\ref{Cbp0r0} indicates the behavior within the interior (i.e.,
middle) of the
film.
}
\label{Cbp0o0}
\end{figure*}

\begin{figure*}[h!]
\begin{center}
\epsfig{file=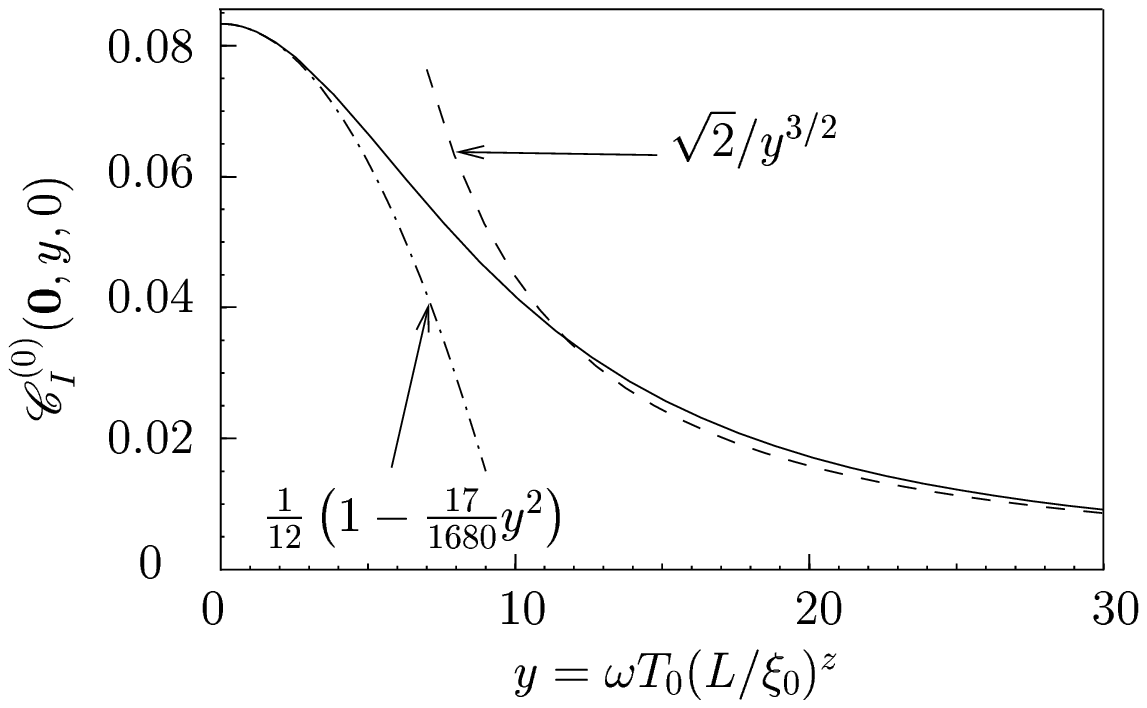,width=0.55\textwidth} 
\end{center}
\caption{Mean-field scaling function $\CC_I^{(0)}({\bf
0},y,0)$ which enters into the expression of the correlation function  
$C_{\rm crit}({\bf p}={\bf 0},x_\bot=L/2,x_\bot=L/2,\omega) =
\hat{\mathfrak{o}}_C(L/\xi_0)^{1-\eta-z}\CC_I({\bf
0},y=\omega \rt_0(L/\xi_0)^z,0)$ for  $T=T_{c,b}$, 
so that $\CC_I^{(0)}({\bf
0},y\rightarrow 0,0) = (1/12)[1- (17/1680) y^2 + O(y^4)]$ and
$\CC_I^{(0)}({\bf 0},y\rightarrow \infty,0) = \sqrt{2}/y^{3/2} + O(1/y^2)$. 
Beyond mean-field theory the asymptotic behavior of 
$\CC_I({\bf 0},y = \omega\rt_0(L/\xi_0)^z,0)$ for 
large $y$ is given by Eq.~\reff{CI0x0}.
}
\label{Cbp0r0}
\end{figure*}

In Fig.~\ref{Cbp0o0} we show 
the functions $\CC^{(0)}_I({\bf\bar p}={\bf
0},\bar\omega=0,\bar L)$ and 
$\CC^{(0)}_I({\bf\bar p},\omega=0,\bar L =0)$ which have (within MFT)
the same scaling function if the
scaling variables are appropriately 
chosen. (In Fig.~\ref{Cbp0o0} the quantities in square brackets
refer to
$\CC^{(0)}_I({\bf\bar p},\omega=0,\bar L =0)$.) 
The limiting behaviors of this
function as given by Eqs.~\reff{CI00x} [\reff{CIx00}] and
\reff{Cbulk} are indicated as dashed and dash-dotted 
lines. In Fig.~\ref{Cbp0r0},
we plot $\CC^{(0)}_I({\bf\bar p}={\bf
0},\bar\omega,\bar L=0)$ and its limiting behaviors given by
Eqs.~\reff{Cbulk} and~\reff{CI0x0}. The comparison between the
scaling functions near the wall (Figs.~\ref{Csp0o0} and \ref{Csp0r0}) and
in the interior of the film (Figs.~\ref{Cbp0o0} and~\ref{Cbp0r0})
shows that they exhibit the same qualitative behaviors. However, the
scaling functions in the interior decay more rapidly for large scaling
variables than their counterparts near the wall. Moreover, their
absolute values scale very differently as functions of $L$.

\section{Nonlinear Behavior}
\label{sec-nlin}

In the previous sections we investigated the linear response and
correlation functions within the Gaussian approximation for the dynamical
functional (Eq.~\reff{msrh}). The response function captures 
the behavior of {\it small} perturbations around the equilibrium
state of the system. The response function analyzed in 
Subsec.~\ref{sec-response} is useful to describe the relaxation process
from an initial state with a small value of the order parameter
such that  nonlinear terms can be neglected.
On the other hand both in experiments and numerical simulations 
this is often not the case given that  
it is much easier  to investigate the response of the system to {\it finite}
changes of the control parameters (such as temperature and external
fields). While typically the former relaxation process
is characterized by exponential
decays in time, the latter leads to power-law behaviors. 
As a concrete example one can think of an experimental protocol in
which the system is initially prepared in an equilibrium state
with a non-zero value of the order parameter. 
This can be realized, for example in magnetic systems, 
by preparing the sample in its low-temperature phase or by applying an
external field. Then the external parameters are changed in a
way that the order parameter in the corresponding equilibrium state
vanishes, as it is the case above or at the critical temperature, in the
absence of external fields. In order 
to describe the ensuing relaxation it is crucial
to account for the effects of nonlinear terms, as will be discussed below.

Let us now consider in more detail the case in which an 
external field $h({\bf x},t)$, 
coupling linearly to the order parameter in the Hamiltonian 
${\mathcal H}$, is present
during the relaxation. The evolution equation for the quantity
$m({\bf x},t) = \langle \varphi({\bf x},t)\rangle$ can be derived in a
standard way~\cite{BJ-76,BEJ-79} and, at the lowest level in the loop
expansion (tree-level), it reads
\begin{equation}
h({\bf x},t) = \partial_t m({\bf x},t) +\Omega[-\Delta + \xi^{-2} +
\frac{g_0}{3!}m^2({\bf x},t)] m({\bf x},t)\;.
\label{eqmotion}
\end{equation}
In the case of confined geometries this equation, which is valid in the
bulk, has to be supplemented with the proper boundary conditions.
For bulk systems with homogeneous external fields the linear
regime is identified as that one in which the nonlinear term in
Eq.~\reff{eqmotion} is negligible compared to $\xi^{-2}$, i.e.,
$\frac{g_0}{3!}m^2({\bf x},t)\ll \xi^{-2}$. In view of
Eq.~\reff{defg0} this means $m({\bf x},t)/\m_0 \ll \xi_0/\xi$. 
(At the ordinary transition, the
equilibrium magnetization close to the surface
vanishes as $\sim (T_{c,b}-T)^{\beta_1}$ for 
$T\rightarrow T_{c,b}^-$, with $\beta_1=1$ within MFT. Therefore,
sufficiently close to $T_{c,b}$, 
the previous inequality is always fulfilled. 
Nevertheless the spatial inhomogeneity caused by the boundary condition at
the surface yields a contribution --~due to the term $-\Delta m$ in
Eq.~\reff{eqmotion}~-- which effectively reduces the one due to
the linear term previously considered. Eventually, the nonlinear term
dominates and the proper nonlinear relaxation is displayed.)
In that regime and in the absence of an external field  
the temporal relaxation from an initial state with
a homogeneous order parameter is exponential. The range of validity of the
linear approximation shrinks 
upon approaching the critical 
point~\cite{R-76,FR-76,BJ-76,BEJ-79}. At $T_{c,b}$
it is no longer possible to neglect the nonlinear
contribution which causes $m$ to relax 
as a function of time according to a
power law. This
general scenario can be modified by the presence of confining walls,
which change the spectrum of the operator $-\Delta$ in Eq.~\reff{eqmotion}. 
Thus, if this
spectrum has a lower bound above zero (as it is the case for
Dirichlet-Dirichlet boundary conditions in a film for which
the zero mode is not allowed) the linear
regime eventually dominates even at bulk criticality
$\xi=\infty$. This is a consequence of the 
critical-point shift in the film geometry. On the other hand, for 
$T < T_{c,b}$, $\xi^{-2}$ has to be replaced by $-\xi^{-2}/2$ (within
MFT) and so, for $T=T_c(L)<T_{c,b}$, the spectrum of the operator
$-\Delta-\xi^{-2}/2$ includes $0$ and the nonlinear contribution is no
longer negligible. Therefore at $T_c(L)$,
$m$ relaxes according to a power law as
a function of time. Beyond MFT this power law is characterized by the
critical exponents of a $d-1$-dimensional bulk system.

Before discussing the nonlinear relaxation in the confined geometry,
we summarize briefly the results of the corresponding analysis of the
semi-infinite geometry, i.e., the effects of a single
surface~\cite{R-82,KO-85,RDW-85}. 
In particular we consider the typical relaxation process, realized 
by applying an external field $h(x_\perp,t) = h(x_\bot)\theta(-t)$ where
$x_\perp$ is the normal distance from the single wall. 
In this case 
the subsequent evolution depends on the resulting initial order parameter
profile $m_0(x_\perp) = m(x_\perp, t = 0)$.  
Within the field-theoretical approach it is possible to compute the
scaling function for the evolving order parameter profile, as carried out in
Refs.~\cite{R-82} and~\cite{RDW-85} for Model A:
\begin{equation}
m(x_\perp, t; \tau) =
\m_0 (t/\rt_0)^{-\beta/\nu z}\Psi((x_\perp/\xi_0)
(t/\rt_0)^{-1/z},t/\rt_R,\{(m_0(x'_\perp)/\m_0) (t/\rt_0)^{\beta/\nu z}\})
\label{scalingsemiinf}
\end{equation}
where the temperature enters via the relaxation time $\rt_R$ (see
 Eq.~\reff{defrt}).
For  $(m_0(x'_\perp)/\m_0)$ $\times(t/\rt_0)^{\beta/\nu z}\gg 1$ 
the behavior of the system becomes independent of the initial
configuration and follows a universal 
scaling function $\Psi(v,w,\infty)$. From a short-distance expansion it
is found that $\Psi(v\rightarrow 0,w,\infty)\sim
v^{(\beta_1-\beta)/\nu}$,
so that at criticality the magnetization close to the surface
decays as $t^{-\beta_1/\nu z}$ for $t\rightarrow\infty$. 
For fixed
$x_\bot$ and at criticality, after some (non-universal)  initial
transient
behavior due to the finite initial magnetization, the order
parameter relaxes as $m \sim t^{-\beta/\nu z}$, i.e., according to the
``bulk'' behavior. (Of course one has $\Psi(\infty,w,\infty)=\Psi_{\rm
bulk}(w)$). As time passes (i.e., $t/\rt_0\gtrsim (x_\bot/\xi_0)^z$) 
the effect of
the surface reaches the point $x_\bot$ and the relaxation crosses over
towards the surface behavior, i.e., $m\sim t^{-\beta_1/\nu z}$.
(Note that for the ordinary surface universality class $\beta_1>\beta$.)
This
crossover is nicely displayed in Monte Carlo simulation data~\cite{KO-85}. 
Off criticality this picture is modified only by the fact that when $m$
becomes sufficiently small, i.e., $t \gg \rt_R$, 
the system enters the linear regime where the decay of $m$ is finally 
exponential. Moreover the influence of the surface penetrates
into the bulk only for a finite distance $\sim \xi$ from the
surface. Well inside the non-critical bulk no crossover is expected
between surface and bulk relaxation. 

Focusing now on the film geometry,
the corresponding nonlinear evolution equation for
$m({\bf x}, t)$ 
is still given by Eq.~\reff{eqmotion} together with the
boundary conditions in Eq.~\reff{BC1}. In order to proceed we introduce
dimensionless quantities via 
\begin{equation}
\bar m({\bf \bar x},\bar t\,)\equiv \sqrt{\frac{g_0}{3!}} L m({\bf \bar
x}L,\bar t L^2/\Omega) = \frac{L}{\xi_0^+}\frac{m({\bf \bar x}L,\bar t
\rt_0 (L/\xi_0)^z)}{\m_0}
\label{adimensional}
\end{equation}
(using Eq.~\reff{defg0}) with ${\bf \bar x}\in {\mathbb R}^{d-1}\times[0,1]$.
Thus in the absence of the external field Eq.~\reff{eqmotion}
turns into
\begin{equation}
\partial_{\bar t} \bar m({\bf\bar x},\bar t\,) + [-\Delta_{\bf\bar x} +
\bar \tau +
\bar m^2({\bf\bar x},\bar t\,)] \bar m({\bf\bar x},\bar t\,) = 0 \qquad \mbox{with}
\qquad \bar m({\bf \bar x}_{\mathcal B},\bar t\,) = 0\;,
\label{eqmotionadim}
\end{equation}
where $\bar \tau \equiv (L/\xi)^2$ for $\tau >0$ and $\bar \tau \equiv
-1/2 (L/\xi)^2$ for $\tau < 0$. (We recall that, within mean-field
theory, $\xi(\tau \rightarrow 0^+) = r_0^{-1/2}$ whereas $\xi(\tau
\rightarrow 0^-) = (-2 r_0)^{-1/2}$.)  Using the scaled variables ${\bf\bar
x}$, $\bar t$, and $\bar \tau$ there is no
explicit dependence of the profile $\bar m({\bf\bar x},\bar t;\bar
\tau)$ on the size $L$ of the layer.
Static solutions of Eq.~\reff{eqmotionadim} have been discussed in
detail in the literature for
various boundary conditions (see, e.g., Refs.~\cite{K-97,FN-81,NF-83} 
and references therein). 
For $\bar\tau\ge \bar\tau_c$ ($\tau_c<0$ determines the
critical-point shift) the asymptotic solution of
Eq.~\reff{eqmotionadim}, 
$\bar m_\infty({\bf\bar x}) =
\lim_{\bar t \rightarrow\infty} \bar m({\bf\bar x},\bar t\,)$, is
$\bar m_\infty({\bf\bar x}) = 0$, whereas in the low-temperature
phase, i.e., 
$\bar\tau<\bar\tau_c<0$,
the order parameter profile is non-trivial and its analytic expression
can be found in Appendix~\ref{app-OPprofile}.
The nonlinear partial differential equation~\reff{eqmotionadim} can be
solved  numerically.
Starting from an arbitrary order parameter profile
$\bar m_0({\bf\bar x})\equiv \bar m({\bf\bar x},\bar t =0)$, with 
$\bar m_0({\bf\bar x}_{\mathcal B})=0$, the
profile evolves according to Eq.~\reff{eqmotionadim}. In analogy
with the results for the semi-infinite
geometry~\cite{KO-85,R-82,RDW-85}, we expect a non-universal behavior
in the early stage of relaxation due to the fact that the
order parameter profile we
start with fulfills the Dirichlet boundary conditions and thus 
takes on {\it small} values near the confining walls, 
while universal properties
are observed in the scaling limit of $\bar m_0({\bf\bar
x})$ being infinitely large. 

We start with discussing the behavior of the system at bulk criticality $\bar
\tau=0$. 
\begin{figure*}
\begin{center}
\epsfig{file=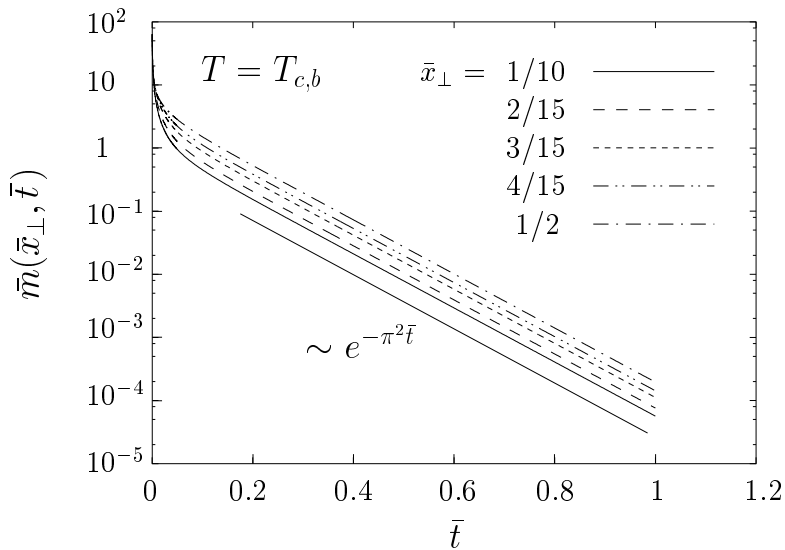,width=0.65\textwidth} 
\end{center}
\caption{%
Late relaxation of the order parameter $\bar m(\bar
x_\bot,\bar t\,)$ at bulk criticality in a film of thickness $L$ 
as a function of the
reduced time variable $\bar t= (t/\rt_0)(\xi_0/L)^z$, for various values of the
distance $\bar x_\bot=x_\bot/L$ from one wall. Already for $\bar t 
\gtrsim 0.2$, 
$\bar m(\bar x_\bot,\bar t\,)$ exhibits, independently of $\bar x_\bot$,
 its asymptotic long-time behavior,
characterized by the exponential decay which is indicated as a straight line
in the figure.}
\label{filmlog}
\end{figure*}
In Fig.~\ref{filmlog} we show the results of the numerical solution of 
Eq.~\reff{eqmotionadim}, evolving from an initial profile 
that is constant in the directions parallel to the confining walls and
has a trapezoidal shape in the transverse direction,
fulfilling the boundary conditions: 
$\bar m(\bar x_\bot,\bar t = 0) = K \bar x_\bot/D$ for $0\le \bar
x_\bot \le D$, $\bar m(\bar x_\bot,\bar t = 0) = K$ for $D < \bar
x_\bot < 1 - D$, and $\bar m(\bar x_\bot,\bar t = 0) = K (1-\bar
x_\bot)/D$ for $1-D\le \bar x_\bot < 1$. 
The parameters $D$ and $K$ have been suitably chosen
to ensure the stability of the numerical solution of the equation
(typical choices are $D \simeq 10^{-2}$ and $K\simeq 10^2$).
Under this
assumption the problem depends on a single space
variable, given by the distance from a wall $0\le \bar x_\bot \le
1/2$, and on the time variable $\bar t$. The solution for $1/2<\bar x_\bot\le
1$ is obtained by taking into account the obvious symmetry of the
problem, i.e., $\bar m(\bar x_\bot) = \bar m(1-\bar x_\bot)$, provided
the initial profile is chosen to share this symmetry.
In Fig.~\ref{filmlog} 
different curves refer to different distances from the wall. These
data demonstrate clearly that asymptotically the
relaxation is {\it exponential} in time, independently of $\bar
x_\bot$, according to the behavior
\begin{equation}
m({\bf x},t\rightarrow\infty) \sim  e^{-\pi^2 (t/\rt_0)(\xi_0/L)^z}\;.
\label{critexpdecay}
\end{equation}
This behavior can be explained by analyzing 
Eq.~\reff{eqmotionadim} for the special case we are discussing, i.e.,
\begin{equation}
\partial_{\bar t} \bar m(\bar x_\bot,\bar t\,) + [-\partial^2_{\bar x_\bot} +
\bar m^2(\bar x_\bot,\bar t\,)] \bar m(\bar x_\bot,\bar t\,) = 0 
\label{critnlin}
\end{equation}
with $\bar m(\bar x_\bot=0,\bar t\,) =  \bar m(\bar x_\bot=1,\bar t\,) = 0$. In
the linear regime (i.e., neglecting $\bar m^2$ in Eq.~\reff{critnlin}) 
this equation can be solved straightforwardly,
leading to
\begin{equation}
 \bar m(\bar x_\bot,\bar t\,) = \sum_{n=1}^\infty \alpha_n
 e^{-\pi^2n^2\bar t} \sin(\pi n \bar x_\bot)
\label{sollinear}
\end{equation}
where the coefficients $\alpha_n$ are determined by the initial
profile. Thus for generic values $\{\alpha_n\}$ the leading 
asymptotic decay
is indeed exponential (see Eq.~\reff{critexpdecay})
and its dependence on $\bar x_\bot$ is given by 
$\sin(\pi\bar x_\bot)$. As already stated at the beginning of this
section, the exponential decay of the order parameter is intimately
related to the fact that in Eq.~\reff{sollinear} the sum contains no
zero mode $n=0$ and that we are considering the problem at the bulk
critical point, located in the disordered phase of the confined
system. 
In the following we consider the behavior at early
times as
shown in the log-log plot in Fig.~\ref{filmloglog}.

\begin{figure*}
\begin{center}
\epsfig{file=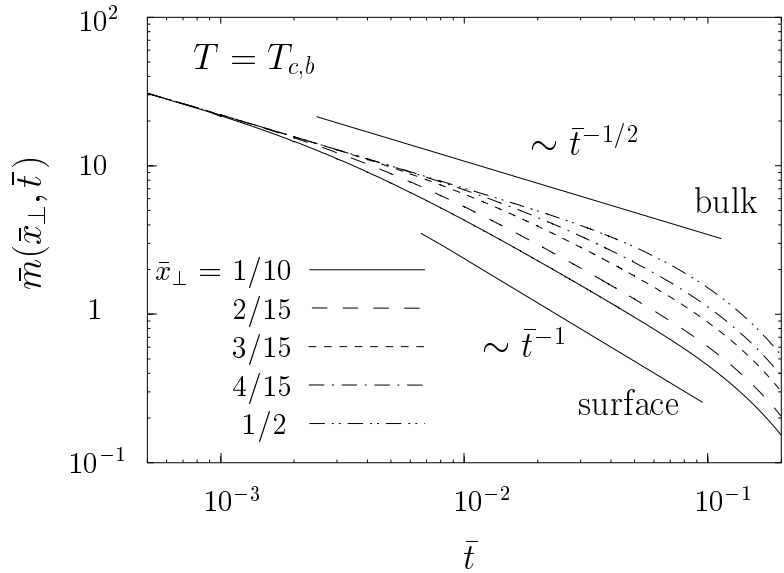,width=0.65\textwidth} 
\end{center}
\caption{Early relaxation of the order parameter $\bar m(\bar
x_\bot,\bar t\,)$ at bulk criticality in a film of thickness $L$ 
as a function of the
reduced time variable $\bar t= (t/\rt_0)(\xi_0/L)^z$, for various values of the
distance $\bar x_\bot = x_\bot/L$ from one wall. This is a
magnification, on a
log-log scale, of the interval $\bar t< 0.2$ in
Fig.~\ref{filmlog}. The two straight lines indicate the power-law
behavior in the bulk (uppermost line) and close to
surfaces, respectively. 
The crossover from bulk to surface and from surface to film
behavior (i.e., exponential decay) 
is evident. For larger values of $\bar x_\bot$ the crossover
occurs later.}
\label{filmloglog}
\end{figure*}

\begin{figure*}
\begin{center}
\epsfig{file=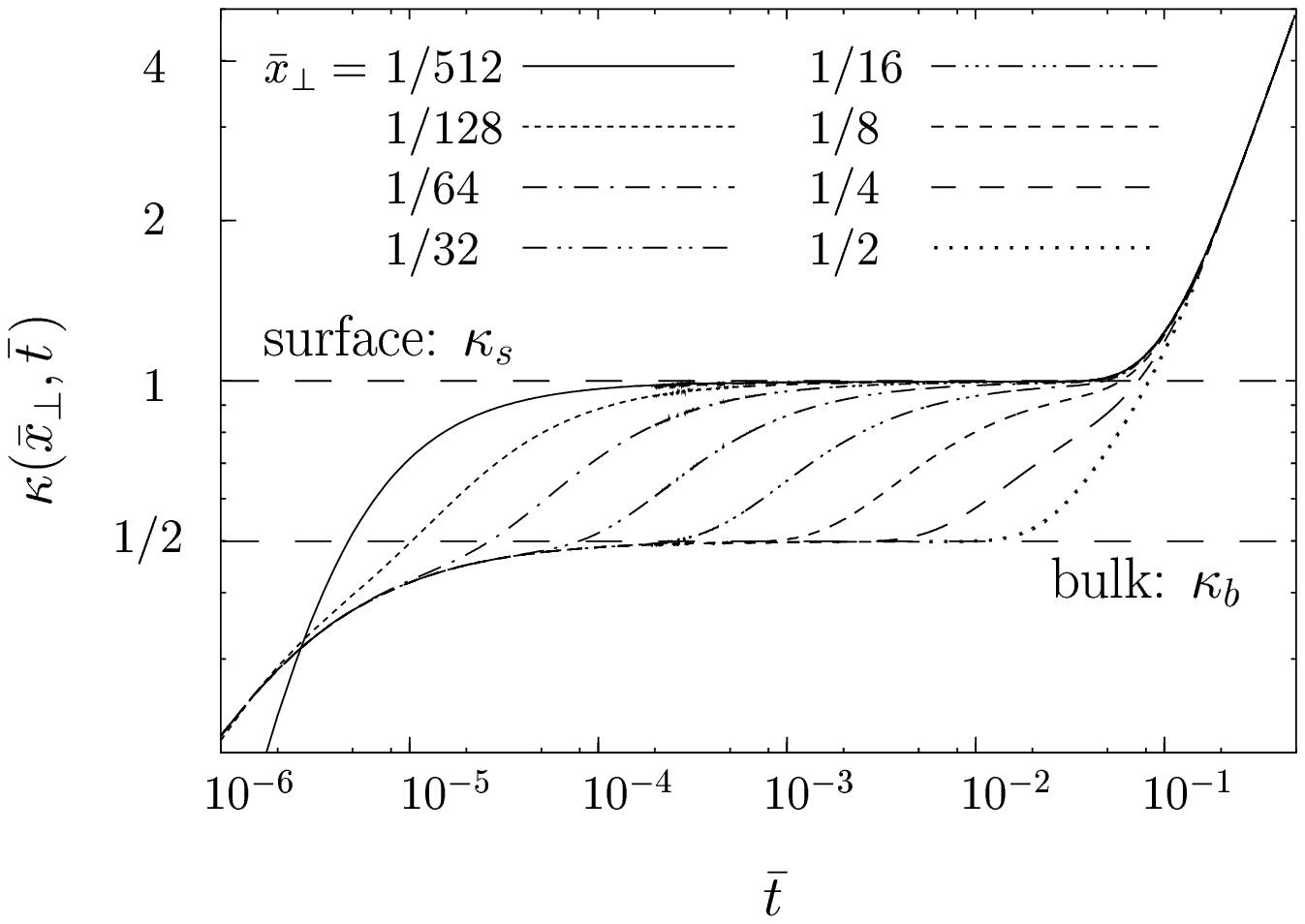,width=0.7\textwidth} 
\end{center}
\caption{Effective exponent $\kappa(\bar x_\bot,\bar t\,)$ for the order
parameter relaxation in a film of thickness $L$ (see
Eq.~\reff{defkappa})
as a function of $\bar t = (t/\rt_0)(\xi_0/L)^z$ and for
various values of $\bar x_\bot=x_\bot/L$. The two horizontal dashed lines
represent the values of the exponent in the bulk ($\kappa=1/2$) and close to a
surface ($\kappa=1$), respectively. For $\bar t < 10^{-4}$ the
non-universal features of initial
relaxation are evident. The linear increase of $\kappa(\bar x_\bot,
\bar t \gtrsim 10^{-1})=\rho \bar t$ indicates an exponential relaxation
$\sim e^{-\rho \bar t}$ with $\rho = \pi^2$.}
\label{filmexp}
\end{figure*}

For small values of $\bar x_\bot$ (lower curves in
 Fig.~\ref{filmloglog})
 we note that the order parameter relaxes according to a 
power law $\sim t^{-\kappa_s}$ until the crossover 
towards exponential decay takes
place. On the other hand, for $\bar x_\bot = 1/2$ (uppermost curve), 
the power law is different and relaxation follows the law $\sim
t^{-\kappa_b}$ with $\kappa_b < \kappa_s$. For intermediate values of
$\bar x_\bot$ there is a crossover between the two power laws. 
In order to elucidate this crossover, we compute the effective exponent of the
relaxation as the logarithmic derivative of the magnetization profile,
i.e.,
\begin{equation}
\kappa(\bar x_\bot,\bar t\,) \equiv - \frac{\partial\log \bar m(\bar
x_\bot,\bar t\,)}{\partial\log \bar t} \; .
\label{defkappa}
\end{equation}
A power-law behavior corresponds to a value of $\kappa$ independent of
$\bar t$, while an exponential decay $\sim e^{-\rho \bar t}$
leads to $\kappa = \rho \bar t$.
The behavior of $\kappa(\bar x_\bot,\bar t\,)$ is shown in
Fig.~\ref{filmexp}.

\begin{figure*}
\begin{center}
\epsfig{file=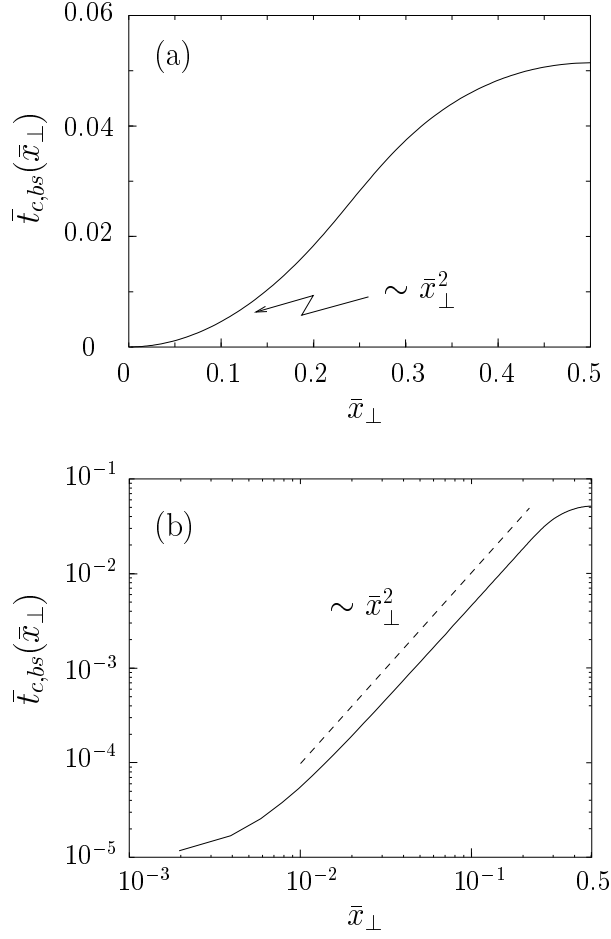,width=0.5\textwidth} 
\end{center}
\caption{(a) Crossover time $\bar t_{c,bs}$ 
(see Eqs.~\reff{tbulksurface} and~\reff{pippo2bis}) between bulk-like and surface-like
behavior of relaxation. 
(b) Log-log plot of the crossover time $\bar t_{c,bs}$. 
The straight line indicates the quadratic law 
expected to hold for small $\bar x_\bot$ within mean-field theory 
according to scaling arguments (see Eq.~\reff{pippo2bis}). For $\bar t \lesssim 10^{-2}$ there are 
deviations due
to the non-universal initial relaxation. Beyond mean-field theory
$\bar t_{c,bs}(\bar x_\bot \rightarrow 0)\sim \bar x_\bot^z$.}
\label{tcbs}
\end{figure*}

In accordance with the relaxation behavior 
in the semi-infinite geometry we expect the
following qualitative picture of the order parameter 
evolution in a film. At the
beginning of the relaxation process the effect of both boundaries on
the relaxation behavior starts to propagate
into the bulk. At a fixed distance from the wall the relaxation process
is characterized by the bulk exponent $\kappa_b$ until the influence 
of
the closest surface reaches this point and changes the exponent into the
surface one $\kappa_s > \kappa_b$. We expect that the final
crossover towards the exponential decay (due to the presence of {\it
two} confining walls with Dirichlet boundary conditions) will take
place via the
intermediate stage of surface relaxation
only if the spatial point under consideration 
can be reached sufficiently in time by the effect of one wall before
the effect of the second wall arrives.
This picture is confirmed by Fig.~\ref{filmexp}, which provides a
quantitative analysis. The two exponents $\kappa$ are readily
determined as $\kappa_s = 1$ and $\kappa_b=1/2$, corresponding to $\beta_1/\nu z$ and $\beta/\nu z$, respectively,
as in the case of the semi-infinite
geometry. 
There is a large degree of freedom in the
definition of times at which crossovers take place. We 
define the following typical times:
\begin{align}
\bar t_{c,b}(\bar x_\bot):\quad& \kappa(\bar t_{c,b},\bar x_\bot) = 0.4 \;,
\label{tbulk}\\
\bar t_{c,bs}(\bar x_\bot):\quad&  \kappa(\bar t_{c,bs},\bar x_\bot) =
0.75\;,
\label{tbulksurface}\\
\bar t_{c,e}(\bar x_\bot):\quad&  \kappa(\bar t_{c,e},\bar x_\bot) = 1.1\;.
\label{texp}
\end{align}
The first one, $\bar t_{c,b}$, is a measure of the time required to relax
from the initial condition. Its dependence on $x_\bot$ carries
non-universal information about the specific initial profile. In the scaling limit of
infinite initial magnetization (in the bulk) $\bar t_{c,b}=0$. The
time $\bar t_{c,bs}$ is a measure of the time required to cross over from
the {\it b}ulk to the {\it s}urface behavior. The value chosen in its definition
is simply half way between $\kappa_b$ and $\kappa_s$. The
corresponding plot is given in Fig.~\ref{tcbs} (a). It is useful to
remark that $\bar t_{c,bs}(\bar x_\bot)$ is expected to be finite
(given that it is still related to ``surface'' behavior) also in the limit
$L\rightarrow\infty$ at fixed $x_\bot$, i.e., in the semi-infinite
geometry. As a consequence, based only on scaling arguments, we can deduce
the behavior of $\bar t_{c,bs}(\bar x_\bot)$ for small $\bar
x_\bot$. 
Since $\kappa$ is dimensionless one expects the scaling behavior
$\kappa(x_\bot,t,L) = F_\kappa^{(3)}(x_\bot/L,(t/\rt_0)(\xi_0/L)^z)$
(compare Eq.~\reff{genscaling}).
Thus, according to the definition Eq.~\reff{tbulksurface}, it follows
\begin{equation}
\frac{t_{c,bs}}{\rt_0}\left(\frac{\xi_0}{L}\right)^z =
F_{c,bs}(x_\bot/L) \;.
\end{equation}
A finite non-trivial limit for $L\rightarrow\infty$ for
fixed $x_\bot$ exists provided
\begin{equation}
F_{c,bs}(y\rightarrow 0) = C_{bs} y^z
\end{equation}
with a constant $C_{bs}$ which implies
\begin{equation}
\frac{t_{c,bs}}{\rt_0}\left(\frac{\xi_0}{L}\right)^z = C_{bs}
\left(\frac{x_\bot}{L}\right)^z \;,
\quad\mbox{for}\quad x_\bot/L \ll 1 \;,
\label{pippo2}
\end{equation}
so that
\begin{equation}
\bar t_{c,bs}(\bar x_\bot) = C_{bs} {\bar x}^z_\bot \;, 
\label{pippo2bis}
\end{equation}
with $z=2$ within MFT.
This is in agreement with Fig.~\ref{tcbs} (b)
(apart from very small values of $\bar x_\bot$ which are numerically still
influenced by the initial relaxation) yielding $C_{bs} \simeq 0.5$.

\begin{figure*}
\begin{center}
\epsfig{file=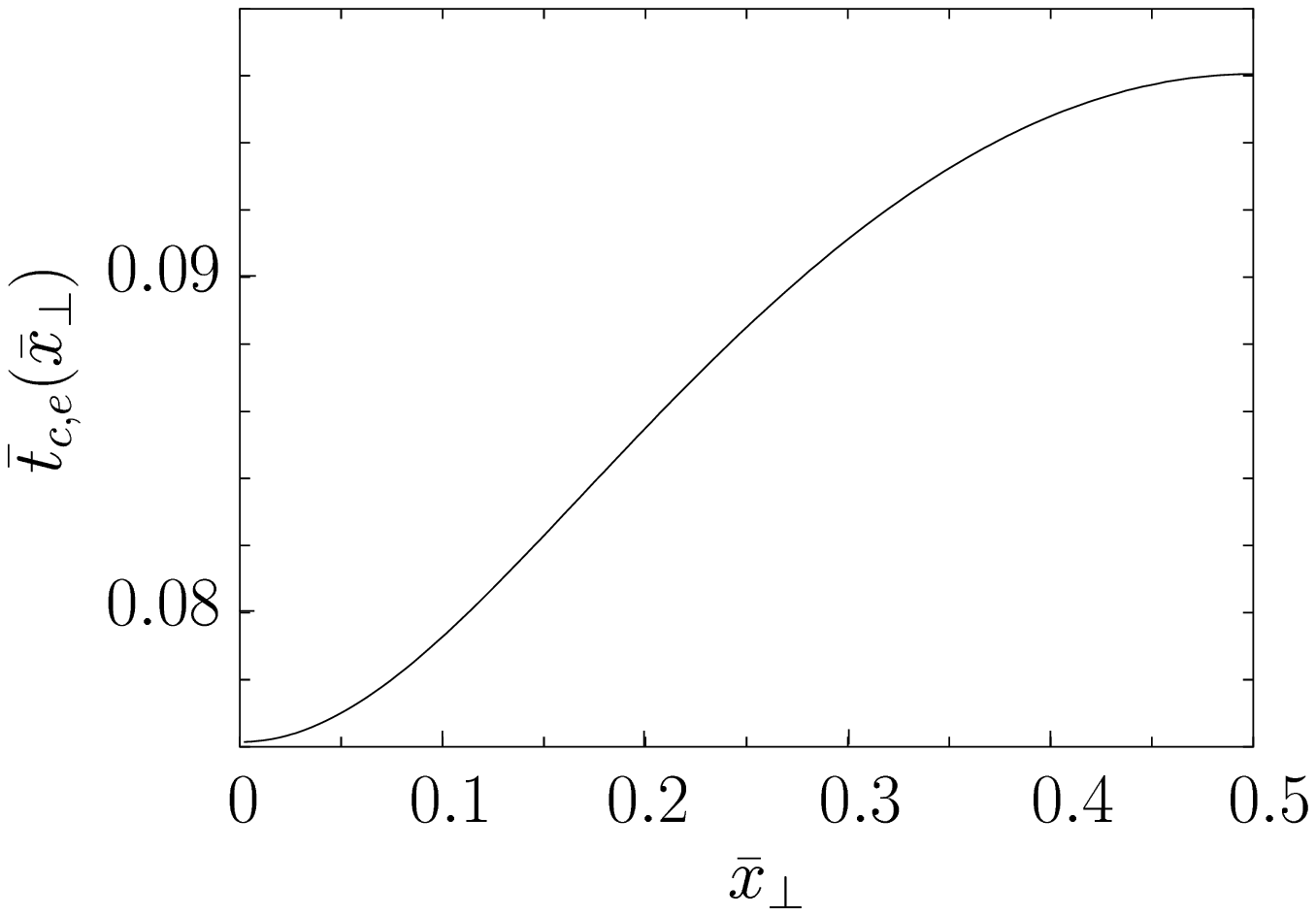,width=0.5\textwidth} 
\end{center}
\caption{Crossover time $\bar t_{c,e}$ (see Eq.~\reff{texp}) 
towards the linear relaxation behavior characterized by an
exponential decay of $\bar m(\bar x_\bot,\bar t\,)$.}
\label{tce}
\end{figure*}

The time $\bar t_{c,e}$ measures the time required to enter
the linear relaxation regime. As can be seen from
Fig.~\ref{tce}, $\bar t_{c,e}(\bar x_\bot)$ attains a nonzero value for
$\bar x_\bot\rightarrow 0$. This means that in the limit
$L\rightarrow\infty$ at fixed $x_\bot$ one has
\begin{equation}
t_{c,e}(x_\bot,L\gg x_\bot)/\rt_0 = D_e (L/\xi_0)^z\;, 
\end{equation}
with $D_e \simeq 0.076$. 
This divergence of the crossover time as function of $L$ is expected because 
in the semi-infinite geometry, i.e., for $L\rightarrow\infty$, 
the crossover towards an exponential decay never takes place.

\section{Summary}
\label{sec-sum}

Within the field-theoretical approach, we have investigated
various universal aspects of the relaxational 
critical dynamics [Model A, Eqs.~\reff{lang}-\reff{lgw}]
in film geometry 
with Dirichlet boundary conditions, corresponding to the so-called 
ordinary surface universality class. 
We have obtained the following main results.

\begin{itemize}
\item[(1)] 
\quad (i) We have provided general scaling properties (see
Sec.~\ref{intro} and Subsec.~\ref{subsec-genscalform}) for the
two-point response $R$ and correlation functions $C$ in the film geometry (Fig.~\ref{images}).
In particular, their Fourier transforms in the directions
parallel to the confining walls (with conjugate momentum $\bf p$)
scale as 
$
R({\bf p},x_{1\bot},x_{2\bot},t) = 
(\hat{\mathfrak{o}}^\pm_R/\rt^\pm_0)
(L/\xi_0^\pm)^{1-\eta-z} {\bar\RR}_\pm(\bar {\bf p} = {\bf
p}L,\bar x_{1\bot} = x_{1\bot}/L,\bar x_{2\bot} = x_{2\bot}/L, \bar t
= (t/\rt_0^\pm)(\xi_0^\pm/L)^z, \bar L = L/\xi)
$
and
$
C({\bf p},x_{1\bot},x_{2\bot},t) = 
(\hat{\mathfrak{o}}_C^\pm/\rt_0^\pm)$ $
(L/\xi_0^\pm)^{1-\eta} {\bar\CC_\pm}(\bar {\bf
p},\bar x_{1\bot}, \bar x_{2\bot},\bar t, \bar L)
$, respectively, in terms of the film thickness $L$, the bulk correlation
length $\xi(\tau =(T-T_{c,b})/T_{c,b} \rightarrow 0^\pm) = \xi_0^\pm |\tau|^{-\nu}$, and the
relaxation time $\rt_R(\tau \rightarrow 0^\pm) = \rt_0^\pm|\tau|^{-\nu
z}$ above and below the {\it bulk} critical temperature $T_{c,b}$, with
the universal ratio $\rt_0^+/\rt_0^- = 3.3(4)$ in spatial dimension
$d=3$ (Appendix~\ref{app-dynampratios}). The semi-infinite limit of $
{\bar\RR}_+$ is discussed in Eqs.~\reff{scalRsemiinf}-\reff{scalingSDRtfinal} and~\reff{RD0}. 
The explicit forms of the associated universal 
scaling functions ${\bar\RR}_+$ and  ${\bar\CC}_+$ have
been computed within Gaussian approximation (denoted by $^{(0)}$). 

\quad (ii) The time evolution of the
mean-field scaling
function $\bar \RR_+^{(0)}$ 
[see Eqs.~\reff{RPsi} and~\reff{defpsi}] for $\bar {\bf p} = 0$ and $T=T_{c,b}$ is shown  in Fig.~\ref{Rreal} for different
values of the scaling variables $\bar x_{i\bot}$ and $\bar t$. 
The curves in Fig.~\ref{Rreal} provide (apart from an amplitude) the
time evolution of the order
parameter profile across the film at $T=T_{c,b}$ after a laterally
homogeneous  
perturbation confined to the plane $x_\bot = x_{1\bot}$ and $\sim
\delta(t)$ 
has been applied.
A qualitative feature of interest is the time $\bar t_I(\bar x_{1\bot})$ 
at which the inflection points of these profiles
reach the boundary  $\bar
x_{2\bot}=0$, leading to a change in the shape of the
profiles (see Fig.~\ref{tflex} and Appendix~\ref{app-tI}). 
\item[(2)] 
\quad (i) We have discussed the {\it dy}namical effects of an external
field \mbox{$h({\bf x}=({\bf x_\|},x_\bot),t)$} 
(conjugate to the order parameter) on the
fluctuation-induced Casimir force $F_{l(r)}({\bf
x}_\|,\tau,L,t,\{h({\bf x},t)\})$ acting at the point ${\bf x}_\|$ 
on the {\it l}eft and {\it
r}ight confining walls.
Its scaling behavior is given by
$
F_{l(r)}({\bf x}_\|, \tau>0,L,t, \{h({\bf x},t)\}) 
= \; L^{-d} \,
\FF^{\rm (dy)}_{l(r)}(\bar {\bf x}_\| = {\bf x}_\|/L,\bar L,  \bar t,
\{(L/\xi_0)^{\beta\delta/\nu} [h(\bar {\bf
x},\bar t)/\h_0]\})$ 
(see also Eq.~\reff{defh0amp}). 
Within Gaussian approximation 
we have provided the expressions for the scaling
function $\FF^{\rm (dy)}_l$ for two specific instances of external
field: 
(W)~$h({\bf x},t) = h_W \delta(x_\bot - x_{1\bot})
\delta(t-t_1)$ and 
(P)~$h({\bf x},t) = h_P \delta({\bf x}_\| - {\bf x}_{1\|})
\delta(x_\bot - x_{1\bot}) \delta(t-t_1)$.
In both cases $\FF^{\rm (dy)}_l$ is the sum
[Eqs.~\reff{casimir-h-spec-wall} and~\reff{casimir-h-spec-point}]
of the {\it st}atic Casimir force $\FF^{\rm (st)}$ corresponding to 
the ordinary-ordinary surface universality class considered here, 
and a dynamic term which is quadratic in
the scaling variables $\hat h_{W,P}$ [see Eqs.~\reff{defhWhat}
and~\reff{defhPhat}] proportional to the amplitudes $h_{W,P}$ 
of the external field.
In particular we focused on the force at $T_{c,b}$ 
where $\FF^{\rm (st)}= (d-1)\Delta < 0$ [Eq.~\reff{deltaOO}].
For both cases W and P we have studied as function of time 
the maximum $A_\Delta^{W,P}$ of the field-induced
contribution $[\FF^{\rm (dy)}_l -
\FF^{\rm (st)}]/\hat h_{W,P}^2$.

\quad (ii) It turns out that $A_\Delta^W$ is attained at $\bar t =
\bar t_I(\bar x_{1\bot})$ discussed in the previous point (i). 
Figure~\ref{plotAmph} shows the dependence of $A^W_\Delta$ 
on the normal distance
$\bar x_{1\bot}$ at which the perturbation has been applied
(see Appendix~\ref{sub-asyA}). Figure~\ref{plotscalh} illustrates
the time dependence of the normalized part
of the field-induced Casimir force $\FF^W_\Delta(\bar x_{1\bot},\bar t\,) = [\FF^{\rm (dy)}_l -
\FF^{\rm (st)}]/[\hat h_W^2 A^W_\Delta(\bar x_{1\bot})]$ which decays
$\sim e^{-2 \pi^2\bar t}$. 

\quad (iii) $A_\Delta^P$ is attained at 
$\bar t = \bar t_M(\delta {\bf x}_\|,\bar x_{1\bot})$
(Fig.~\ref{plottMpoint}) which depends additionally on the scaled
lateral distance  
$\delta \bar {\bf x}_\|
\equiv ({\bf x}_\| - {\bf x}_{1\|})/L$ between the action of the force
and the epicenter of the perturbation. 
The maximum of the force on the walls spreads
with an asymptotically constant radial velocity which decreases with
the film thickness (Eq.~\reff{speed}). 
The force
decreases monotonically for increasing lateral distances $|\delta \bar
{\bf x}_\||$. 
Figures~\ref{plotAx0point} and~\ref{plotRatioA3d} show the dependence
of the corresponding maximum $A^P_\Delta(\delta {\bf
x}_\|,\bar x_{1\bot})$ 
(attained at $\bar t = \bar t_M(\delta {\bf
x}_\|,\bar x_{1\bot})$) on $|\delta \bar {\bf x}_\||$ and $\bar
x_{1\bot}$. 
The time dependence of the normalized part
of the Casimir force $\FF^P_\Delta(\delta \bar {\bf x}_\|,\bar x_{1\bot},\bar t\,) = [\FF^{\rm (dy)}_l -
\FF^{\rm (st)}]/[\hat h_P^2 A^P_\Delta(\delta \bar {\bf x}_\|,\bar
x_{1\bot})]$ is
reported in Fig.~\ref{scalTT3d}; it decays $\sim \bar t^{-(d-1)}
e^{-2\pi^2\bar t}$.
\item[(3)] We have computed the universal 
scaling function $\bar \CC^{(0)}_+$ of the dynamical two-point
correlation function [see the previous point (1)]
within Gaussian approximation.
In Fig.~\ref{Crealcrit} 
the time evolution of 
$\bar \CC_+^{(0)}$  [see Eq.~\reff{defC0Psi}]
for $\bar {\bf p} = 0$ and $T=T_{c,b}$ is shown for different values
of the scaling variables $\bar x_{i\bot}$ and 
$\bar t$. Figure~\ref{Creal} refers,
instead, to the cases in which $\bar t_0 = 1/(\bar {\bf p}^2 + \bar
L^2)$ is finite, i.e., if $T>T_{c,b}$ or ${\bf p} \neq 0$. 

\item[(4)] 
\quad (i) In view of the possibility to probe the spatial structure of
correlations by means of neutron or X-rays scattering
under grazing incidence, we have investigated the
behavior of the frequency- and momentum-dependent correlation function
in planes parallel to the surface of the film, i.e., of 
$C({\bf
p},x_{1\bot} = x_\bot, x_{2\bot} = x_\bot,\omega) = \hat{\mathfrak{o}}^\pm_C
(L/\xi_0^\pm)^{1-\eta+z} {\CC_\pm}(\bar {\bf
p},\bar x_\bot,\bar x_\bot,
\bar \omega = \omega \rt_0^\pm(L/\xi_0^\pm)^z,\bar L)
$ [see Eq.~\reff{scalComega}]. 

\quad (ii) Near the walls the surface
behavior
$
\CC_+(\bar {\bf
p},\bar x_\bot\rightarrow 0,\bar x_\bot\rightarrow 0, \bar \omega,\bar
L) = \bar
x_\bot^{2(\beta_1-\beta)/\nu}\CC_W(\bar {\bf p},\bar \omega,\bar L)
$ [see Eq.~\reff{scalComegaWall}] is recovered.
The various asymptotic behaviors of $\CC_W$ have been discussed in
Subsec.~\ref{subsec-genscalform} [see Eqs.~\reff{defA}--\reff{CCWx00}
and~\reff{ampAWinf0}--(\ref{ampCWinf0}),(\ref{ampAW0})].
Within Gaussian approximation the scaling function
$\CC^{(0)}_W$ is shown and discussed 
Figs.~\ref{Csp0o0} and~\ref{Csp0r0}.
In contrast to the
semi-infinite geometry, in films
$\CC^{(0)}_W$ attains a finite value at the origin of
the $(\xi^{-1},{\bf p},\omega)$-space which 
diverges for $L\rightarrow \infty$.

\quad (iii) The correlation function in the
middle of the film is characterized by the scaling function 
$\CC_I(\bar {\bf p},\bar\omega,\bar L) \equiv 
\CC_+(\bar {\bf p},1/2,1/2,\bar\omega,\bar L)$. Its asymptotic behaviors
are discussed in Subsec.~\ref{subsec-genscalform} [see  
Eqs.~\reff{CI000}--\reff{CCIx00}
and~\reff{ampAIinf0}--(\ref{ampCIinf0}),(\ref{ampAI0})]. 
Figures~\ref{Cbp0o0} and~\ref{Cbp0r0} show the shape 
of the scaling function
$\CC_I^{(0)}$ 
along the axes of the
$(\xi^{-1},{\bf p},\omega)$-space. 

\quad (iv) Comparing the scaling functions $\CC^{(0)}_W$ and
$\CC^{(0)}_I$ it turns out that, although their shapes are 
similar, the latter exhibits more rapid algebraic decays 
for $\bar L$, $\bar {\bf p}$, $\bar \omega \rightarrow\infty$.

\item[(5)] 
\quad (i) We have considered the nonlinear relaxation from an
initially ordered state by solving numerically the evolution equation
for the scaled order parameter profile $\bar m(\bar {\bf x},\bar t\,)$
[see Eqs.~\reff{adimensional} and~\reff{eqmotionadim}], which is
proportional to the time-dependent mean value of the order parameter
$\langle\varphi({\bf x},t)\rangle$ across the film. 
In particular we have analyzed the relaxation at the {\it bulk} critical
point $T=T_{c,b}$ from an initial profile that is laterally constant 
and has a symmetric shape in the
transverse direction which fulfills the Dirichlet boundary conditions. The
universal aspects of the relaxation process are 
independent of the actual shape of the initial profile. In view of this
symmetry the scaled order parameter profile $\bar m(\bar {\bf x},\bar t)$ depends only on $\bar x_\bot$ and
$\bar t$. 

\quad (ii) In Fig.~\ref{filmlog} the late stage of the relaxation of the order
parameter is shown as a function of $\bar t$ for various values of
$\bar x_\bot$. For all $\bar x_\bot$, $\bar m(\bar x_\bot,\bar t
\rightarrow\infty)$
displays an {\it exponential} decay due to the critical point shift in
the film geometry. On the other hand, during the early
stage of 
relaxation, i.e., for $\bar t \lesssim 0.1$, a crossover between an algebraic 
surface- and bulk-like behavior clearly emerges (see
Fig.~\ref{filmloglog}). 
Close to the surface one observes $\bar m(\bar
x_\bot,\bar t\,) \sim \bar t^{-\beta_1/(\nu z)}$ ($\beta_1/(\nu z) =
1$ within mean-field theory), whereas upon moving inside the film 
$\bar m(\bar
x_\bot,\bar t\,) \sim \bar t^{-\beta/(\nu z)}$ ($\beta/(\nu z) =
1/2$ within mean-field theory). This crossover is
clearly detected by the time dependence of the effective exponent $\kappa(\bar
x_\bot,\bar t\,) \equiv - \partial \log \bar m(\bar x_\bot,\bar
t)/\partial \bar t$ shown in Fig.~\ref{filmexp}. 
We have also defined (Eq.~\reff{tbulksurface}) and studied
(Fig.~\ref{tcbs}) the 
typical time $\bar t_{c,bs}(\bar x_\bot )$ at which the
crossover from the {\it b}ulk to the {\it s}urface behavior takes
place for a given distance 
$\bar x_\bot$ from the near wall, and that one for the
crossover to the ultimate {\it e}xponential decay $\bar t_{c,e}(\bar
x_\bot)$
(Eq.~\reff{texp} and Fig.~\ref{tce}).

\quad (iii) Thus the nonlinear relaxation of the order parameter in film
geometry is characterized by a cascade of four clearly identifyable,
successive decay modes: nonuniversal initial relaxation dominated by
the initial profile, bulk-like power-law decay $\sim t^{-\beta/\nu
z}$, surface-like power-law decay $\sim t^{-\beta_1/\nu z}$, and
exponential decay $\sim \exp(-\pi^2\bar t)$. The initial relaxation
lasts, in the present case, up to $\bar t \lesssim 10^{-5}$ and the
exponential decay prevails for $\bar t \gtrsim 10^{-1}$. The crossover
time between the two power laws depends on the distance of the point
of observation from the near wall.
\end{itemize}

\bigskip\bigskip

\section*{\small Acknowledgments}

A.G. is indebted to Frank Schlesener for many helpful discussions.

\appendix

\section{Universal dynamic amplitude ratios}
\label{app-dynampratios}

In this appendix we determine the universal
amplitude ratio $\rt_0^+/\rt_0^-$ as introduced in Sec.~\ref{intro}. 
We consider the model described in
Sec.~\ref{sec-model} in the absence of any confining wall (bulk
behavior) and for the specific case of a one-component field
(i.e., Ising universality class, $N=1$). The linear response of this model
in the presence of a finite external field $h$, 
and thus of a non-vanishing magnetization $m$, has been
studied in Refs.~\cite{BJ-76} and~\cite{BEJ-79} within a loop expansion up to
one loop. The results obtained there
allow one to compute the 
relaxation time of the system (a) in the high-temperature phase 
($\tau > 0$, with $h = 0$ and thus $m=0$) and (b) in the
low-temperature phase ($\tau < 0$ and thus $m = M_0\neq 0$ also for 
$h = 0$). In particular the value of the spontaneous
magnetization $M_0$ for $\tau<0$ and
$h = 0$ can be determined from the equation of state
reported in Eq.~(5.1) of Ref.~\cite{BJ-76}:
\begin{equation}
M_0(\tau<0) = \sqrt{\frac{3}{u^*}}\mu^{(d-2)/2}
(-2 \tau)^\beta \left[1+\frac{\epsilon}{6} + O(\epsilon^2)\right]
\label{eqstatelowT}
\end{equation}
where $\epsilon = 4 - d$, 
$\beta = (1-\epsilon/3)/2 + O(\epsilon^2)$ is the critical exponent
of the spontaneous magnetization $M_0$, 
$\tau = (T - T_{c,b})/T_{c,b}$,
$\mu$ 
is a momentum scale (see below), and $u^*$ is the
fixed-point value of the renormalized coupling constant (whose actual
value is irrelevant for the rest of the computation). The true
correlation length in the high-temperature phase is given by 
\begin{equation}
\xi(\tau \rightarrow 0^+) = \mu^{-1} \tau^{-\nu}\left[1+\frac{\epsilon}{12}+O(\epsilon^2)\right]
\end{equation}
with $\nu = (1+\epsilon/6)/2 + O(\epsilon^2)$. Accordingly, $\mu$ can
be expressed in terms of the nonuniversal amplitude $\xi_0^+$ as $\mu
= (\xi_0^+)^{-1}[1 + \epsilon/12 + O(\epsilon^2)]$.

The linear response
function can be computed from Eq.~(5.2) in Ref.~\cite{BJ-76}, taking
into account the different values of $M_0$ in the cases (a) ($M_0=0$)
and (b) (see Eq.~\reff{eqstatelowT}). This equation expresses the
deviation $m_1(t)$ of the magnetization from the constant value
$M_0$ in terms of the small magnetic field $h_1(t)$ that gives
rise to it, and thus provides an expression for the linear
susceptibility $\chi$ defined through 
\begin{equation}
m_1(t) = \int_{-\infty}^t \dd s \;\chi(t-s) h_1(s) \;.
\end{equation}
It is straightforward to express the Fourier transform 
$\hat\chi_{\pm}(\omega) = \int_0^{+\infty}\!\dd t\;e^{i\omega t}
\chi_\pm(t)$ of $\chi_\pm(t)$ for 
$\tau\gtrless0$ as
\begin{equation}
\hat\chi^{-1}_+(\omega) = \omega_0^+ \left[ -
i\frac{\omega}{\omega_0^+} + 1 - \frac{\epsilon}{6} \right] +
O(\epsilon^2) \; 
\end{equation}
and
\begin{equation}
\hat\chi^{-1}_-(\omega) =  \omega_0^- \left[ -
i\frac{\omega}{\omega_0^-} + 1 + \frac{\epsilon}{3} +
\frac{\epsilon}{2} F_R(i \omega/\omega_0^-) \right] + O(\epsilon^2)\;,
\end{equation}
where 
\begin{equation}
\omega_0^+ \equiv \Omega \mu^2 \tau^{\nu
z} \; \label{omegap}
\end{equation}
and
\begin{equation}
\omega_0^- \equiv \Omega \mu^2 (-2 \tau)^{\nu
z} \; ,\label{omegam}
\end{equation}
with $z = 2 + O(\epsilon^2)$ and
\begin{equation}
F_R(x) = \int_0^\infty \!\! \dd u \;
\frac{1+ u - e^u}{u^2} e^{- 2 u/x} = -1 + \left(1-\frac{2}{x}\right) \ln\left(1-\frac{x}{2}\right)
\end{equation}
which has a branch cut on the real axis for $x > 2$. The kinetic
coefficient $\Omega$ in Eqs.~\reff{omegap} and~\reff{omegam} can be
expressed in terms of the non-universal amplitudes $T_{0,{\rm ac}}^+$
or $T_{0,{\rm exp}}^+$, and $\xi_0^+$ through, c.f., Eq.~\reff{resultTf} 
or~\reff{resultTexp}. For instance, $\Omega = (\xi_{0,{\rm
exp}}^+)^2/\rt^+_{0,{\rm exp}} [1 + O(\epsilon^2)]$.

As it is the case for the (bulk) correlation length, there are
different possible definitions of the relaxation time $\rt_R$. 
For example it can be
defined as the inverse characteristic frequency (see, e.g.,
Eq.~(3.17) in Ref.~\cite{HH-77}), i.e.,
\begin{equation}
\rt_{R,{\rm ac}}(\tau) = \frac{i}{\hat \chi^{-1}(\omega=0)} \left.\frac{\partial\hat
\chi^{-1}(\omega)}{\partial\omega}\right|_{\omega=0}
\label{Trel}
\end{equation}
where the subscript {\rm ac} stands for ``autocorrelation''. Indeed 
using the FDT (see, e.g., Eq.~\reff{fdtomega}) one can show that 
$\rt_{R,{\rm ac}}$ coincides with the so-called
integrated autocorrelation time of the magnetization:
\begin{equation}
\rt_{R,{\rm ac}} = \int_0^{\infty}\!\!\dd t \;
\frac{C_M(t)}{C_M(0)}\; ,
\end{equation}
where $C_M(t)$ is the two-time dynamic
(auto)correlation function of thermal equilibrium  fluctuations (around $M_0$)
of the magnetization;  it is given via the
FDT (see, e.g., Eq.~\reff{fdtint}) by
$C_M(t)=\int_{|t|}^{+\infty} \!\! \dd s \,\chi(s)$.
Another possible definition of $\rt_R$ introduces
the ``true'' relaxation time determined by the position
of the frequency pole in $\chi(\omega)$ closest to the real axis:
\begin{equation}
\hat\chi^{-1}(\omega=-i\rt^{-1}_{R,{\rm exp}}(\tau)) = 0 \;.
\label{Ttrue}
\end{equation}
From Eq.~(\ref{Ttrue}) it follows that 
$\rt_{R,{\rm exp}}$ governs the
long-time exponential decay $\sim \exp(-t/T_{R,{\rm exp}})$ of the
correlation and linear response function away from $T_c$. 

Using the previous expressions one finds that
\begin{equation}
T_{R,{\rm ac}} = \left\{
\begin{array}{lc} 
(\omega_0^+)^{-1} \left(1+\frac{1}{6}\epsilon\right) + O(\epsilon^2)\;, & \quad\tau > 0 \;, \\
(\omega_0^-)^{-1} \left(1-\frac{5}{24}\epsilon \right) + O(\epsilon^2)\;, & \quad\tau < 0 \;,
\end{array}
\right.
\label{resultTf}
\end{equation}
which yields the universal amplitude ratio
\begin{equation}
\frac{T_{0,{\rm ac}}^+}{T_{0,{\rm ac}}^-} 
= 2^{\nu z}\left( 1 + \frac{3}{8}\epsilon\right) +
O(\epsilon^2) \;.
\end{equation}
Instead, for $T_{R,{\rm exp}}$ one finds
\begin{equation}
T_{R,{\rm exp}} = \left\{
\begin{array}{lc} 
(\omega_0^+)^{-1} \left(1+\frac{1}{6}\epsilon\right) + O(\epsilon^2)\;, & \quad\tau > 0 \;, \\
(\omega_0^-)^{-1} \left[1+(\frac{1}{6}-\frac{\ln
2}{2})\epsilon \right] + O(\epsilon^2)\;, & \quad\tau < 0 \;,
\end{array}
\right.
\label{resultTexp}
\end{equation}
so that the corresponding universal amplitude ratio is given by
\begin{equation}
\frac{T_{0,{\rm exp}}^+}{T_{0,{\rm exp}}^-} = 2^{\nu z}\left( 1 +
\frac{\ln 2}{2}\epsilon\right) +
O(\epsilon^2) \;.
\label{ratioexp}
\end{equation}

In Ref.~\cite{CMPV-03} the purely relaxational dynamics of the Ising
model has been studied both analytically and numerically 
(in three dimensions). In particular the generalizations of $T_{R,{\rm
ac}}$ and
$T_{R,{\rm exp}}$ to the case of a non-vanishing
wavevector have been considered for a generic point in the
$(\tau,M_0)$-plane, and their universal scaling functions
have been computed (see Eqs.~(15), (16), and (19) therein). 
These results provide predictions also for the universal ratios 
$T^+_{0,{\rm exp}}/T_{0,{\rm ac}}^+$ and 
$T^-_{0,{\rm exp}}/T_{0,{\rm ac}}^-$. Using
the notations of Ref.~\cite{CMPV-03}, 
$T^+_{0,{\rm exp}}/T_{0,{\rm ac}}^+ = 
\mathcal{T}_{\rm exp}(0;\infty) = 1 + O(\epsilon^2)$ 
and $T^-_{0,{\rm exp}}/T_{0,{\rm ac}}^- = 
\mathcal{T}_{\rm exp}(0;-1)= 1 + (3/8-\ln2/2)\epsilon +
O(\epsilon^2)$, respectively. It is straightforward to check that 
the results reported in Eqs.~\reff{resultTf} and~\reff{resultTexp} 
are in accordance with these predictions.


\section{Static correlation function}
\label{app-staticcorr}

The static (i.e., equal-time) correlation function $C^{(0)}_{\rm st}$ 
can be obtained from
Eq.~\reff{reprC4} by integrating $C^{(0)}({\bf q}_n,\omega)$ over $\omega$
(see Eqs.~\reff{corr} and~\reff{reprC3}):
\begin{equation}
C^{(0)}_{\rm st}({\bf p},x_{1\bot},x_{2\bot}) = \sum_{n=1}^\infty
\Phi_n(x_{1\bot};L)\Phi_n(x_{2\bot};L)\frac{1}{{\bf q}_n^2 + r_0}\;.
\end{equation}
Using the relations (see Eq.~\reff{eigenf})
\begin{equation}
\Phi_n(x_1;L)\Phi_n(x_2;L) = \frac{1}{L}\left\{\cos[\frac{\pi
n}{L}(x_1-x_2)]-\cos[\frac{\pi n}{L}(x_1+x_2)]\right\} \;,
\label{Phiprod}
\end{equation}
${\bf q}_n^2 = {\bf p}^2 + \pi^2n^2/L^2$, and the identity
(see, e.g., \S5.4.5-1 in Ref.~\cite{IS})
\begin{equation}
\sum_{n=1}^\infty \frac{\cos(\pi n z)}{\pi^2 n^2 + \alpha^2} = \left\{ \frac{1}{2\alpha}\frac{\cosh
[\alpha(|z|-1)]}{\sinh\alpha} - \frac{1}{2\alpha^2}\right\}\;,\quad
\mbox{for}\quad -2<z\le2 \;,
\end{equation}
(outside the range $|z|\le 2$ this
expression has to be extended periodically in $z$ with period $2$),
one finds
\begin{equation}
C^{(0)}_{\rm st}({\bf p},x_{1\bot},x_{2\bot}) = \frac{1}{2 a} \frac{\cosh[a(|x_{1\bot}-x_{2\bot}|-L)]-\cosh[a(x_{1\bot}+x_{2\bot}-L)]}{\sinh(aL)}
\label{Cstat}
\end{equation}
where $a^2 \equiv {\bf p}^2 + \xi^{-2}$ (see footnote~\ref{footnote3}). 
By applying addition
formulae for hyperbolic functions, Eq.~\reff{Cstat} can be expressed
as
\begin{equation}
C^{(0)}_{\rm st}({\bf p},x_{1\bot},x_{2\bot}) = 
L\frac{\sinh[aL(x_\bot^</L)]\sinh[aL(1- x_\bot^>/L)]}{aL \sinh(aL)} \;,
\label{Cstatstand}
\end{equation}
where $x_\bot^> = {\rm max}\{x_{1\bot},x_{2\bot}\} = (x_{1\bot}+x_{2\bot}+|x_{1\bot}-x_{2\bot}|)/2$ 
and $x_\bot^< = {\rm min}\{x_{1\bot},x_{2\bot}\} =  (x_{1\bot}+x_{2\bot}-|x_{1\bot}-x_{2\bot}|)/2$,
in agreement with Eq.~(B4) in Ref.~\cite{KED-95} (and with Eq.~(4.1) of
Ref.~\cite{Diehl-86} in the case $c_0=+\infty$). This result can also
be found directly from Eq.~\reff{Romega} 
by using the FDT (see Eq.~\reff{fdtCR}): $C^{(0)}_{\rm st}({\bf
p},x_{1\bot},x_{2\bot}) = C^{(0)}({\bf p},x_{1\bot},x_{2\bot},t=0)= R^{(0)}({\bf p},x_{1\bot},x_{2\bot},\omega=0)$.


\section{Response function and its asymptotic time dependence}
\label{app-formulas}
With a view of the scaling function $\Psi$ for the response function
$R^{(0)}$ (see Eqs.~(\ref{Rgeneral}) and (\ref{Rgenpsi}))
we consider the sum
\begin{equation}
\sum_{n=1}^\infty \Phi_n(x_1;L)\Phi_n(x_2;L)
e^{-\Omega (\pi n/L)^2 t} \equiv \frac{1}{L}\Psi(x_1/L,x_2/L,\Omega t/L^2)
\label{somma}
\end{equation}
with the eigenfunctions $\Phi_n(x;L)$ defined in
Eq.~\reff{eigenf}. 
Using Eq.~\reff{Phiprod}, Eq.~\reff{somma} reduces to the evaluation of
expressions of the form
\begin{equation}
\sum_{n=1}^\infty \cos(\pi n \mu) e^{-\pi^2n^2\tau} = -\frac{1}{2} +
\frac{1}{2}\sum_{n=-\infty}^{+\infty} e^{i\pi n \mu - \pi^2n^2\tau} 
\end{equation}
which can be expressed, 
according to standard definitions (see, e.g., \S16.27.3 in  
Ref.~\cite{AS}), 
in terms of Jacobi's theta
function
\begin{equation}
\vartheta_3(z,q) = \sum_{n=-\infty}^{+\infty} q^{n^2} e^{2 n i z} 
\end{equation}
so that with $\bar t = \Omega t/L^2$ (see Eq.~\reff{omegascal}) and $\bar
x_i = x_i/L$ 
\begin{equation}
\Psi(\bar x_1,\bar x_2,\bar t\,) =
\frac{1}{2}\left[\vartheta_3(\pi\frac{\bar x_1-\bar
x_2}{2},e^{-\pi^2\bar t}) -
\vartheta_3(\pi\frac{\bar x_1+\bar x_2}{2},e^{-\pi^2\bar t}) \right]\;.
\label{defpsi}
\end{equation}
By using Poisson's resummation formula [i.e., $\sum_{n=-\infty}^\infty
e^{2\pi i n x} = \sum_{n=-\infty}^\infty \delta(x-n)$] we
obtain 
\begin{equation}
\vartheta_3(\pi\bar x,e^{-\pi^2\bar t}) = \frac{1}{\sqrt{\pi\bar t}}
\sum_{n=-\infty}^{+\infty} e^{-(n-\bar x)^2/\bar t} 
\label{poisson}
\end{equation}
so that
\begin{equation}
\begin{split}
&\Psi(\bar x_1,\bar x_2,\bar t\,) =
\frac{1}{\sqrt{4 \pi \bar t}}
\Bigg\{e^{-(\bar x_1-\bar x_2)^2/(4\bar t\,)} 
- e^{-(\bar x_1+ \bar x_2)^2/(4\bar t\,)} \\
&\hspace{1cm}+ \sum_{n=1}^\infty\left[ e^{-(\frac{\bar
x_1-\bar x_2}{2}-n)^2/\bar t} 
+ e^{-(\frac{\bar x_1-\bar x_2}{2}+n)^2/\bar t} -
e^{-(\frac{\bar x_1+\bar x_2}{2}-n)^2/\bar t} -
e^{-(\frac{\bar x_1+\bar x_2}{2}+n)^2/\bar t} \right]\Bigg\}\;.
\end{split}
\label{onewall}
\end{equation}
This expression is useful for discussing the semi-infinite limit of the
response function (see Eq.~\reff{Ronewall}).

The long-time limit $\bar t = \Omega t/L^2\rightarrow\infty$ 
of Eq.~\reff{somma} 
follows from 
\begin{equation}
\vartheta_3(\pi\bar x,e^{-\pi^2\bar t}) = 1 + 2 e^{-\pi^2\bar t}
\cos(2\pi\bar x) + O(e^{-4 \pi^2\bar t})
\end{equation}
so that
\begin{equation}
\Psi(\bar x_1,\bar x_2,\bar t\rightarrow\infty) = e^{-\pi^2\bar
t}[\cos\pi(\bar x_1-\bar x_2) -\cos\pi(\bar x_1+\bar x_2) ] + O(e^{-4
\pi^2\bar t})
\end{equation}
as expected also from Eq.~\reff{somma} because in that limit the sum
in Eq.~\reff{somma} is dominated by its first term so that
\begin{equation}
\Psi(\bar x_1,\bar x_2,\bar t\rightarrow\infty) = L \Phi_1(x_1;L)\Phi_1(x_2;L)
e^{-\pi^2\bar t}
+ O(e^{-4 \pi^2\bar t}).
\end{equation}


\section{Analytic expression for the mean-field order parameter profile}
\label{app-OPprofile}

The rescaled mean-field order
parameter profile $\bar m_\infty({\bf \bar x})$ is the stationary solution
of Eq.~\reff{eqmotionadim}:
\begin{equation}
[-\Delta_{\bf\bar x} +
\bar \tau +
\bar m^2_\infty({\bf\bar x})] \bar m_\infty({\bf\bar x}) = 0 
\label{eqmotionadim-stat}
\end{equation}
with the Dirichlet boundary conditions $\bar m_\infty({\bf \bar
x}_{\mathcal B}) = 0$. Thus 
$\bar m_\infty({\bf\bar x})$ is 
an extremum of the functional~\reff{lgw}, which can be
rewritten as
\begin{equation}
{\cal H}[m] = L^{d-4} \m_0^2\xi_0^2 \int_{\bar V}\!\dd^d \bar x \left[
\frac{1}{2} (\nabla_{\bf\bar x} \bar m({\bf\bar x}))^2 + 
\frac{1}{2} \bar \tau \bar m({\bf\bar x})^2
+\frac{1}{4}\bar m({\bf\bar x})^4 \right] \label{lgwadim}
\end{equation} 
by taking into account Eqs.~\reff{omegascal} and~\reff{adimensional},
the definitions ${\bf\bar x} = {\bf x}/L$, 
${\bar \tau} = (L/\xi)^2 = \bar L^2$ for $\tau >0$, $\bar\tau = - 1/2 (L/\xi)^2 =
-\bar L^2/2$
for $\tau<0$,
and $\bar V = {\mathbb R}^{d-1}\times [0,1]$ as well as $\bar m({\bf \bar
x}_{\mathcal B}) = 0$. 
We recall that, within mean-field theory, $\xi(\tau \rightarrow 0^+) =
r_0^{-1/2}$ whereas $\xi(\tau \rightarrow 0^-) = (-2 r_0)^{-1/2}$
(Eq.~\reff{lgw}); in
the main text we use also the abbreviation $L/\xi = \bar L$.
Here and in the following $\xi$ means $\xi_-$ for
$\tau <0$ and $\xi_+$ for $\tau >0$.
In view of the translational symmetry in all
directions parallel to the confining walls, finding the equilibrium
profile reduces to solving a one-dimensional differential equation 
for $\psi(\bar x_\bot) = \bar
m(\ve{\bar x})$ where
$\ve{\bar x} = (\ve{\bar x}_\|,\bar x_\bot)$: 
\begin{equation}
\begin{split}
-\psi''(\bar x_\bot) + \bar\tau \psi(\bar x_\bot) + \psi^3(\bar x_\bot) = 0& \;,\\
\psi(0) = \psi(1) = 0& \;,
\end{split}
\label{eqpsi}
\end{equation}
with the symmetry $\psi(\bar x_\bot) = \psi (1-\bar x_\bot)$ so that for
regular solutions $\psi'(1/2)=0$. Accordingly a set of equivalent boundary
conditions for Eq.~\reff{eqpsi} is $\psi(0) = 0$ and $\psi'(1/2)=0$.
Equation~\reff{eqpsi} has always 
the trivial solution $\psi(\bar x_\bot) \equiv 0$. Note that
Eq.~\reff{eqpsi} describes the one-dimensional closed motion of a particle 
with coordinate $\psi$ in a potential $V(\psi) = -\bar\tau\psi^2/2
-\psi^4/4$ as a function of ``time'' $\bar x_\bot$.
Using this analogy one can show
that
the non-trivial solution $\psi(\bar x_\bot)$ of Eq.~\reff{eqpsi} 
is bounded by the non-trivial
solution $\bar\psi$ of Eq.~\reff{eqpsi} with free boundaries
resembling the bulk solution,
\begin{equation}
\bar\psi = \left\{
\begin{array}{lcl}
0 &\mbox{for}& \bar\tau > 0 \;,\\ 
\sqrt{-\bar \tau}&\mbox{for}& \bar\tau \le 0 \;,
\end{array}
\right.
\label{bound}
\end{equation} 
in the sense that $|\psi(\bar x_\bot)| \le
\bar\psi$~\footnote{%
For $\bar\tau<0$, $\bar \psi$ is the position of the
maximum of $V(\psi)$ for $\psi>0$. A particle starting from the
point $\psi = 0$ at the ``time'' $\bar x_\bot=0$ with $\psi'(\bar x_\bot)>0$
cannot perform a closed motion (and then return to the starting point
at the ``time'' $\bar x_\bot = 1$) if its initial kinetic energy is 
large enough
to overcome the potential barrier $V(\bar\psi)$, which allows for 
$\psi>\bar\psi$. As a consequence, for a closed motion one 
has $|\psi|\le\bar\psi$.
}.
In view of this,
in the following we
consider $\bar\tau < 0$. Equation~\reff{eqpsi} can be integrated once
by using the boundary condition $\psi(0) = 0$:
\begin{equation}
\psi'^2(\bar x_\bot) = \psi'^2(0) + \bar\tau \psi^2(\bar x_\bot) +
\frac{1}{2} \psi^4(\bar x_\bot) \;.
\label{eqpsi2}
\end{equation}
The solution of 
Eq.~\reff{eqpsi2} 
(together with the boundary condition $\psi(0)=0$) depends on $\psi'(0)$ which
has to be determined as a function of $\bar\tau$ in order to fulfill
the second boundary condition $\psi'(1/2)=0$.
It is convenient to introduce the scale transformation 
$\psi(\bar x_\bot) = \sqrt{2} k \zeta
\, \sigma(\zeta \bar x_\bot)$, with $\sigma(0)=0$ and $\sigma'(0) = 1$.  
In terms of $\sigma(w=\zeta \bar x_\bot)$  
Eq.~\reff{eqpsi2} turns into
\begin{equation}
\sigma'^2 = (1 - \sigma^2) (1 - k^2 \sigma^2)\;,
\label{eqpsi3}
\end{equation}
where
\begin{equation}
-\bar\tau = \zeta^2(k^2+1)\;.
\label{tau} 
\end{equation}
The two parameters $k>0$ and
$\zeta>0$ just introduced have to be determined as a function of
$\bar\tau$ in such a way that
$\sigma'(\zeta/2) = 0$, corresponding to the condition 
$\psi'(1/2)=\sqrt{2}k\zeta^2 \sigma'(\zeta/2) = 0$, and
that Eq.~\reff{tau} is satisfied.
Equation~\reff{eqpsi3} is solved by the Jacobian
elliptic integral of the first kind with modulus $k$ (assuming that
$\psi'(\bar x_\bot) = \sqrt{2} k \zeta^2 \sigma'(\zeta \bar x_\bot)>0$
for $0<\bar x_\bot<1/2$),
\begin{equation}
\int_0^{\sigma(w)}\!\dd s
\frac{1}{\sqrt{(1-s^2)(1-k^2s^2)}} = w\;,
\label{elliptic-int}
\end{equation}
 and thus (see, e.g., 8.144.1 in Ref.~\cite{GR})
\begin{equation}
\sigma(w) = \sn(w;k)\;.
\end{equation}
The parameters $k$ and $\zeta$ are related via Eq.~\reff{tau}. They
are fixed by imposing the boundary condition $\sigma'(\zeta/2) = 0$. Thus
Eq.~\reff{eqpsi3} renders two possible values: $\sigma(\zeta/2) = 1$
and $\sigma(\zeta/2) = 1/k^2$. On the other hand, 
Eq.~\reff{bound} implies that 
$\psi(1/2) = \sqrt{\bar \tau} = \sqrt{2} k \zeta\, \sigma(\zeta/2) \le \bar \psi = \zeta
\sqrt{k^2+1}$ (where we used Eq.~\reff{tau}), so that
$\sigma(\zeta/2)\le \sqrt{(k^2+1)/(2 k^2)}$, i.e., 
(a) $\sigma(\zeta/2) = 1$ for
$0<k<1$ and (b) $\sigma(\zeta/2)=1/k^2$ for $k>1$. We first consider
case (a). Using $\sigma(\zeta/2) =  1$ in Eq.~\reff{elliptic-int} one finds the
following relation between $\zeta$ and $k$ (see, e.g., 8.111.2 and
8.112.1 in Ref.~\cite{GR}):
\begin{equation}
K(k) = \frac{\zeta}{2}\;,
\end{equation}
where $K(k)$ is the complete elliptic integral of the first kind. This
allows one to replace the variable $\zeta$ in Eq.~\reff{tau}:
\begin{equation}
-\bar\tau = [2 K(k)]^2(k^2+1)\;.
\label{detk}
\end{equation}
This is an implicit equation  $k = k(\bar\tau)$ 
for the modulus in
terms of the physical variable $\bar\tau = - 1/2(L/\xi)^2$ (as defined
for $\bar\tau < 0$).
Thus the solution of Eqs.~\reff{eqmotionadim-stat} and~\reff{eqpsi} 
is given by 
\begin{equation}
\psi(\bar x_\bot) = 2\sqrt{2} k K(k) \, \sn(2 K(k) \bar x_\bot; k)\;. 
\label{ntprofile}
\end{equation}
Since $K(k)$ is a monotonicly increasing function with 
$K(0)=\pi/2$, one has $[2K(k)]^2(k^2+1)\ge \pi^2$ and therefore there is
a non-trivial solution $\psi(\bar x_\bot)\neq 0$ only for 
\begin{equation}
\bar \tau \le -
\pi^2\;. 
\label{shift}
\end{equation}
This determines the well-known mean-field 
critical point shift in the film
geometry with Dirichlet boundary conditions on both 
sides~\cite{KO-72,K-72,FN-81,NF-83} 
($\nu = 1/2$):
\begin{equation}
\tau_c = -\pi^2 \left(\frac{\xi_0^-}{L}\right)^{1/\nu}\;.
\end{equation}
The discussion of case (b) proceeds accordingly. 
Using the property $\sn(k u;1/k) = k \,\sn (u;k)$ (see, e.g.,
formula 106.01 in Ref.~\cite{BD}) one finds that the solution for case
(b) is the same as for case (a) provided the modulus $k$ is replaced
by $1/k$.

These results can also be obtained from the findings in 
Ref.~\cite{K-97}, in which the mean-field order parameter profile is
analytically determined for the case of $(+,+)$ and $(+,-)$ boundary
conditions (see Ref.~\cite{K-97} for details). 
Alternatively, Eq.~\reff{eqpsi} can be
integrated so that
\begin{equation}
\psi'^2(\bar x_\bot) = \bar\tau \psi^2(\bar x_\bot) + \frac{1}{2}
\psi^4(\bar x_\bot) - \bar\tau
\psi^2(1/2) - \frac{1}{2} \psi^4(1/2)\;,
\label{eqpsicoll}
\end{equation}
using $\psi'(1/2)=0$ which is valid both for 
Dirchlet-Dirichlet and $(+,+)$ boundary conditions. 
The function $u(\bar x_\bot)$ defined as
\begin{equation}
u(\bar x_\bot) = \frac{A}{\psi(\bar x_\bot)}\;
\label{inversion}
\end{equation}
satisfies Eq.~\reff{eqpsicoll}, provided that $A$ is chosen according
to
\begin{equation}
\frac{A^2}{2} = - \bar\tau \psi^2(1/2) - \frac{1}{2}\psi^4(1/2) \;,
\end{equation}
which is equivalent to $- \bar\tau u^2(1/2) - 1/2 u^4(1/2)
= A^2/2$. Therefore both $\psi(\bar x_\bot)$ and $u(\bar x_\bot)$
solve Eq.~\reff{eqpsicoll}. If $\psi$ satisfies $(+,+)$
boundary conditions, then $u$ satisfies Dirichlet-Dirichlet boundary
conditions and vice versa. Thus it is possible to take advantage of
the result reported in Ref.~\cite{K-97} for the former case in order to solve
the latter we are presently interested in. (Note that the
normalizations used here are different from those used in Ref.~\cite{K-97}, 
see Eq.~\reff{eqpsi} here and
Eq.~(A3) therein.) In Ref.~\cite{K-97} two different parameterizations
are provided for the solution denoted as $m_{+,+}$: one for 
$\tau L^2 > -\pi^2$ in Eq.~(A13) therein, and the other for 
$\tau L^2 \le -\pi^2$ in Eq.~(A15) therein. One can see that in the former
case the value of $A^2$ is negative, i.e., the
corresponding solution for Dirichlet-Dirichlet boundary conditions is
not real, and one is left only with the trivial solution which is
identically zero. Instead, in
the latter case, taking into account the different normalizations,
\begin{equation}
\psi_{+,+}(z) = 2\sqrt{2} K(k) \frac{1}{\sn(2K(k)z;k)}\;,
\end{equation}
where $k$ is determined according to Eq.~\reff{detk}. Thus
$A^2 = 64 k^2 [K(k)]^4$ and the corresponding function
$u_{D,D}(z)=A/\psi_{+,+}(z)$ is identical to the solution $\psi(z)$
given in Eq.~\reff{ntprofile}.

From Eqs.~\reff{ntprofile} and~\reff{detk} it is possible to recover
the result for the mean-field profile in the semi-infinite geometry
by considering the limit of large $L$ for fixed $\xi_-$ and $x_\bot$ which
corresponds to $|\bar\tau| \gg 1$, $\bar x_\bot \ll 1$. For $|\bar\tau|
\gg 1$ one has $k\rightarrow 1^-$ and thus one can use the approximation
(see, e.g., formula 17.3.26 in Ref.~\cite{AS})
\begin{equation}
K(k) = -\frac{1}{2} \ln\left(\frac{1-k^2}{16}\right)(1+O(1-k))\;.
\label{Kapprox}
\end{equation}
Given that $\sn(u,1) = \tanh\, u$ (see, e.g., formula 127.02 in 
Ref.~\cite{BD}) one easily finds at leading order
\begin{equation}
\psi(\bar x_\bot) \simeq \bar\psi\, \tanh
\left(\sqrt{\frac{-\bar\tau}{2}}\bar x_\bot\right)\quad\mbox{for}\quad
\bar x_\bot \ll 1, |\bar \tau| \gg 1\;.
\label{profsemiinf}
\end{equation}
Using Eqs.~\reff{adimensional} and \reff{bound} and the fact
that within mean-field theory $\xi_0^+/\xi_0^- = \sqrt{2}$,
one can
express the profile as
\begin{equation}
m({\bf x},\tau) = 
\m_0 \frac{\xi_0^-}{\xi}\tanh
\left(\frac{1}{2}\frac{x_\bot}{\xi}\right)
\mathop{\longrightarrow}_{x_\bot\rightarrow \infty} \m_0
\tau^{1/2}(1-2 e^{-x_\bot/\xi_-})\;,
\end{equation}
where here $\xi \equiv \xi(\tau<0)=\xi_-$.
\begin{figure*}
\begin{center}
\epsfig{file=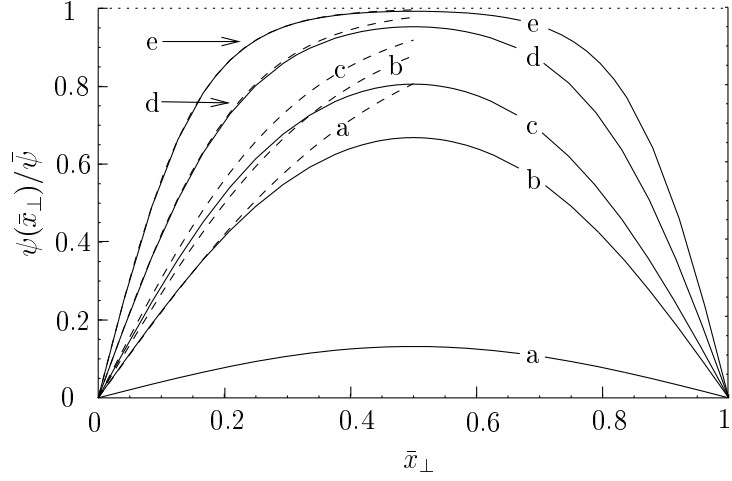,width=0.6\textwidth} 
\end{center}
\caption{Order parameter profile $\psi(\bar x_\bot)$
(normalized to the corresponding bulk value $\bar\psi$, see Eq.~\reff{bound})
across the film, for some values of 
$\bar\tau = - 1/2(L/\xi)^2$: $-10$ (a), $-15$ (b), $-20$
(c), $-40$ (d), $-80$ (e). The order parameter vanishes for $\bar\tau \ge
-\pi^2 \simeq -9.87$ (Eq.~\reff{shift}). 
The dashed lines for $\bar x_\bot \le 1/2$
represent the corresponding profiles (normalized to the bulk value
$\bar \psi$) in the semi-infinite geometry (given by
Eq.~\reff{profsemiinf}) in which the order parameter vanishes only for
$\bar\tau \ge 0$. In
the cases (d) and (e) these provide good approximations to the
actual order parameter profile for $\bar x_\bot\lesssim 1/2$; for (e) the
differences are barely visible.}
\label{profiloparag}
\end{figure*}
This expression agrees with the well-known result for the
semi-infinite geometry~\cite{K-74,LR-75}. In Fig.~\ref{profiloparag} the
order parameter profile $\bar\psi(\bar x_\bot)/\bar \psi$ (normalized
to the corresponding bulk value $\bar\psi$, see Eq.~\reff{bound}) is
shown for some values of $\bar\tau$. The comparison with the
profile in the semi-infinite geometry 
(Eq.~\reff{profsemiinf}) is also
shown.

From Fig.~\ref{profiloparag} one can infer that for 
$\bar\tau\rightarrow -\infty$ the order parameter
profile in the middle of the film $\bar x_\bot=1/2$ 
approaches rapidly the bulk value.
Indeed, defining $\delta\psi \equiv \psi(\bar x_\bot = 1/2)-\bar\psi$
one finds
\begin{equation}
-\frac{\delta\psi}{\bar \psi} = 1 - \frac{\sqrt{2} k}{\sqrt{1+k^2}}\;,
\end{equation}
where $k$ is determined by Eq.~\reff{detk}. Using
Eq.~\reff{Kapprox} one finds that for $|\bar\tau| \gg 1$
(i.e., in the limit of large film thickness $L$ at a fixed temperature)
\begin{equation}
-\frac{\delta\psi}{\bar \psi} = 4 e^{-\sqrt{-\bar\tau/2}}
 (1+O(e^{-\sqrt{-\bar\tau/2}})) \;,
\end{equation}
i.e., 
in the middle of the film the deviation
of the order parameter profile from the bulk value
due to the distant confining walls decays $\sim \exp[-L/(2\xi)]$.


\section{Relaxation of the response function}
\label{app-tI}

In the main text $\bar t_I = F_{t_I}(\bar x_{1\bot})$ has been
defined 
as the reduced 
time at which the inflection point of the scaling function 
$\Psi(\bar x_{1\bot},\bar
x_{2\bot},\bar t\,)$ for the mean-field 
response function (as a function of $\bar x_{2\bot}$, see Fig.~\ref{Rreal}) 
close to 
$\bar x_\bot=0$ disappears. The position of this point
$\bar x_{2\bot}^I = X_I(\bar x_{1\bot},\bar t\,)$  is determined by
the condition
\begin{equation}
\Psi^{(0,2)}(\bar x_{1\bot},\bar
x_{2\bot}^I,\bar t\,) = 0 \; ,
\label{D2Psi}
\end{equation}
where, here and in the following, we use the notation
$\Psi^{(n,m)}(\bar x_{1\bot},\bar
x_{2\bot},\bar t\,) \equiv \partial^n_{\bar x_{1\bot}} \partial^m_{\bar
x_{2\bot}} \Psi(\bar x_{1\bot},\bar
x_{2\bot},\bar t\,)$.
Accordingly, $\bar t_I$ is given by the time at which the inflection
point reaches the surface at $\bar x_\bot = 0$, i.e., it is implicitly
determined by the condition
$X_I(\bar x_{1\bot},\bar t_I) = 0$. This provides an equation
for the function $F_{t_I}$ (see Eq.~\reff{defFI}): 
\begin{equation}
X_I(\bar x_{1\bot}, F_{t_I}(\bar x_{1\bot})) = 0\; .
\label{implFtI}
\end{equation}
The function shown in Fig.~\ref{tflex} has
been determined by solving numerically Eqs.~\reff{D2Psi}
and~\reff{implFtI}. Here we determine analytically the behavior of 
$F_{t_I}(\bar x_{1\bot})$ for $\bar x_{1\bot} \rightarrow 0$ and $\bar
x_{1\bot} \rightarrow 1$. From Eq.~\reff{defpsi} it follows
that
\begin{equation}
\Psi(\bar x_{1\bot},\bar x_{2\bot},\bar t\,) = 
\Psi(\bar x_{2\bot},\bar x_{1\bot},\bar t\,), \quad
\Psi(\bar x_{1\bot}, \bar x_{2\bot},\bar t\,) = 
- \Psi(\bar x_{1\bot},-\bar x_{2\bot},\bar t\,)\;,
\label{proppsi1}
\end{equation}
and
\begin{equation}
\Psi(\bar x_{1\bot}= 0,\bar x_{2\bot},\bar t\,) = 
\Psi(\bar x_{1\bot}= 1,\bar x_{2\bot},\bar t\,) = 0\;.
\label{proppsi2}
\end{equation}
Accordingly $\Psi^{(0,m)}(\bar x_{1\bot}= 0,\bar
x_{2\bot},\bar t\,) = \Psi^{(0,m)}(\bar x_{1\bot}= 1,\bar
x_{2\bot},\bar t\,) = 0$ and $\Psi^{(0,2m)}(\bar x_{1\bot},\bar
x_{2\bot}=0,\bar t\,)=0$. (Eq.~\reff{proppsi1} implies that $\Psi$ is an
odd function of $\bar x_{2\bot}$ if analytically continued to negative
values of $\bar x_{2\bot}$ by using Eq.~\reff{defpsi}.)
Therefore Eq.~\reff{D2Psi} is always trivially  satisfied 
for $\bar x_{2\bot} = 0$, whatever the values of $\bar x_{1\bot}$ and
$\bar t$ are. The function $F_{t_I}$ is given by the nontrivial
solution of 
\begin{equation}
\Psi^{(0,2)}(\bar x_{1\bot},\bar x_{2\bot}\rightarrow 0, F_{t_I}(\bar x_{1\bot})) = 0 \;.
\end{equation}
This solution can be found by considering the series expansion of
$\Psi^{(0,2)}$ for $\bar x_{2\bot} \rightarrow 0$, i.e.,
\begin{equation}
\Psi^{(0,2)}(\bar x_{1\bot},\bar x_{2\bot}\rightarrow 0, F_{t_I}(\bar
x_{1\bot})) = \Psi^{(0,3)}(\bar x_{1\bot},0, F_{t_I}(\bar
x_{1\bot})) \; \bar x_{2\bot} + O(\bar x_{2\bot}^2) \; .
\end{equation}
Therefore
$\bar t_I = F_{t_I}(\bar x_{1\bot})$ 
is
determined 
by the
condition
\begin{equation}
\Psi^{(0,3)}(\bar x_{1\bot},0, \bar t_I) = 0 \;.
\label{impleqtI}
\end{equation}

Let us consider the limiting case $\bar x_{1\bot}\rightarrow 0$,
i.e., the behavior of $F_{t_I}(y\rightarrow 0)$.
As discussed
in Subsec.~\ref{sec-response} (see Eq.~\reff{tIorig}), in this limit 
one expects
$t_I \rightarrow 0$. Using Eqs.~\reff{defpsi} and~\reff{poisson}, 
one realizes
that in the limit $\bar t_I \rightarrow 0$ 
the terms of the sums with $n=0$ are the
leading contributions to $\Psi^{(0,3)}(\bar x_{1\bot},0, \bar
t_I)$. (The terms with $n\neq 0$
give rise to exponentially small corrections.) Thus Eq.~\reff{impleqtI}
reduces to
\begin{equation}
(\bar x_{1\bot}^2 - 6 \bar t_I)\bar x_{1\bot} = 0\;,
\end{equation} 
that, apart from the expected trivial solution $\bar x_{1\bot}=0$,
is solved by
\begin{equation}
t_I = F_{t_I}(\bar x_{1\bot}\rightarrow 0) = \frac{\bar x_{1\bot}^2}{6}
\; .
\end{equation}
The full numerical solution shown in  Fig.~\ref{tflex} is in
accordance with this analytic result.

We now consider the case $\bar x_{1\bot}\rightarrow 1$, i.e., 
$F_{t_I}(1)$. Expressing Eq.~\reff{impleqtI} as before, 
one
realizes that the leading contributions to $\Psi^{(0,3)}(\bar
x_{1\bot}\rightarrow 1,0, \bar
t_I)$ stem from the terms in the sums with $n=0,1$ (Eqs.~\reff{defpsi} and
\reff{poisson}). 
Indeed
Fig.~\ref{tflex} shows that $\bar t_I$ is significantly smaller than $1$
also 
for $\bar x_{1\bot}  = 1$, allowing one to neglect terms that are
suppressed by a factor $\sim
\exp(-1/\bar t_I)$ compared to the leading one. 
This leads to
\begin{equation}
\bar x_{1\bot}(\bar x_{1\bot}^2 -  6\bar t_I) 
- (2 - \bar x_{1\bot})\left[(2 - \bar
x_{1\bot})^2 - 6\bar t_I\right]\exp\left(\frac{\bar x_{1\bot}-1}{\bar t_I}\right) = 0 \;,
\end{equation}  
which has the expected trivial solution $\bar x_{1\bot}=1$, 
whereas the nontrivial one for $\bar x_{1\bot}\rightarrow 1$ is
given by
\begin{equation}
\bar t_I^2 - \bar t_I + \frac{1}{12} = 0 \;,
\label{quadratic}
\end{equation}
yielding
\begin{equation}
\bar t_I = F_{t_I}(\bar x_{1\bot} = 1) \simeq \frac{1}{2} -
\frac{1}{\sqrt{6}} \simeq 0.09176.
\end{equation}
This provides a very accurate approximation of the actual value 
$F_{t_I}(1)$. Note that the second solution of the quadratic 
equation~\reff{quadratic}
can be discarded because it is inconsistent with the assumption
$\bar t_I \ll 1$, under which Eq.~\reff{quadratic} has been derived. 
The actual value of $F_{t_I}(1)$ can be computed from
Eq.~\reff{impleqtI} after having discarded the trivial solution $\bar
x_{1\bot} = 1$. This can be done considering the expansion of
$\Psi^{(0,3)}$ around $\bar x_{1\bot}=1$, retaining only the leading
term. This yields the implicit equation
\begin{equation}
\Psi^{(1,3)}(1,0,\bar t_I) = 0\;.
\end{equation}
Here we do not report the corresponding expression that can be easily
worked out. The corresponding 
solution can be numerically determined and is given by
\begin{equation}
F_{t_I}(1) \simeq 0.0917918.
\end{equation}

\section{Computation of the Casimir force}
\label{app-casimir}

\subsection{General expression}

According to Eq.~\reff{casim-st} it is possible to compute the force
exerted on the confining walls 
in terms of the stress-tensor $T_{\bot\bot}$. In order to determine
the effect of a time-dependent external field on this force
we first note that according
to Eq.~\reff{Rdefbis} within Gaussian approximation,
due to $\langle \varphi \rangle_0 = 0$ and
$\langle [\varphi]^n[\tilde\varphi]^m\rangle_0 = 0$ for $m > n$, one
has
\begin{equation}
\begin{split}
\langle\varphi({\bf x}_1,t_1)\varphi({\bf x}_2,t_2)\rangle_h = &
\;\langle\varphi({\bf x}_1,t_1)\varphi({\bf x}_2,t_2)\rangle_0\\ 
&+
\int\!\!{\rm d}V'{\rm d}t'{\rm d}V''{\rm d}t''\,
h({\bf x}',t') h({\bf x}'',t'') \\
&\quad\quad\times 
\Omega^2 
\langle \tilde\varphi({\bf x}',t')\varphi({\bf x}_1,t_1)\rangle_0 
\langle \tilde\varphi({\bf x}'',t'')\varphi({\bf x}_2,t_2)\rangle_0 
\end{split}
\label{corr-pert}
\end{equation}
(higher order terms in $h$ are zero because $n=2$) 
where $\langle \cdot\rangle_h$ has been introduced after
Eq.~\reff{Rdef}. 
Using Eq.~\reff{corr-pert} one finds
\begin{equation}
\begin{split}
\langle T_{\bot\bot}({\bf x},t)\rangle_h|_{{\bf x}\in \partial V} &=\; \frac{1}{2}
\left.\partial_{x_{1\bot}}\partial_{x_{2\bot}} 
\langle\varphi({\bf x}_1,t)\varphi({\bf x}_2,t)\rangle_h \right|_{{\bf
x}_1={\bf x}_2 = {\bf x}\in \partial V}\\
&=\; \langle T_{\bot\bot}\rangle_0|_{\partial V} 
+\frac{1}{2}\left[\int{\rm d}V'{\rm d}t'\; h({\bf x}',t')\,\partial_{x_\bot}R^{(0)}({\bf
x}',t';{\bf x},t)|_{{\bf x}\in\partial V}\right]^2
\end{split}
\label{casimir-h}
\end{equation}
where the response function $R$ is defined in Eq.~\reff{Rdefbis}.
The previous equation provides the expression for the force density
$F_{l(r)} ({\bf x}_\|)$ acting on the left (right) plate (depending on
which part of the two boundaries 
$\partial V$ is used in Eq.~\reff{casimir-h}) for a
general external field. The previous expression can be expressed
in terms of the corresponding dimensionless scaling variables (see
Eqs.~\reff{scalRt}, \reff{amplitudeR}, \reff{RPsi}, \reff{scal_cas},
and \reff{scal_cas_dyn}):
\begin{equation}
\begin{split}
\FF^{(\rm dy)(0)}_l(\bar L,\bar t,\{\hat h\}) = \FF^{(\rm st)(0)}(0,\bar L) 
+& \frac{1}{8}
\Big[\int  \frac{{\rm d}^{d-1}\bar{\bf p}}{(2
\pi)^{d-1}} 
{\rm d}\bar
x_{1\bot}{\rm d}\bar t_1\; \hat h(\bar {\bf p},\bar x_{1\bot},\bar
t_1)\\
&\times e^{-i\bar{\bf p}\bar x_{2\|}} \partial_{\bar x_{2\bot}}\bar \RR^{(0)}(\bar
{\bf p},\bar x_{1\bot},\bar x_{2\bot}, \bar t -\bar t_1,\bar L)|_{\bar x_{2\bot}=0}
\Big]^2 
\end{split}
\label{casimir-h-bis}
\end{equation}
where the 
scaling variable $\hat h$ is
defined by
\begin{equation}
\hat h(\bar {\bf p},\bar x_{1\bot},\bar
t_1) \equiv  \xi_0^{(d+2)/2} (L/\xi_0)^{\beta\delta/\nu} L^{-(d-1)} h(\bar {\bf p}/L,\bar x_{1\bot} L,\bar
t_1 \rt_0 (\xi_0/L)^{-z})\;
\label{scalh-hat}
\end{equation}
with
\begin{equation}
h({\bf p},x_\bot, t) = \int{\rm d}^{d-1} x_\|\; h({\bf x}_\|,
x_\bot,t) \,{\rm e}^{i {\bf p}\cdot{\bf x}_\|}\;.
\end{equation}
$\hat h$ is the
analogue of $\bar h$ introduced after Eq.~\reff{scal_cas_dyn}. Note
that  the former carries, compared to the latter, an extra factor
$L^{-(d-1)}$, stemming from the Fourier transform in
the $d-1$ parallel spatial directions. Moreover, within
the Gaussian model ($g_0=0$) it is {\it  not} possible to define,
in accordance with Eq.~\reff{defh0}, the scale factor 
$\h_0$ for the external field $h$, 
whose engineering 
dimension  $\xi_0^{-(d+2)/2}$ follows from the Gaussian action.

Note that, according to Eq.~\reff{casimir-h}, the dependence of the
Casimir force on the applied field is quadratic and therefore the
effect of a sum of fields $h_1+h_2$ is {\it not} equal to
the sum of the separate effects of each field.

\subsection{Specific cases}

\label{sub-asyA}

In Subsec.~\ref{sec-casimir} we consider two different instances of
externally applied fields: a perturbation $h({\bf x},t) = h_W
\delta(x_\bot - x_{1\bot})\delta(t-t_1)$ which is spatially constant
in the
plane $x_\bot = x_{1\bot}$ 
parallel to the confining walls and a perturbation $h({\bf x},t) = h_P  \delta({\bf x}_\| - {\bf x}_{1\|}) \delta(x_\bot
- x_{1\bot})\delta(t-t_1)$ that is
localized at a point ${\bf x} = ({\bf x}_{1\|},x_{1\bot})$ within the
film.
In the following we present the details of the
computation of some relevant quantities determining the response to
these external fields.

\subsubsection{Planar perturbation}

Here we discuss the asymptotic behaviors of the amplitude
$A^W_\Delta(\bar x_{1\bot})$ defined in Sec.~\ref{sec-casimir} (see
Eq.~\reff{AAh}), which determines the Casimir force maximum  $(d-1)\Delta +
\hat h_W^2 A^W_\Delta(\bar x_{1\bot})$. 
In terms of the notation introduced after Eq.~\reff{D2Psi}, 
$A^W_\Delta$ is given by
\begin{equation}
A^W_\Delta(\bar x_{1\bot}) = \frac{1}{2}\left[\Psi^{(0,1)}(\bar
x_{1\bot},0,\bar t_I(\bar x_{1\bot}))\right]^2 \;.
\label{AAh-app}
\end{equation}
Using Eqs.~\reff{defpsi} and~\reff{poisson} one finds an expression of
$\Psi^{(0,1)}$ which is suited to discuss some asymptotic behaviors
at short times:
\begin{equation}
\Psi^{(0,1)}(\bar
x_{1\bot},0,\bar t\,) = -\frac{1}{\sqrt{\pi} \bar t^{3/2}}
\sum_{n=-\infty}^{+\infty} \left(n-\frac{\bar
x_{1\bot}}{2}\right)\exp\left[ -\frac{1}{\bar t}\left(n-\frac{\bar
x_{1\bot}}{2} \right)^2\right] \;.
\end{equation}
Accordingly,
\begin{equation}
\begin{split}
\Psi^{(0,1)}(\bar
x_{1\bot}\rightarrow 1,0,\bar t\,) =& 
\frac{(1-\bar x_{1\bot})}{\sqrt{\pi}\bar t^{3/2}} \\
&\times\sum_{n=-\infty}^{+\infty} \left[ -\frac{1}{2}+ \frac{(n-1/2)^2}{\bar t}
\right] \exp\left[ -\frac{(n-1/2)^2}{\bar t}\right] +
O((1-\bar x_{1\bot})^2)
\end{split}
\label{asy1}
\end{equation}
and
\begin{equation}
\Psi^{(0,1)}(\bar
x_{1\bot}\rightarrow 0,0,\bar t\,) = \frac{\bar x_{1\bot}}{2
\sqrt{\pi}\bar t^{3/2}} \left\{\exp[-\bar x_{1\bot}^2/(4\bar t\,)] +
O(e^{-1/\bar t})\right\} \;.
\label{asy2}
\end{equation}
(Note that the approximation provided by this expression deteriorates
upon increasing $\bar t$.)
In Appendix~\ref{app-tI} we found that, at leading orders,
 $\bar t_I(\bar x_{1\bot} \rightarrow 1) = F_{t_I}(1)\simeq 0.0918$ and
$\bar t_I(\bar x_{1\bot} \rightarrow 0) = \bar x_{1\bot}^2/6$. From
Eqs.~\reff{AAh-app} and \reff{asy1} one has 
\begin{equation}
A^W_\Delta(\bar x_{1\bot}\rightarrow 1) = a_1(1-\bar x_{1\bot})^2
\end{equation}
with
\begin{equation}
a_1 = {\mathcal L}_1(\bar t_I(1)) \simeq 17.535 
\label{def-a1}
\end{equation}
and where
\begin{equation}
{\mathcal L}_d(\bar t\,)= \frac{1}{(4\pi \bar t\,)^{d-1}}\frac{1}{2 \pi
\bar t^3} \left\{
\sum_{n=-\infty}^{+\infty} \left[ -\frac{1}{2}+ \frac{(n-1/2)^2}{\bar t}
\right] \exp\left[ -\frac{(n-1/2)^2}{\bar t}\right] \right\}^2 \;;
\label{def-L}
\end{equation}
this function will be useful also for the discussion of the localized
perturbation below.
In the opposite limit $\bar x_{1\bot}\rightarrow 0$ Eqs.~\reff{AAh-app}
and~\reff{asy2} yield
\begin{equation}
A^W_\Delta(\bar x_{1\bot}\rightarrow 0) = {\mathcal K}_1/\bar x_{1\bot}^4
\label{AdeltaWx0}
\end{equation}
with
\begin{equation}
{\mathcal K}_1 \simeq 0.427888 \;,
\label{def-a0}
\end{equation}
where we have introduced the constant
\begin{equation}
{\mathcal K}_d  = \frac{1}{2}\frac{[2(d+2)]^{d+2}}{(4\pi)^d}
e^{-(d+2)} \;.
\label{defKd}
\end{equation}
Moreover, the asymptotic expressions for $\Psi^{(0,1)}$ in
Eqs.~\reff{asy1} and~\reff{asy2}
provide the corresponding ones for the function $\FF^W_\Delta$ (see
Eq.~\reff{def-Fdelta}) because
\begin{equation}
\FF^W_\Delta(\bar x_{1\bot},\bar t\,) = \left[\frac{\Psi^{(0,1)}(\bar
x_{1\bot},0,\bar t\,)}{\Psi^{(0,1)}(\bar
x_{1\bot},0,\bar t_I(\bar x_{1\bot}))}\right]^2 \;.
\end{equation}
For $\bar x_{1\bot}\rightarrow 1$, Eq.~\reff{asy1}
gives
\begin{equation}
\FF^W_\Delta(\bar x_{1\bot}\rightarrow 1,\bar t\,) = \frac{{\mathcal L}_1(\bar t\,)}{{\mathcal L}_1(\bar t_I(1))}\;,
\label{FWx1}
\end{equation}
whereas, for $\bar x_{1\bot}\rightarrow 0$, Eq.~\reff{asy2}
and $\bar t_I(\bar x_{1\bot}\rightarrow 0) = \bar x_{1\bot}^2/6$ 
render
\begin{equation}
\FF^W_\Delta(\bar x_{1\bot}\rightarrow 0,\bar t\,) =
\left(\frac{e}{s}\right)^3 e^{-3/s} \;, \quad s = \bar t/\bar t_I(\bar
x_{1\bot}\rightarrow 0) \;.
\label{FWx0}
\end{equation}
These asymptotic behaviors are indicated in Fig.~\ref{plotscalh}.
Equation~\reff{Psi01asy-largex} 
allows one to determine the long-time behavior of
$\FF^W_\Delta(\bar x_{1\bot},\bar t\,)$ 
for fixed $\bar x_{1\bot}$:
\begin{equation}
\FF^W_\Delta(\bar x_{1\bot},\bar t\rightarrow\infty) \sim e^{-
2\pi^2\bar t} [1 + O(e^{-3\pi^2\bar t})] \;,
\label{FFWexpdecay}
\end{equation}
which clearly displays the expected exponential decay for $\bar t \gg
1/(3 \pi^2)$.

\subsubsection{Localized perturbation}

In Subsec.~\ref{sec-casimir} we have introduced the amplitude
$A^P_\Delta(\delta\bar{\bf x}_\| = {\bf x}_\| - {\bf x}_{1\|},\bar x_{1\bot})$ associated with the
response to a point-like field (see Eq.~\reff{AAhpoint}):
\begin{equation}
A^P_\Delta(\delta\bar{\bf
x}_\|,\bar x_{1\bot})  = \frac{1}{2}
\frac{e^{-(\delta\bar{\bf x}_\|)^2/[2
\bar t_M(\delta\bar{\bf
x}_\|, \bar
x_{1\bot})]}}{[4 \pi \, \bar t_M(\delta\bar{\bf
x}_\|, \bar
x_{1\bot})]^{d-1}}
\left[\Psi^{(0,1)} (\bar x_{1\bot},\bar t_M(\delta\bar{\bf
x}_\|, \bar
x_{1\bot}))|_{\bar x_{2\bot}=0}
\right]^2\;.
\label{AAhpoint-app}
\end{equation}
Here we discuss its asymptotic behavior 
for $|\delta\bar{\bf x}_\|| \rightarrow 0$.
The typical time $\bar t_M(\delta\bar{\bf
x}_\|,\bar x_{1\bot})$ when the effect of the perturbation
attains its maximum at a point with $|\delta \bar {\bf x}_\|| \ll 1$
 is expected to be of the same order as
that in the case of a planar perturbation at the same $\bar
x_{1\bot}$, which is
given
by the function $\bar t_I(\bar x_{1\bot})$ shown
in Fig.~\ref{tflex}. Accordingly, $\bar t_M(\delta\bar{\bf
x}_\| \rightarrow  0,\bar x_{1\bot}) \lesssim 0.1$ and therefore, in
Eq.~\reff{AAhpoint-app} we can use the expressions given in
Eqs.~\reff{asy1} and~\reff{asy2} for $\Psi^{(0,1)}$ in order to obtain the
asymptotic behaviors for $\bar x_{1\bot} \rightarrow 1$ and $\bar
x_{1\bot}\rightarrow 0$, respectively. We expect that, as it is the
case for $A^W_\Delta$ in Fig.~\ref{plotAmph}, they provide good
approximations of the actual dependence. 
Taking into account that $\bar t_M(\delta\bar{\bf
x}_\| \rightarrow 0,\bar x_{1\bot}\rightarrow 0) = [\bar x_{1\bot}^2 +
(\delta\bar{\bf
x}_\|)^2]/[2(d+2)]$ (Eq.~\reff{tMx0})
and using Eq.~\reff{asy2} one finds
\begin{equation}
A^P_\Delta(\delta\bar{\bf
x}_\| \rightarrow 0,\bar x_{1\bot} \rightarrow 0) = {\mathcal K}_d
\frac{\bar x_{1\bot}^2}{[\bar x_{1\bot}^2 +
(\delta\bar{\bf
x}_\|)^2]^{d+2}} \;,
\label{Apsmallxpxt}
\end{equation}
and in particular
\begin{equation}
A^P_\Delta(\delta\bar{\bf
x}_\| = 0,\bar x_{1\bot} \rightarrow 0) = {\mathcal K}_d/\bar
x_{1\bot}^{2(d+1)} \;.
\label{Apsmallxp0}
\end{equation}
For $d=3$ one has ${\mathcal K}_3 \simeq 0.169773$ 
whereas ${\mathcal K}_4 \simeq 0.148406$.
For $\bar x_{1\bot}\rightarrow 1$ we consider only the case
$\delta\bar{\bf
x}_\| = 0$. From
Eq.~\reff{asy1}, one finds 
\begin{equation}
A^P_\Delta(\delta\bar{\bf
x}_\| = 0,\bar x_{1\bot} \rightarrow 1) = 
a_d (1 -\bar x_{1\bot})^2 \;,
\label{Apxt1xp0}
\end{equation}
where (see Eq.\reff{def-L})
\begin{equation}
a_d \equiv {\mathcal L}_d(\bar t_M(\delta\bar{\bf x}_\| = 0,
\bar x_{1\bot}\rightarrow 1)) \;.
\label{defad}
\end{equation}
Taking into account that [see the discussion after Eq.~\reff{tMx1}]
$
\bar t_M(\delta\bar{\bf x}_\| = 0,\bar x_{1\bot} \rightarrow 1) = (4 -
\sqrt{11})/10 \simeq  0.0683375
$
for $d=3$ and
$
\bar t_M(\delta\bar{\bf x}_\| = 0,\bar x_{1\bot} \rightarrow 1) = (9 -
\sqrt{57})/20 \simeq  0.0604236
$
for $d=4$, one has
$
a_3 = {\mathcal L}_3(\bar t_M(\delta\bar{\bf
x}_\| = 0,\bar x_{1\bot}\rightarrow 1)) \simeq 17.927
$
and
$
a_4 = {\mathcal L}_4(\bar t_M(\delta\bar{\bf
x}_\| = 0,\bar x_{1\bot}\rightarrow 1)) \simeq  19.416$.

Let us discuss the behavior of $A^P_\Delta(\delta\bar{\bf
x}_\|,\bar x_{1\bot})$ for large $|\delta\bar{\bf
x}_\||$ and $0 < \bar x_{1\bot} < 1$ fixed. According to
Eq.~\reff{tM-large-xp} one has, in leading order, $\bar t_M(\delta\bar{\bf
x}_\| \rightarrow \infty,\bar x_{1\bot}) = |\delta\bar{\bf
x}_\||/(2\pi) - (d-1)/(4\pi^2) + O(1/|\delta\bar{\bf
x}_\||)$ 
and therefore one can use in
Eq.~\reff{AAhpoint-app} the asymptotic behavior of
$\Psi^{(0,1)}$ reported in Eq.~\reff{Psi01asy-largex}, finding
\begin{equation}
A^P_\Delta(\delta\bar{\bf
x}_\| \rightarrow \infty,\bar x_{1\bot}) = 2\pi^2 \frac{e^{- 2\pi
|\delta\bar{\bf x}_\||}}{(2|\delta\bar{\bf x}_\||)^{d-1}} \sin^2(\pi
\bar x_{1\bot})\;.
\label{RatioAlargexp}
\end{equation}
For $|\delta\bar{\bf x}_\||,\bar x_{1\bot} \rightarrow 0$ 
one obtains from Eq.~\reff{Apsmallxpxt}
\begin{equation}
\frac{A^P_\Delta(\delta\bar{\bf
x}_\|,\bar x_{1\bot})}{A^P_\Delta(\delta\bar{\bf
x}_\| = 0 ,\bar x_{1\bot})} = \frac{1}{\left[1+ (\delta \bar {\bf x}_\|)^2/\bar x_{1\bot}^2\right]^{d+2}}\;,
\label{Ratiosmallxpxt}
\end{equation}
which indeed provides a very good approximation of the actual curves
already for $|\delta\bar{\bf x}_\||,\bar x_{1\bot} \lesssim 0.5$ (see Fig.~\ref{plotRatioA3d}).

\section{Expansion of the in-plane correlation function}
\label{app-expansion}

In Subsec.~\ref{sec-correlation} we discuss 
the mean-field expression for the correlation function for 
points located within a 
plane
parallel to the confining walls, i.e., $C^{(0)}({\bf
p},x_\bot,x_\bot,\omega)$, with the corresponding scaling function given in
Eq.~\reff{Cinplane}. This expression depends actually on two variables:
$\bar a\equiv a L$ and $\bar x_\bot \equiv x_\bot/L$ ($0\le \bar
x_\bot \le 1$).  In order to discuss the behavior of
Eq.~\reff{Cinplane} for $|\bar a| \ll 1$ (so that $|\bar a (1-2\bar
x_\bot)|\ll 1$), we write
\begin{equation}
\frac{\cosh \bar a - \cosh \bar a(1-2 \bar x_\bot)}{\bar a\sinh \bar a} =
\sum_{n=0}^\infty \PP_n(\bar x_\bot) {\bar a}^{2 n}
\label{Cexp}
\end{equation}
where $\PP_n$ is a polynomial of degree $2n+2$, has the symmetry
$\PP_n(\bar x_\bot) = \PP_n(1-\bar x_\bot)$, and $\PP_n(0)=0$. From the series
expansion of the l.h.s. of Eq.~\reff{Cexp} one easily finds that
\begin{align}
\PP_0(x) &= 2 x(1-x) \;, \\
\PP_1(x) &= -\frac{2}{3} x^2 (1-x)^2 \;,\\
\PP_2(x) &= \frac{2}{45} x^2(1-x)^2(1+2 x - 2 x^2) \;,\\
\PP_3(x) &= -\frac{4}{945} x^2 (1-x)^2\left(1+ 2 x -\frac{x^2}{2} - 3 x^3
+\frac{3}{2}x^4\right)\;.
\end{align}
Taking into account that $\bar a^2 \equiv {\bar {\bf p}}^2 + {\bar L}^2 -
i\bar \omega$ (see Eq.~\reff{abar}), one finds that
\begin{equation}
\begin{split}
\Im &\frac{\cosh\bar a - \cosh\bar a(1-2 \bar x_\bot)}{\bar a\sinh\bar a} =\\ 
&\quad\quad =
\sum_{n=1}^\infty \PP_n(\bar x_\bot) \sum_{k=0}^{2k+1 \le n} (-1)^{k+1}
\left( \!{n \atop 2k+1}\!
\right) \left({\bar {\bf p}}^2 +{\bar L}^2\right)^{n-(2k+1)}{\bar \omega}^{2k+1} \;,
\end{split}
\end{equation}
and thus 
\begin{equation}
\begin{split}
&\CC^{(0)}(\bar {\bf p},\bar x_\bot,\bar x_\bot,\bar \omega,\bar L) = \\
&\quad\quad = 2 \sum_{n=1}^\infty \PP_n(x_\bot/L) \sum_{k=0}^{2k+1 \le n} (-1)^{k+1}
\left( \!{n \atop 2k+1}\!
\right) \left({\bar {\bf p}}^2 +{\bar L}^2\right)^{n-(2k+1)}{\bar \omega}^{2k} \;. 
\end{split}
\label{Cinplaneexp}
\end{equation}



\end{document}